\documentclass[journal,onecolumn]{IEEEtran}
\usepackage{amsfonts}
\usepackage{mathrsfs}
\usepackage{amsmath}
\usepackage{amssymb}
\usepackage[pdftex]{graphicx}
\usepackage{epic}
\usepackage{epstopdf}
\usepackage{color}
\usepackage{cite}
\usepackage{bbm}
\usepackage{multirow}
\usepackage{makecell}
\usepackage{mathtools}
\usepackage{enumerate}
\usepackage{lineno}
\usepackage{url}
\usepackage{float}
\usepackage{color}
\usepackage{bm}
\usepackage{booktabs}
\usepackage{tikz}
\usepackage{comment}
\usepackage[colorlinks, linkcolor=black, urlcolor = blue, citecolor = blue]{hyperref}

\newtheorem{theorem}{Theorem}
\newtheorem{definition}{Definition}
\newtheorem{lemma}{Lemma}

\newtheorem{remark}{Remark}
\newtheorem{observation}{Observation}
\newtheorem{corollary}{Corollary}
\newtheorem{claim}{Claim}

\begin{document}

\title{Binary Codes for Correcting Two Edits}

\author{Yubo~Sun and Gennian~Ge%
\thanks{The research of G. Ge was supported by the National Key Research and Development Program of China under Grant 2020YFA0712100, the National Natural Science Foundation of China under Grant 12231014, and Beijing Scholars Program.}
\thanks{Y. Sun ({\tt 2200502135@cnu.edu.cn}) and G. Ge ({\tt gnge@zju.edu.cn}) are with the School of Mathematical Sciences, Capital Normal University, Beijing 100048, China. ({\em Corresponding author:
Gennian Ge.})}
}

\maketitle

\begin{abstract}
    An edit refers to a single insertion, deletion, or substitution.
    This paper aims to construct binary codes that can correct two edits.
    To do  this, a necessary and sufficient condition for a code to be two-edit correctable is provided, showing that a code is a two-edit correcting code if and only if it can correct two deletions, up to two substitutions, and one deletion and up to one substitution, separately.
    This criterion allows for the construction of two-edit correcting codes leveraging these three types of error correcting codes.
    In the field of constructing codes for correcting two deletions, we present a construction with $4\log n+O(\log\log n)$ redundant bits that can be viewed as a subcode proposed by Guruswami and H{\aa}stad, and provide an alternative proof.
    Moreover, our two-deletion correcting codes can also correct up to two substitutions after making a slight modification. 
    In the field of constructing codes for correcting one deletion and up to one substitution, we present a construction with $4 \log n+O(\log\log n)$ redundant bits, which outperforms the best previously known results $6 \log n+O(1)$.
    Leveraging these codes, we obtain a construction of two-edit correcting codes with $6 \log n+O(\log\log n)$ redundant bits.
    This outperforms the best previously known result, which requires at least $8\log n$ redundant bits.
    Moreover, we also consider the list-decoding problem under the two-edit channel and construct a two-edit list-decodable code with a list size of two employing $4 \log n+O(\log\log n)$ redundant bits.
\end{abstract}


\IEEEpeerreviewmaketitle

\section{Introduction}
\IEEEPARstart{I}{nsertions}, deletions, and substitutions are common types of errors in DNA-based storage systems \cite{Heckel-19-background}. According to a recent study conducted by Organick et al. \cite{Organick-18-background}, the rates of insertions, deletions, and substitutions were reported as $5.4 \times 10^{-4}$, $1.5 \times 10^{-3}$, and $4.5 \times 10^{-3}$, respectively. 
These rates represent the probabilities of these different types of errors occurring within a given sequence, as observed in the experiment.
Therefore, investigating the construction of codes that can correct a combination of insertions, deletions, and substitutions is of great significance.
It is well known that correcting deletions alone is already a challenging task, hence the task becomes even more complicated when attempting to correct deletions alongside other types of errors.

A code is referred to as a $t$-edit correcting code if it can correct a combination of insertions, deletions, and substitutions, as long as the total number of errors is at most $t$. In 1966, Levenshtein initiated the study of constructing edit-correcting codes and proposed a single-edit correcting code with $\log n+1$ redundant bits using the standard VT syndrome \cite{Levenshtein-66-SPD-1D}. In the same paper, Levenshtein showed that the optimal redundancy of codes capable of correcting $t$ deletions is lower bounded by $(t+o(1))\log n$. Remarkably, any $t$-edit correcting code must be able to correct $t$ deletions, implying that the optimal redundancy of $t$-edit correcting codes is also lower bounded by $(t+o(1))\log n$.
Therefore, for $t=1$, Levenshtein's construction is asymptotically optimal.
However, when $t \geq 2$, explicitly constructing (asymptotically) optimal $t$-edit correcting codes remains an open problem. 
Several constructions with low redundancy have been presented in \cite{Haeupler-19-FOCS-kE,Cheng-18-FOCS-kE,Sima-20-ISIT-tD}.
In particular, the code presented in \cite{Sima-20-ISIT-tD} represents the current best-known result in this area, having a redundancy of at least $4t \log n$.
There is still a significant gap between the best-known construction and the lower bound, even for the special case when $t=2$. 
This motivates further efforts to improve the construction of two-edit correcting codes.

In this paper, we focus on studying two-edit correcting codes over the binary alphabet. 
Specifically, we are interested in designing codes with the smallest possible redundancy, ensuring that when a codeword $\boldsymbol{x}$ of length $n$ is transmitted through a two-edit channel, the receiver can effectively reconstruct the original codeword from the received corrupted version $\boldsymbol{y}$.
An important observation is that two edits can consist of six different types of errors: namely two insertions, one insertion and up to one substitution, one insertion and one deletion, up to two substitutions, one deletion and up to one substitution, and two deletions. 
If the specific type of errors occurred in $\boldsymbol{x}$ can be determined from (the length of) $\boldsymbol{y}$, we can investigate these six types of error-correcting codes separately and construct two-edit correcting codes by simply taking the intersection of these codes.
However, when the length of $\boldsymbol{y}$ equals $n$, it becomes indeterminate whether the type of error is one insertion and one deletion or up to two substitutions.
This ambiguity makes designing two-edit correcting codes challenging.
In this paper, we present an unexpected result, demonstrating that this ambiguity in error types does not affect the unique reconstruction of the original sequence when the code can correct one deletion and up to one substitution.
This finding allows us to handle these errors independently.
Levenshtein \cite{Levenshtein-66-SPD-1D} showed that a code can correct two deletions if and only if it can correct a combination of insertions and deletions, as long as the total number of errors is at most two, and Smagloy et al. \cite{Smagloy-20-ISIT-DS} showed that a code can correct one deletion and up to one substitution if and only if it can correct one insertion and up to one substitution.
Therefore, in designing two-edit correcting codes, it is  sufficient to focus on codes that can correct three types of errors separately: namely two deletions, up to two substitutions, and one deletion and up to one substitution.

A code that can correct $t$ deletions is referred to as a $t$-deletion correcting code.
In the field of constructing $t$-deletion correcting codes, a recent breakthrough by Brakensiek et al. \cite{Brakensiek-18-IT-kD} has sparked a series of advancements in this area \cite{Sima-21-IT-kD,Sima-20-ISIT-tD,Song-22-IT-DS,Li-23-DS,Gabrys-19-IT-2D,Sima-20-IT-2D,Guruswami-21-IT-2D}.
The codes presented in \cite{Gabrys-19-IT-2D,Sima-20-IT-2D,Guruswami-21-IT-2D} exclusively focus on correcting two deletions, which is also one of the main focus of this paper. 
In particular, the code presented in \cite{Guruswami-21-IT-2D} represents the current best-known result in this area, having a redundancy of $4\log n+O(\log\log n)$. 
Recently, Khuat et al. \cite{Khuat-23-2D} claimed that they constructed codes capable of correcting one insertion and one deletion, which are also capable of correcting two deletions, with $2\log n+O(1)$ redundant bits. 
However, there is a flaw in their code construction that may not be fixed.
A counterexample is that the sequences $001001$ and $000110$ can appear in their code simultaneously as these two sequences share a common subsequence $0000$, which contradicts the ability of the code to correct two deletions.
Our main contribution in this field is presenting a construction with $4\log n+O(\log\log n)$ redundant bits that can be viewed as a subcode proposed by Guruswami and H{\aa}stad \cite{Guruswami-21-IT-2D}, and providing an alternative proof.
Since our approach uses a different technique from the one in \cite{Guruswami-21-IT-2D}, it may be of independent interest.

A code that can correct $t_1$ deletions and $t_2$ substitutions is referred to as a $t_1$-deletion $t_2$-substitution correcting code.
The field of constructing codes for correcting a combination of deletions and substitutions is initiated by Smagloy et al. \cite{Smagloy-20-ISIT-DS,Smagloy-23-IT-DS}, they investigated single-deletion single-substitution correcting codes and presented a construction with $6\log n+O(1)$ redundant bits. 
Later Song et al. \cite{Song-22-IT-DS} generalized Smagloy et al.'s construction and constructed single-deletion $t$-substitution correcting codes with $(t+1)(2t+1) \log n+O(1)$ redundant bits. 
In the same paper, Song et al. constructed $t_1$-deletion $t_2$-substitution correcting codes with $(4t_1+3t_2+o(1)) \log n$ redundant bits by using the syndrome compression technique with a precoding technique.
Another variant, which relaxes the decoding requirement, is the list decoding problem under the deletions and substitutions channel. This problem has also been explored in the papers \cite{Gabrys-23-IT-DS} and \cite{Song-22-arXiv-DS}.
Specifically, Gabrys et al. \cite{Gabrys-23-IT-DS} and Song et al. \cite{Song-22-arXiv-DS} constructed single-deletion single-substitution list-decodable codes with a list size of two employing $4 \log n+O(1)$ and $3 \log n+O(1)$ redundant bits, respectively.
Our main contribution in this field is constructing single-deletion single-substitution correcting codes with $4 \log n+O(\log\log n)$ redundant bits by using the ideas developed in the context of our two-deletion correcting codes.

The task of correcting substitutions is relatively easier than correcting errors related to deletions. We show that our two-deletion correcting codes can also correct up to two substitutions after making a slight modification. Therefore, by combining the constructions of our two-deletion correcting codes and single-deletion single-substitution correcting codes, we construct two-edit correcting codes with $6 \log n+O(\log\log n)$ redundant bits. This outperforms the best previously known result, which requires at least $8\log n$ redundant bits.
Moreover, we also consider the list-decoding problem under the two-edit channel and construct two-edit list-decodable codes with a list size of two employing $4\log n+O(\log\log n)$ redundant bits.

The rest of this paper is organized as follows. Section \ref{sec:pre} introduces the relevant notations,  definitions, and important tools used throughout the paper as well as the previous related results.
Section \ref{sec:condition} presents a necessary and sufficient condition for a code to be two-edit correctable.
Sections \ref{sec:2D}, \ref{sec:2S}, \ref{sec:1D1S}, and \ref{sec:2E} focus on the construction of codes that can correct two deletions, up to two substitutions, one deletion and up to one substitution, and two edits, respectively.
Finally Section \ref{sec:conclusion} concludes the paper.

\section{preliminaries}\label{sec:pre}
\subsection{Notations}

Let $\Sigma$ denote the alphabet set $\{0, 1\}$, and $\Sigma^n$ denote the set of all sequences of length $n$ over $\Sigma$. We represent sequences using bold letters, such as $\boldsymbol{x}$ and $\boldsymbol{y}$, and symbols within a sequence using plain letters, such as $x_1$ and $y_1$. The $i$-th entry of a sequence $\boldsymbol{x} \in \Sigma^n$ is denoted as $x_i$, and the sequence $\boldsymbol{x}$ can be expressed as either $(x_1, x_2, \ldots, x_n)$ or $x_1 x_2 \cdots x_n$.
The \emph{complement} of $\boldsymbol{x}$ is denoted as $\overline{\boldsymbol{x}} \triangleq \overline{x}_1 \overline{x}_2 \cdots \overline{x}_n$, where $\overline{x}_i=1-x_i$ for $1\leq i\leq n$, and the \emph{reverse} of $\boldsymbol{x}$ is denoted as $R(\boldsymbol{x}) \triangleq x_n x_{n-1} \cdots x_1$.
If $\boldsymbol{y}\in \Sigma^n$ is also a sequence, denote $\boldsymbol{xy}$ as their \emph{concatenation} $x_1 \cdots x_n y_1 \cdots y_n$.
For two positive integers $i$ and $j$, $[i, j]$ refers to the set of integers $\{i, i+1, \ldots, j\}$ when $j\geq i$, and the empty set when $j<i$. Additionally, $[i, j]$ can be written as $[j]$ and $[[j]]$ when $i = 1$ and $i=0$, respectively.
For a set of indices $\mathcal{I} = \{i_1, i_2, \ldots, i_t\}$ with $1 \leq i_1 < i_2 < \cdots < i_t \leq n$, $\boldsymbol{x}_{\mathcal{I}}$ represents the projection of $\boldsymbol{x}$ onto the indices of $\mathcal{I}$. In other words, $\boldsymbol{x}_{\mathcal{I}} = x_{i_1}x_{i_2}\cdots x_{i_t}$. In this case, $\boldsymbol{x}_{\mathcal{I}}$ is referred to as a \textit{$t$-subsequence} of $\boldsymbol{x}$, and $\boldsymbol{x}$ is called an \textit{$n$-supersequence} of $\boldsymbol{x}_{\mathcal{I}}$. When $i_{j+1} = i_j + 1$ for $1 \leq j \leq t-1$, i.e., $\mathcal{I}$ forms an interval, $\boldsymbol{x}_{\mathcal{I}}$ is called a \textit{$t$-consecutive-subsequence} of $\boldsymbol{x}$.

\subsection{Error Models and Codes}

We begin by describing the types of errors considered in this paper.
Given a sequence $\boldsymbol{x} \in \Sigma^n$, the operations of insertions, deletions, substitutions, and edits can transform it into different sequences. 
Specifically, one insertion transforms it into one of its $(n+1)$-supersequences, one deletion transforms it into one of its $(n-1)$-subsequences, one substitution transforms it into a sequence in $\Sigma^n$ that differs from $\boldsymbol{x}$ in exactly one position, and one edit refers to any operation of one insertion, deletion, or substitution.

Now, we define the concept of a correcting code for these types of errors. Given a sequence $\boldsymbol{x}$, suppose we have non-negative integers $t_1$, $t_2$, and $t_3$, we define $\mathcal{B}_{t_1,t_2,t_3}(\boldsymbol{x})$ as the set of sequences obtained from $\boldsymbol{x}$ by performing exactly $t_1$ insertions, exactly $t_2$ deletions, and up to $t_3$ substitutions, where these errors can occur in any order. 
Moreover, for a positive integer $t$, we define $\mathcal{B}_t(\boldsymbol{x})$ as the set of sequences obtained from $\boldsymbol{x}$ by performing up to $t$ edits, where these errors can occur in any order. 
A code $\mathcal{C} \subseteq \Sigma^n$ is referred to as a \emph{$t_1$-insertion $t_2$-deletion $t_3$-substitution correcting code} if, for any two distinct sequences $\boldsymbol{x}, \boldsymbol{y} \in \mathcal{C}$, it holds that $\mathcal{B}_{t_1,t_2,t_3}(\boldsymbol{x}) \cap \mathcal{B}_{t_1,t_2,t_3}(\boldsymbol{y}) = \emptyset$. In other words, the code ensures that any pair of distinct sequences in $\mathcal{C}$ have disjoint sets of sequences that can be obtained from them by applying the allowed operations. This property is essential for unique decoding.
If one or more of the values $t_1$, $t_2$, or $t_3$ is zero, the corresponding type of error is not considered by the code and will be omitted from the name of the code. In such cases, the name of the code reflects the error correction capability related to the remaining types of errors. 
For example, if $t_1 = 0$, the code will be referred to as a \emph{$t_2$-deletion $t_3$-substitution correcting code}, indicating that it can only correct deletions and substitutions but not insertions. 
Similarly, if $t_2 = 0$, it will be called a \emph{$t_1$-insertion $t_3$-substitution correcting code}, and if $t_3 = 0$, it will be called a \emph{$t_1$-insertion $t_2$-deletion correcting code}.
Moreover, $\mathcal{C}$ is called a \emph{$t$-edit correcting code} if for any two distinct sequences $\boldsymbol{x}, \boldsymbol{y} \in \mathcal{C}$, it holds that $\mathcal{B}_t(\boldsymbol{x}) \cap \mathcal{B}_{t}(\boldsymbol{y})= \emptyset$.

In addition to unique decoding, we also consider the list decoding capability of the code.
For a positive integer $L$, a code $\mathcal{C}$ is referred to as a \emph{$t_1$-insertion $t_2$-deletion $t_3$-substitution list-decodable code with a list size of $L$}, if for any $L+1$ distinct sequences $\boldsymbol{x}_1, \boldsymbol{x}_2, \ldots, \boldsymbol{x}_{L+1} \in \mathcal{C}$, it holds that $\cap_{i=1}^{L+1} \mathcal{B}_{t_1,t_2,t_3}(\boldsymbol{x}_i) = \emptyset$, and $\mathcal{C}$ is called a \emph{$t$-edit list-decodable code with a list size of $L$}, if for any $L+1$ distinct sequences $\boldsymbol{x}_1, \boldsymbol{x}_2, \ldots, \boldsymbol{x}_{L+1} \in \mathcal{C}$, it holds that $\cap_{i=1}^{L+1} \mathcal{B}_t(\boldsymbol{x}_i)= \emptyset$.
Similarly, if one or more of the values $t_1$, $t_2$, or $t_3$ is zero, the corresponding type of error is not considered by the code and will be omitted from the name of the code.

To evaluate a code $\mathcal{C}$, we consider its \emph{redundancy} $r(\mathcal{C}) \triangleq n-\log|\mathcal{C}|$, where the logarithm base is $2$ and $|\mathcal{C}|$ denotes the \emph{cardinality} of $\mathcal{C}$.

\subsection{Useful Tools}

In this subsection, we introduce the main tools used throughout the paper.

\subsubsection{High Order VT Syndromes}
For non-negative integer $k$, the \textit{$k$-th order VT syndrome} of $\boldsymbol{x} \in \Sigma^n$ is denoted as $\mathrm{VT}^{(k)}(\boldsymbol{x})=\sum_{i=1}^n c_i^{(k)} x_{i}$, where $c_i^{(k)}= \sum_{j=1}^i j^{k-1}$ if $k \geq 1$ and $c_i^{(k)}= 1$ if $k=0$. 

VT syndrome is a powerful tool in designing codes for correcting errors related to deletions. 
It has been extensively used in various research lines for designing deletion-correcting codes.
One notable example of these is the work of Levenshtein, who showed that binary sequences with the same first-order VT syndrome modulo $n+1$ form a single-deletion correcting code \cite{Levenshtein-66-SPD-1D}.
Furthermore, recent works  \cite{Sima-21-IT-kD,Sima-20-ISIT-tD,Song-22-IT-DS,Sima-20-IT-2D,Guruswami-21-IT-2D} have demonstrated the effectiveness of using higher-order VT syndromes in constructing multiple-deletion correcting codes.

\subsubsection{Differential Sequences}
For any sequence $\boldsymbol{x} \in \Sigma^n$, set $x_0=0$ and $x_{n+1}=0$, its \textit{differential sequence} is defined as $\psi(\boldsymbol{x})= (\psi(x)_1, \psi(x)_2, \ldots, \psi(x)_{n+1}) \in \Sigma^{n+1}$, where $\psi(x)_i= x_i \oplus x_{i-1} \triangleq x_{i}- x_{i-1} \pmod{2}$, for any $1 \leq i \leq n+1$.
Clearly, each sequence is uniquely determined by its differential sequence and vice-versa.

Simply applying VT syndrome directly to a binary sequence may not always be enough to correct errors related to deletions. 
In such cases, it is important to explore alternative approaches.
In \cite{Sun-23-IT-DR}, we have demonstrated the effectiveness of applying VT syndrome on differential sequence derived from the original binary sequence to construct codes for correcting two deletions in the reconstruction model.
In this paper, we will further show that applying VT syndrome on differential sequence can also be used to construct two-deletion correcting codes as well as single-deletion single-substitution correcting codes.

Below we determine the types of errors that occur in $\psi(\boldsymbol{x})$ as well as the number of ones changed in $\psi(\boldsymbol{x})$ when $\boldsymbol{x}$ suffers from errors.

\begin{observation}\label{obs:deletion} 
  Suppose $\boldsymbol{x} \in \Sigma^n$ suffers one deletion at the $i$-th position, i.e., $\cdots x_{i-1} x_i x_{i+1} \cdots \rightarrow \cdots x_{i-1} x_{i+1} \cdots$, it holds that $\psi(x)_i \psi(x)_{i+1}$ is replaced by $\psi(x)_i \oplus \psi(x)_{i+1}$ and the other entries in $\psi(\boldsymbol{x})$ remain unchanged, i.e., $\cdots \psi(x)_i \psi(x)_{i+1} \cdots \rightarrow \cdots \psi(x)_i \oplus \psi(x)_{i+1} \cdots$.
  Therefore, when $\boldsymbol{x}$ suffers one deletion, the number of ones in $\psi(\boldsymbol{x})$ either decreases by two or does not change. More precisely,
  \begin{itemize}
    \item if one deletion in $\boldsymbol{x}$ does not change the number of ones in $\psi(\boldsymbol{x})$, then the type of error occurs in $\psi(\boldsymbol{x})$ is either $00 \rightarrow 0$, $10 \rightarrow 1$, or $01 \rightarrow 1$;
    \item if one deletion in $\boldsymbol{x}$ decreases the number of ones in $\psi(\boldsymbol{x})$ by two, then the type of error occurs in $\psi(\boldsymbol{x})$ is $11 \rightarrow 0$.
  \end{itemize}
  Moreover, suppose $\boldsymbol{x} \in \Sigma^n$ suffers two deletions, where the erroneous positions are $i$ and $i+1$, then $\psi(x)_i \psi(x)_{i+1} \psi(x)_{i+2}$ is replaced by $\psi(x)_i \oplus \psi(x)_{i+1} \oplus \psi(x)_{i+2}$ and the other entries in $\psi(\boldsymbol{x})$ remain unchanged, i.e., $\cdots \psi(x)_i \psi(x)_{i+1} \psi(x)_{i+2} \cdots \rightarrow \cdots \psi(x)_i \oplus \psi(x)_{i+1} \oplus \psi(x)_{i+2} \cdots$. In this case, the number of ones in $\psi(\boldsymbol{x})$ either decreases by two or does not change.
\end{observation}

\begin{observation}\label{obs:substitution}
  Suppose $\boldsymbol{x} \in \Sigma^n$ suffers one substitution at the $i$-th position, i.e., $\cdots x_{i-1} x_i x_{i+1} \cdots \rightarrow \cdots x_{i-1} \overline{x}_i x_{i+1} \cdots$, it holds that $\psi(x)_i \psi(x)_{i+1} \rightarrow \overline{\psi(x)_i \psi(x)_{i+1}}$ and the other entries in $\psi(\boldsymbol{x})$ remain unchanged, i.e., $\cdots \psi(x)_i \psi(x)_{i+1} \cdots \rightarrow \cdots \overline{\psi(x)_i \psi(x)_{i+1}} \cdots$. 
  Therefore, when $\boldsymbol{x}$ suffers one substitution, the number of ones in $\psi(\boldsymbol{x})$ either decreases by two, does not change, or increases by two. More precisely,
  \begin{itemize}
    \item if one substitution in $\boldsymbol{x}$ increases the number of ones in $\psi(\boldsymbol{x})$ by two, then the type of error occurs in $\psi(\boldsymbol{x})$ is $00 \rightarrow 11$;
    \item if one substitution in $\boldsymbol{x}$ does not change the number of ones in $\psi(\boldsymbol{x})$, then the type of error occurs in $\psi(\boldsymbol{x})$ is either $01 \rightarrow 10$ or $10 \rightarrow 01$;
    \item if one substitution in $\boldsymbol{x}$ decreases the number of ones in $\psi(\boldsymbol{x})$ by two, then the type of error occurs in $\psi(\boldsymbol{x})$ is $11 \rightarrow 00$.
  \end{itemize}
\end{observation}

\subsubsection{\texorpdfstring{$P$}{}-Bounded Error-Correcting Codes}
Given non-negative integers $t_1$, $t_2$, $t_3$, and $P$ (typically, $P$ is set to be of order $\log n$), $\mathcal{C} \subseteq \Sigma^n$ is referred to as a \emph{$P$-bounded $t_1$-insertion $t_2$-deletion $t_3$-substitution correcting code} if, for any two distinct sequences $\boldsymbol{x}=\boldsymbol{u} \tilde{\boldsymbol{x}} \boldsymbol{v}, \boldsymbol{y}= \boldsymbol{u} \tilde{\boldsymbol{y}} \boldsymbol{v} \in \mathcal{C}$ with $|\tilde{\boldsymbol{x}}|= |\tilde{\boldsymbol{y}}|\leq P$, it holds that $\mathcal{B}_{t_1,t_2,t_3}(\tilde{\boldsymbol{x}}) \cap \mathcal{B}_{t_1,t_2,t_3}(\tilde{\boldsymbol{y}})= \emptyset$.
If one or more of the values $t_1$, $t_2$, or $t_3$ is zero, the corresponding type of error is not considered by the code and will be omitted from the name of the code.


Leveraging positional information allows for the construction of $P$-bounded error-correcting codes with reduced redundancy compared to general error-correcting codes. 
Specifically, these codes may be constructed using only $O(\log P)$ bits of redundancy.
For example, Schoeny et al. \cite{Schoeny-17-IT-BD} proposed a construction of $P$-bounded single-deletion correcting codes with $\log P + 1$ bits of redundancy, which is a significant improvement compared to the general lower bound of $\log n$ for single-deletion correcting codes.
The low cost of using $P$-bounded error-correcting codes makes them applicable in various fields \cite{Sun-23-IT-DR, Sun-23-IT-BD, Sun-23-IT-BDR}. 
In this paper, we use $P$-bounded error-correcting codes to assist in constructing error-correcting codes.

\subsubsection{Strong-Locally-Balanced Sequences}

Let $\ell$ be a positive integer and $0<\epsilon< \frac{1}{2}$. A sequence $\boldsymbol{x} \in \Sigma^n$ is referred to as an \textit{$(\ell,\epsilon)$-locally-balanced sequence} if the number of ones in each $\ell$-consecutive-subsequence of $\boldsymbol{x}$ falls within the interval $[(\frac{1}{2}-\epsilon) \ell, (\frac{1}{2}+\epsilon) \ell]$. 
In other words, for any $1 \leq i \leq n-\ell+1$, we have $(\frac{1}{2}-\epsilon) \ell \leq \mathrm{VT}^{(0)}(\boldsymbol{x}_{[i,i+\ell-1]}) \leq (\frac{1}{2}+\epsilon) \ell$, where $\mathrm{VT}^{(0)}(\boldsymbol{x}_{[i,i+\ell-1]})$ denotes the number of ones in the subsequence $\boldsymbol{x}_{[i,i+\ell-1]}$.
A code $\mathcal{C} \subseteq \Sigma^n$ is referred to as an \textit{$(\ell,\epsilon)$-locally-balanced code} if any sequence $\boldsymbol{x} \in \mathcal{C}$ is an $(\ell,\epsilon)$-locally-balanced sequence.
Moreover, $\boldsymbol{x} \in \Sigma^n$ is referred to as a \emph{strong-$(\ell,\epsilon)$-locally-balanced sequence} if it is an $(\ell',\epsilon)$-locally-balanced sequence for any $\ell'\geq \ell$, and $\mathcal{C} \subseteq \Sigma^n$ is referred to as a \textit{strong-$(\ell,\epsilon)$-locally-balanced code} if any sequence $\boldsymbol{x} \in \mathcal{C}$ is a strong-$(\ell,\epsilon)$-locally-balanced sequence.

Strong-locally-balanced codes have been shown to be suitable candidates for various applications, such as simultaneous energy and information transfer \cite{Tandon-16-IT-SCC,Nguyen-21-IT-LBCC} and DNA storage \cite{Wang-22-ISIT-LBCC,Gabrys-20-ISIT-LBCC}, to enhance the reliability and accuracy of information transmission.
Instead of correcting errors, constrained codes are commonly used to reduce the probability of errors. 
In this paper, we demonstrate the potential of strong-locally-balanced codes in constructing error-correcting codes.
The next lemma shows that the usage of strong-locally-balanced codes incurs a negligible cost.

\begin{lemma}[Claim 4 of \cite{Gabrys-18-IT-D_T}]
  Suppose $n \geq 10$ and $b \geq 5$, set $\ell= b^4 \log n$ and $\epsilon= \frac{1}{3b}$, there exists a strong-$(\ell,\epsilon)$-locally-balanced code of size at least $2^{n}(1-2n^{2-\frac{2}{9} b^2 \log e})$.
\end{lemma}

\begin{corollary}\label{cor:balance}
   Suppose $n\geq 10$ and $b=6$, set $\ell= b^4 \log n= 1296 \log n$, and $\epsilon= \frac{1}{3b}= \frac{1}{18}$,
   the size of $\mathcal{S} \triangleq \{ \boldsymbol{x}\in \Sigma^n: \boldsymbol{x} \text{ and } \psi(\boldsymbol{x}) \text{ are strong-$(\ell,\epsilon)$-locally-balanced sequences}\}$ is at least
   $2^{n-1}$. 
\end{corollary}

\begin{IEEEproof}
    Let $\mathcal{S}_1 \triangleq \{ \boldsymbol{x}\in \Sigma^n: \boldsymbol{x} \text{ is a strong-$(\ell,\epsilon)$-locally-balanced sequence}\}$ and $\mathcal{S}_2 \triangleq \{\boldsymbol{x}\in \Sigma^n: \psi(\boldsymbol{x})\in \mathcal{S}_1\}$, we have $|\mathcal{S}_1|= |\mathcal{S}_2|\geq 2^{n}(1-2n^{2-\frac{2}{9} b^2 \log e})$.
    Then we may compute
    \begin{align*}
        |\mathcal{S}|
        = |\mathcal{S}_1 \cap \mathcal{S}_2|
        = |\mathcal{S}_1|+ |\mathcal{S}_2|- |\mathcal{S}_1 \cup \mathcal{S}_2| 
        \geq 2^{n}\left(2-4n^{2-\frac{2}{9} b^2 \log e}\right)- 2^n
        = 2^{n}\left(1-4n^{2-\frac{2}{9} b^2 \log e}\right)
        \geq 2^{n-1}.
    \end{align*}
\end{IEEEproof}

Below we introduce a possible way to encode the code $\mathcal{S}$ (defined in the above corollary) with only one bit of redundancy. 
We first establish the connection between locally-balanced codes and strong locally-balanced codes.

\begin{lemma}\label{lem:link}
    Let $0< \tilde{\epsilon}< \frac{1}{2}$ and $s\geq 1$ be such that $0< \tilde{\epsilon} + \frac{1- 4\tilde{\epsilon}^2}{4s} \leq \epsilon<\frac{1}{2}$. If $\boldsymbol{x} \in \Sigma^n$ is an $(\tilde{\ell},\tilde{\epsilon})$-locally-balanced sequence, then it is also a strong-$(\ell,\epsilon)$-locally-balanced sequence when $\ell \geq s\tilde{\ell}$.
\end{lemma}

\begin{IEEEproof}
    Assume $\boldsymbol{y}$ is a consecutive subsequence of $\boldsymbol{x}$ of length $\ell'\geq \ell$, let $t\equiv \ell' \pmod{\tilde{\ell}} \in [[\tilde{\ell}-1]]$.
    Since $\boldsymbol{x}$ is an $(\tilde{\ell},\tilde{\epsilon})$-locally-balanced sequence, we may compute 
      \begin{align*}
        \mathrm{VT}^{(0)}(\boldsymbol{y}) 
        &\geq \left( \frac{1}{2}-\tilde{\epsilon} \right) (\ell'-t)+ \max \left\{ 0,t-\left( \frac{1}{2}+\tilde{\epsilon} \right) \tilde{\ell} \right\} \\
        &\geq \left( \frac{1}{2}-\tilde{\epsilon} \right) \left(\ell'- \left( \frac{1}{2}+\tilde{\epsilon} \right)\tilde{\ell} \right) \\
        &\geq \left( \frac{1}{2}-\tilde{\epsilon} \right) \left(\ell'- \left( \frac{1}{2}+\tilde{\epsilon} \right)\frac{\ell'}{s} \right)\\
        &= \left( \frac{1}{2}-\tilde{\epsilon} - \frac{1-4\tilde{\epsilon}^2}{4s} \right)\ell'\\
        &\geq \left( \frac{1}{2}-\epsilon\right) \ell',
      \end{align*}
      and
      \begin{align*}
        \mathrm{VT}^{(0)}(\boldsymbol{y}) 
        &\leq \left( \frac{1}{2}+\tilde{\epsilon} \right)(\ell'-t)+ \min\left\{t,\left( \frac{1}{2}+\tilde{\epsilon} \right)\tilde{\ell} \right\}\\
        &\leq \left( \frac{1}{2}+\tilde{\epsilon} \right) \left(\ell'- \left( \frac{1}{2}+\tilde{\epsilon} \right)\tilde{\ell} \right)+ \left( \frac{1}{2}+\tilde{\epsilon} \right)\tilde{\ell} \\
        &\leq \left( \frac{1}{2}+\tilde{\epsilon} \right) \left(\ell'+ \left( \frac{1}{2}-\tilde{\epsilon} \right)\frac{\ell'}{s} \right)\\
        &= \left( \frac{1}{2}+\tilde{\epsilon}+ \frac{1-4\tilde{\epsilon}^2}{4s} \right)\ell'\\
        &\leq \left( \frac{1}{2}+\epsilon\right) \ell'.
      \end{align*}
      Then the conclusion follows.
\end{IEEEproof}

\begin{lemma}[Section 3.3.4 of \cite{Lev-23-arXiv-CC}]\label{lem:encoder}
    There exists an encoder for $(\tilde{\ell},\tilde{\epsilon})$-locally-balanced codes with only one bit of redundancy, where $0<\tilde{\epsilon}<\frac{1}{2}$, $\tilde{\ell} \geq \frac{\ln n}{\tilde{\epsilon}^2}$, and $n$ represents the length of sequences in the code.
\end{lemma}

By Lemmas \ref{lem:link} and \ref{lem:encoder}, we obtain an encoder for strong-$(\ell,\epsilon)$-locally-balanced codes with one bit of redundancy, where $0<\epsilon<\frac{1}{2}$ and $\ell\geq \min\{ \frac{s}{\tilde{\epsilon}^2} \ln n: 0<\tilde{\epsilon}< \frac{1}{2}, s\geq 1, \tilde{\epsilon}+ \frac{1-4\tilde{\epsilon}^2}{4s}\leq \epsilon\}$.
When $\epsilon= \frac{1}{18}$, we may compute $\ell \geq 6784.1091 \log n$ by choosing $s\approx 13.4752$ and $\tilde{\epsilon}\approx 0.0371$.
Moreover, using the method developed in Construction 3 of \cite{Lev-23-arXiv-CC}, it is possible to encode the code $\mathcal{S}$ (defined in Corollary \ref{cor:balance}) with one bit of redundancy for some $\ell \geq 6784.1091 \log n$.
Notice that this condition is stronger than that in Corollary \ref{cor:balance}. We leave the problem of encoding $\mathcal{S}$ with only one bit of redundancy and smaller $\ell$ when fixing $\epsilon$ as an open problem.

\subsection{Relevant Results}

The following list of constructions is not intended to be a comprehensive survey of all known results, but rather focuses on those that are more closely related to our constructions.

\subsubsection{Single-Edit Correcting Codes}
\begin{lemma}[Theorem 2 of \cite{Levenshtein-66-SPD-1D}]\label{lem:levenstein}
    Let $m\geq 2n$ be an integer and $a\in [[m-1]]$, $\mathcal{C}\triangleq \{\boldsymbol{x}\in \Sigma^n: \mathrm{VT}^{(1)}(\boldsymbol{x})\equiv a \pmod{m}\}$ is a single-edit correcting code.
\end{lemma}

\subsubsection{Two-Deletion Correcting Codes}

\begin{lemma}[Lemma 1 of \cite{Levenshtein-66-SPD-1D}]\label{lem:eq_2D}
    A code $\mathcal{C} \subseteq \Sigma^n$ can correct two deletions if and only if it can correct a combination of insertions and deletions, as long as the number of errors is at most two.
\end{lemma}

\begin{definition}
    For two symbols $a$ and $b$ in $\Sigma$, the \emph{$ab$-indicator sequence} of $\boldsymbol{x} \in \Sigma^n$ is defined as $\mathbbm{1}_{ab}(\boldsymbol{x}) \in \Sigma^{n-1}$ where $\mathbbm{1}_{ab}(x)_i= 1$ if $x_ix_{i+1}= ab$ and $\mathbbm{1}_{ab}(x)_i= 0$ otherwise, for any $1 \leq i \leq n-1$.
\end{definition}

\begin{definition}
  For fixed integers $n$ and $P$ with $n>P$, set $n'= \lfloor n/P \rfloor$. Given any sequence $\boldsymbol{x} \in \Sigma^n$, let $\boldsymbol{x}^{(P,i)}= \boldsymbol{x}_{[(i-1)P+1,iP]}$ if $1 \leq i \leq n'$ and $\boldsymbol{x}^{(P,n'+1)}= \boldsymbol{x}_{[n'P+1,n]}$.
\end{definition}

\begin{lemma}[Theorem 6 of \cite{Sun-23-IT-DR}]\label{lem:2D_bounded}
For fixed integers $a_1,a_1' \in [[4P-1]]$, $a_2,a_2' \in [[4P^2-1]]$, $a_3,a_3' \in [[ 8P^3-1]]$, $a_4,a_4' \in [[2]]$ and $a_5,a_5' \in [[4P-1]]$, we define the code $\mathcal{C}_{2D}^P \subseteq \Sigma^n$ in which each sequence $\boldsymbol{x}$ satisfies the following constraints:
\begin{itemize}
  \item $\sum_{i} \mathrm{VT}^{(1)}(\mathbbm{1}_{01}(\boldsymbol{x}^{(P,2i-1)} \boldsymbol{x}^{(P,2i)})) \equiv a_1 \pmod{4P}$ and $\sum_{i} \mathrm{VT}^{(1)}(\mathbbm{1}_{01}(\boldsymbol{x}^{(P,2i)} \boldsymbol{x}^{(P,2i+1)})) \equiv a_1' \pmod{4P}$;
  \item $\sum_{i} \mathrm{VT}^{(2)}(\mathbbm{1}_{01}(\boldsymbol{x}^{(P,2i-1)} \boldsymbol{x}^{(P,2i)})) \equiv a_2 \pmod{4P^2}$ and $\sum_{i} \mathrm{VT}^{(2)}(\mathbbm{1}_{01}(\boldsymbol{x}^{(P,2i)} \boldsymbol{x}^{(P,2i+1)})) \equiv a_2' \pmod{4P^2}$;
  \item $\sum_{i} \mathrm{VT}^{(3)}(\mathbbm{1}_{01}(\boldsymbol{x}^{(P,2i-1)} \boldsymbol{x}^{(P,2i)})) \equiv a_3 \pmod{8P^3}$ and $\sum_{i} \mathrm{VT}^{(3)}(\mathbbm{1}_{01}(\boldsymbol{x}^{(P,2i)} \boldsymbol{x}^{(P,2i+1)})) \equiv a_3' \pmod{8P^3}$;
  \item $\sum_{i } \mathrm{VT}^{(0)}(\mathbbm{1}_{10}(\boldsymbol{x}^{(P,2i-1)} \boldsymbol{x}^{(P,2i)})) \equiv a_4 \pmod{3}$ and $\sum_{i} \mathrm{VT}^{(0)}(\mathbbm{1}_{10}(\boldsymbol{x}^{(P,2i)} \boldsymbol{x}^{(P,2i+1)})) \equiv a_4' \pmod{3}$;
  \item $\sum_{i} \mathrm{VT}^{(2)}(\mathbbm{1}_{10}(\boldsymbol{x}^{(P,2i-1)} \boldsymbol{x}^{(P,2i)})) \equiv a_5 \pmod{4P}$ and $\sum_{i} \mathrm{VT}^{(2)}(\mathbbm{1}_{10}(\boldsymbol{x}^{(P,2i)} \boldsymbol{x}^{(P,2i+1)})) \equiv a_5' \pmod{4P}$.
\end{itemize}
$\mathcal{C}_{2D}^P$ is a $P$-bounded two-deletion correcting code.
\end{lemma}

\begin{definition}
  For any $\boldsymbol{x} \in \Sigma^n$, set $x_{n+1}=0$, its \emph{run sequence} is defined as $r^{\boldsymbol{x}}= r^{\boldsymbol{x}}_1 r^{\boldsymbol{x}}_2 \cdots r^{\boldsymbol{x}}_{n+1}$, where $r^{\boldsymbol{x}}_1= x_1$ and for any $1 \leq i \leq n$, $r^{\boldsymbol{x}}_{i+1}= r^{\boldsymbol{x}}_{i}$ if $x_{i+1}= x_{i}$ and $r^{\boldsymbol{x}}_{i+1}= r^{\boldsymbol{x}}_{i}+1$ otherwise.
\end{definition}

\begin{remark}
    In \cite{Guruswami-21-IT-2D}, the original definition of a run sequence requires $x_{n+1}=1$, which is intended to ensure that one deletion in a sequence decreases the value of the last symbol of its run sequence by either $0$ or $2$. 
    Notice that setting $x_{n+1}$ to $0$ does not change this property. 
    We make this modification to facilitate establishing the connection between run sequences and differential sequences, as that to be explained later in Lemma \ref{lem:con}.
\end{remark}

\begin{lemma}[Section II.B and Lemmas 5, 6, and 9 of \cite{Guruswami-21-IT-2D}]\label{lem:2D}
  Let $d \geq 7$, $\boldsymbol{x} \in \Sigma^n$ is referred to as a \emph{$d$-regular sequence} if each consecutive subsequence of $\boldsymbol{x}$ of length at least $d \log n$ contains both two consecutive ones and two consecutive zeros.
  For fixed integers $a_1 \in [[4]]$, $a_2 \in [[4n+4]]$, $b_1 \in [[2]]$, $b_2 \in [[2n]]$, and $b_3 \in [[2n^2]]$, we define the code $\mathcal{C} \subseteq \Sigma^n$ in which each sequence $\boldsymbol{x}$ satisfies the following constraints:  
  \begin{itemize}
      \item $\boldsymbol{x}$ is a $d$-regular sequence;
      \item $r^{\boldsymbol{x}}_{n+1}
      \equiv a_1 \pmod{5}$ and $\mathrm{VT}^{(0)}(r^{\boldsymbol{x}}) \equiv a_2 \pmod{4n+5}$;
      \item $\mathrm{VT}^{(0)}(\boldsymbol{x}) \equiv b_0 \pmod{3}$, $\mathrm{VT}^{(1)}(\boldsymbol{x}) \equiv b_1 \pmod{2n+1}$, and $\mathrm{VT}^{(2)}(\boldsymbol{x})- \mathrm{VT}^{(1)}(\boldsymbol{x})\equiv b_3 \pmod{2n^2+1}$.
  \end{itemize}
Suppose $\boldsymbol{x} \in \mathcal{C}$ suffers two deletions and the resultant subsequence is $\boldsymbol{z}$, then from $\boldsymbol{z}$, we can either recover $\boldsymbol{x}$ or find an interval of length at most $2d\log n$ containing the two erroneous positions.
\end{lemma}

\begin{IEEEproof}
    Based on the final three paragraphs of Section II.B of \cite{Guruswami-21-IT-2D}, we can determine the values of $r_{n+1}^{\boldsymbol{x}}$, $\mathrm{VT}^{(0)}(r^{\boldsymbol{x}})$, and $\mathrm{VT}^{(k)}(\boldsymbol{x})$ for $k \in \{0,1,2\}$ from $\boldsymbol{z}$ using the corresponding values of $a_1$, $a_2$, and $b_k$ for $k\in \{0,1,2\}$, respectively.
    Moreover, we have $\mathrm{VT}^{(0)}(\boldsymbol{x})- \mathrm{VT}^{(0)}(\boldsymbol{z}) \in \{0,1,2\}$ and $r_{n+1}^{\boldsymbol{x}}-r_{n-1}^{\boldsymbol{z}} \in \{0,2,4\}$.
    
    If $\mathrm{VT}^{(0)}(\boldsymbol{x})- \mathrm{VT}^{(0)}(\boldsymbol{z}) \in \{0,2\}$, Lemma 5 of \cite{Guruswami-21-IT-2D} showed that we can recover $\boldsymbol{x}$ from $\boldsymbol{z}$ using $\mathrm{VT}^{(1)}(\boldsymbol{x})$ and $\mathrm{VT}^{(2)}(\boldsymbol{x})$.
    
    If $\mathrm{VT}^{(0)}(\boldsymbol{x})- \mathrm{VT}^{(0)}(\boldsymbol{z})=1$, we further consider the value of $r_{n+1}^{\boldsymbol{x}}-r_{n-1}^{\boldsymbol{z}}$. If $r_{n+1}^{\boldsymbol{x}}-r_{n-1}^{\boldsymbol{z}} \in \{0,4\}$, Lemma 6 of \cite{Guruswami-21-IT-2D} showed that we can recover $\boldsymbol{x}$ from $\boldsymbol{z}$ using $\mathrm{VT}^{(1)}(\boldsymbol{x})$ and $\mathrm{VT}^{(0)}(r^{\boldsymbol{x}})$. 
    If $r_{n+1}^{\boldsymbol{x}}-r_{n-1}^{\boldsymbol{z}}= 2$, Lemma 9 of \cite{Guruswami-21-IT-2D} showed that we can either recover $\boldsymbol{x}$ or find an interval of length at most $2d\log n$ containing the two erroneous positions from $\boldsymbol{z}$ using $\mathrm{VT}^{(1)}(\boldsymbol{x})$, $\mathrm{VT}^{(2)}(\boldsymbol{x})$, and $\mathrm{VT}^{(0)}(r^{\boldsymbol{x}})$.
\end{IEEEproof}

\begin{remark}
  Our main focus in this paper is on the principal term coefficients of redundancy.
  Let $P \geq 2d\log n$, simply intersecting the codes constructed in Lemmas \ref{lem:2D_bounded} and \ref{lem:2D} yields a construction of two-deletion correcting codes with $4\log n + 14\log\log n + O(1)$ redundant bits. 
  It is worth noting that one can refine this approach, as demonstrated in \cite{Guruswami-21-IT-2D}, where they present a construction with $4\log n + 10\log\log n + O(1)$ redundant bits. 
  
\end{remark}

\subsubsection{Single-Deletion Single-Substitution Correcting Codes}

\begin{lemma}[Lemma 1 of \cite{Smagloy-20-ISIT-DS}]\label{lem:eq_DS}
    $\mathcal{C} \subseteq \Sigma^n$ can correct one deletion and up to one substitution if and only if it can correct one insertion and up to one substitution.
\end{lemma}

\begin{lemma}[Corollary 1 of \cite{Song-22-IT-DS}]\label{lem:DS}
For fixed integers $a_1 \in [[3n-1]]$, $a_2 \in [[3n^2-1]]$, and $a_3 \in [[3n^3-1]]$, we define the code $\mathcal{C} \subseteq \Sigma^n$ in which each sequence $\boldsymbol{x}$ satisfies $\mathrm{VT}^{(k)}(\boldsymbol{x}) \equiv a_k \pmod{3n^k}$, for $k \in \{1,2,3\}$.
$\mathcal{C}$ is a single-deletion single-substitution correcting code.
\end{lemma}

\begin{lemma}\label{lem:bounded-DS}
Let $P \leq n$ be a positive integer.
For fixed integers $c_1,c_1' \in[[6P-1]]$, $c_2,c_2' \in [[12P^2-1]]$, and $c_3,c_3' \in [[24P^3-1]]$, we define the code $\mathcal{C}_{DS}^P \subseteq \Sigma^n$ in which each sequence $\boldsymbol{x}$ satisfies the following constraints:
\begin{itemize}
  \item $\sum_{i} \mathrm{VT}^{(k)}(\boldsymbol{x}^{(P,2i-1)} \boldsymbol{x}^{(P,2i)}) \equiv c_k \pmod{3(2P)^k}$, for $k \in \{1,2,3\}$;
  \item $\sum_{i} \mathrm{VT}^{(k)}(\boldsymbol{x}^{(P,2i)} \boldsymbol{x}^{(P,2i+1)}) \equiv c_k' \pmod{3(2P)^k}$, for $k \in \{1,2,3\}$.
\end{itemize}
$\mathcal{C}_{DS}^P$ is a $P$-bounded single-deletion single-substitution correcting code.
\end{lemma}

\begin{IEEEproof}
    For any two distinct sequences $\boldsymbol{x}=\boldsymbol{u} \tilde{\boldsymbol{x}} \boldsymbol{v}, \boldsymbol{y}=\boldsymbol{u} \tilde{\boldsymbol{y}} \boldsymbol{v} \in \mathcal{C}_{DS}^P$ with $|\tilde{\boldsymbol{x}}|= |\tilde{\boldsymbol{y}}|\leq P$, there exists an integer $t$ such that $\boldsymbol{x}^{(P,t')}= \boldsymbol{y}^{(P,t')}$ for every $t' \notin \{t,t+1\}$ and $\tilde{\boldsymbol{x}}$ and $\tilde{\boldsymbol{y}}$ are consecutive subsequences of $\boldsymbol{x}^{(P,t)} \boldsymbol{x}^{(P,t+1)}$ and $\boldsymbol{y}^{(P,t)} \boldsymbol{y}^{(P,t+1)}$, respectively.
    Now we show that $\mathrm{VT}^{(k)}(\boldsymbol{x}^{(P,t)} \boldsymbol{x}^{(P,t+1)})= \mathrm{VT}^{(k)}(\boldsymbol{y}^{(P,t)} \boldsymbol{y}^{(P,t+1)}) \pmod{3(2P)^k}$, for $k \in \{1,2,3\}$
    by using either the first or second condition in our code construction, depending on the parity of $t$. Without loss of generality assume $t$ is odd, the case when $t$ is even can be analyzed similarly. Then by the first condition in our code construction, we may compute
    \begin{align*}
      \mathrm{VT}^{(k)}(\boldsymbol{x}^{(P,t)} \boldsymbol{x}^{(P,t+1)})
        & \equiv c_k - \sum\limits_{i \neq \frac{t+1}{2}} \mathrm{VT}^{(k)}(\boldsymbol{x}^{(P,2i-1)} \boldsymbol{x}^{(P,2i)}) \\
        & \equiv c_k - \sum\limits_{i \neq \frac{t+1}{2}}  \mathrm{VT}^{(k)}(\boldsymbol{y}^{(P,2i-1)} \boldsymbol{y}^{(P,2i)}) \\
        &\equiv \mathrm{VT}^{(k)}(\boldsymbol{y}^{(P,t)} \boldsymbol{y}^{(P,t+1)}) \pmod{3(2P)^k} .
    \end{align*}
     By Lemma \ref{lem:DS}, we have $\mathcal{B}_{0,1,1}(\boldsymbol{x}^{(P,t)} \boldsymbol{x}^{(P,t+1)}) \cap \mathcal{B}_{0,1,1}(\boldsymbol{y}^{(P,t)} \boldsymbol{y}^{(P,t+1)})= \emptyset$, implying that $\mathcal{B}_{0,1,1}(\tilde{\boldsymbol{x}}) \cap \mathcal{B}_{0,1,1}(\tilde{\boldsymbol{y}})= \emptyset$.
     Therefore, $\mathcal{C}_{DS}^P$ is a $P$-bounded single-deletion single-substitution correcting code.
\end{IEEEproof}

\begin{lemma}\label{cla:1}
   Suppose $\boldsymbol{x} \in \Sigma^n \setminus \{0^n,1^n\}$, let $\boldsymbol{x}(d,e)$ be the error-sequence obtained from $\boldsymbol{x}$ by deleting $x_{d}$ and substituting $x_{e}$, where $d\in [n]$ and $e\in [[n]]$ ($d=e$ or $e=0$ indicates that $\boldsymbol{x}(d,e)$ is obtained from $\boldsymbol{x}$ by deleting $x_{d}$).
   Then it is reasonable to assume that $1 \leq d \neq e \leq n$, and the deletion always occurs after the substitution.
\end{lemma}

\begin{IEEEproof}
    When $d= e$ or $e=0$, $\boldsymbol{x}(d,e)$ can be obtained from $\boldsymbol{x}$ by deleting $x_{d}$. 
    Since $\boldsymbol{x} \notin \{0^n, 1^n\}$, let $d'$ be the position closest to $d$ such that $x_{d'} \neq x_{d}$. 
    Moreover, let $e'= d'+1$ if $d'< d$ and $e'= d'-1$ otherwise.
    It holds that $\boldsymbol{x}(d',e')= \boldsymbol{x}(d,e)$.
    For example, let $\boldsymbol{x}= 10111001$, then $\boldsymbol{x}_{[8] \setminus \{6\}}= \boldsymbol{x}(5,6)= 1011101$.
    Now, $\boldsymbol{x}(d',e')$ ($\boldsymbol{x}(d,e)$) can be obtained from $\boldsymbol{x}$ by first substituting $x_{e'}$ and then deleting $x_{d'}$ where $d' \neq e'$, which completes the proof.
\end{IEEEproof}

Similarly, when $\boldsymbol{x} \in \Sigma^n \setminus \{0^n,1^n\}$ suffers one insertion and up to one substitution, we can assume that the insertion always occurs after the substitution (if it exists).

\begin{lemma}[Definition 1, Theorem 1, and Lemma 1 of \cite{Song-22-arXiv-DS}]\label{lem:DS_L}
For fixed integers $b_0 \in [[3]]$, $b_1 \in [[2n-1]]$, and $b_2 \in [[2n^2-1]]$, we define the code $\mathcal{C}_{DS}^L \subseteq \Sigma^n \setminus \{0^n,1^n\} $ in which each sequence $\boldsymbol{x}$ satisfies the following constraints:  
\begin{itemize}
  \item $\mathrm{VT}^{(0)}(\boldsymbol{x}) \equiv b_0 \pmod{4}$;
  \item $\mathrm{VT}^{(k)}(\boldsymbol{x}) \equiv b_k \pmod{2n^k}$, for $k \in \{1,2\}$.
\end{itemize}
$\mathcal{C}_{DS}^L$ is single-deletion single-substitution list-decodable code with a list size of two.
Moreover, let $\boldsymbol{x}(d,e)$ be the error-sequence received from $\boldsymbol{x}$ by first substituting $x_e$ and then deleting $x_d$. If $\boldsymbol{x}, \boldsymbol{y} \in \mathcal{C}_{DS}^L$ with $\boldsymbol{x}(d_x,e_x)= \boldsymbol{y}(d_y,e_y)$ and $d_x \leq d_y$, then $d_x< e_x \leq d_y$ and $d_x \leq e_y< d_y$.
\end{lemma}

\begin{lemma}\label{lem:IS_L}
  Let $\mathcal{C}_{DS}^{L}$ be the code constructed in Lemma \ref{lem:DS_L}, $\mathcal{C}_{DS}^L$ is also a single-insertion single-substitution list-decodable code with a list size of two.
\end{lemma}

\begin{figure*}
\center
\begin{tikzpicture}[line width=1pt]
  \draw (-3,0) node []{$\boldsymbol{y}$}
        (-2,0)   node []{$\cdots$}
        (-1,0)   node []{$d_x$}
        (0,0)    node []{$\cdots$}
        (1.5,0)    node []{$\cdots$}
        
        (-3,1) node []{$\boldsymbol{x}$}
        (-2,1)   node []{$\cdots$}
        (-0.5,1)    node []{$\cdots$}
        (0.5,1)    node []{$d_y$}
        (1.5,1)    node []{$\cdots$};

  \filldraw       (-0.5,0)  circle [radius=1pt]
                  (-1,1)    circle [radius=1pt]
                  (0,1)    circle [radius=1pt]
                  (-1.5,0)  circle [radius=1pt]
                  (-2.5,0)  circle [radius=1pt]
                  (0.5,0)   circle [radius=1pt]
                  (1,0)   circle [radius=1pt]
                  (2,0)   circle [radius=1pt]
                  (1,1)     circle [radius=1pt]
                  (-1.5,1)  circle [radius=1pt]
                  (-2.5,1)  circle [radius=1pt]
                  (2,1)   circle [radius=1pt];

  \draw        
               (-2.5,0.9)--(-2.5,0.1)
               (-2,0.9)--(-2,0.1)
               (-1.5,0.9)--(-1.5,0.1)
               (-1,0.9)--(-0.5,0.1)
               (-0.5,0.9)--(0,0.1)
               (0,0.9)--(0.5,0.1)
               (1.5,0.9)--(1.5,0.1)
               (2,0.9)--(2,0.1)
               (1,0.9)--(1,0.1);
\end{tikzpicture}
\caption{Illustrations of $\boldsymbol{x}$ and $\boldsymbol{y}$ when $E(\boldsymbol{x},d_x-1)\cap E(\boldsymbol{y},d_y)\neq \emptyset$ or $\boldsymbol{x}_{[n]\setminus \{d_y\}}= \boldsymbol{y}_{[n] \setminus \{d_x\}}$, where $1 \leq d_x\leq d_y \leq n$.}
Each pair of bits connected by a solid line are of equal value.
\label{fig:insdel}
\end{figure*}
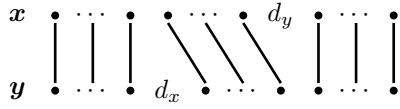

\begin{IEEEproof}
    Let $\boldsymbol{x}(e)$ be the error-sequence received from $\boldsymbol{x}$ by substituting $x_e$, where $e \in [[n]]$ and $e=0$ indicates that no substitution occurs, and $E(\boldsymbol{x},d)$ be the set of error-sequences received from $\boldsymbol{x}$ by inserting a symbol between $x_{d}$ and $x_{d+1}$, where $d=0$ and $d=n$ indicate that the insertion occurs before $x_1$ and after $x_n$, respectively.
    When $1 \leq d_x\leq d_y \leq n$, by Figure \ref{fig:insdel}, we have $E(\boldsymbol{x},d_x-1)\cap E(\boldsymbol{y},d_y)\neq \emptyset$ if and only if $\boldsymbol{x}_{[n]\setminus \{d_y\}}= \boldsymbol{y}_{[n] \setminus \{d_x\}}$.
    
    For any three distinct sequences $\boldsymbol{x}, \boldsymbol{y}, \boldsymbol{z} \in \mathcal{C}_{DS}^{L}$, if $\mathcal{B}_{1,0,1}(\boldsymbol{x}) \cap \mathcal{B}_{1,0,1}(\boldsymbol{y}) \cap \mathcal{B}_{1,0,1}(\boldsymbol{z}) \neq \emptyset$, there exist $d_x, e_x, d_y, e_y, d_z,e_z \in [[n]]$, such that $E(\boldsymbol{x}(e_x),d_x)\cap E(\boldsymbol{y}(e_y),d_y) \cap E(\boldsymbol{z}(e_z),d_z) \neq \emptyset$.
    Without loss of generality assume $d_x \leq d_y \leq d_z$.
    Since sequences in $\mathcal{C}_{DS}^{L}$ have the same first-order VT syndrome modulo $2n$, it follows by Lemma \ref{lem:levenstein} that $\mathcal{C}_{DS}^{L}$ is a single-edit correcting code.
    Therefore, it is reasonable to assume $d_x \neq d_y \neq d_z$.
    Then we have $d_x< d_y< d_z$.
    \begin{itemize}
      \item Since $E(\boldsymbol{x}(e_x),d_x)\cap E(\boldsymbol{y}(e_y),d_y)\neq \emptyset$, we have $\boldsymbol{x}(e_x)_{[n] \setminus \{d_y\}}= \boldsymbol{y}(e_y)_{[n] \setminus \{d_x+1\}}$, i.e., $\boldsymbol{x}(d_y,e_x)=\boldsymbol{y}(d_x+1,e_y)$. Then by Lemma \ref{lem:DS_L}, $d_x+1< e_y\leq  d_y$.
      \item Since $E(\boldsymbol{y}(e_y),d_y) \cap E(\boldsymbol{z}(e_z),d_z) \neq \emptyset$, we have $\boldsymbol{y}(e_y)_{[n] \setminus \{d_z\}}= \boldsymbol{z}(e_z)_{[n] \setminus \{d_y+1\}}$, i.e., $\boldsymbol{y}(d_z,e_y)=\boldsymbol{z}(d_y+1,e_z)$. Then by Lemma \ref{lem:DS_L}, $d_y+1\leq e_y<  d_z$.
    \end{itemize}
    Now we obtain $d_y+1 \leq e_y \leq d_y$, which is impossible.
    Therefore, $\mathcal{B}_{1,0,1}(\boldsymbol{x}) \cap \mathcal{B}_{1,0,1}(\boldsymbol{y}) \cap \mathcal{B}_{1,0,1}(\boldsymbol{z})= \emptyset$, that is, $\mathcal{C}_{DS}^{L}$ is a single-insertion single-substitution list-decodable code with a list size of two.
\end{IEEEproof}

\section{A necessary and sufficient condition for a code to be two-edit correctable}\label{sec:condition}

In this section, our main contribution is to demonstrate that the ability to correct three specific types of errors separately, namely two deletions, up to two substitutions, and one deletion and up to one substitution, is a necessary and sufficient condition for a code to be two-edit correctable. 
While it is clear that this condition is necessary, we will now proceed to show that it is also sufficient.

For any code $\mathcal{C} \subseteq \Sigma^n$, assume $\boldsymbol{x}, \boldsymbol{y} \in \mathcal{C}$ and $\boldsymbol{x} \neq \boldsymbol{y}$.
\begin{itemize}
  \item If $\mathcal{C}$ is a two-deletion correcting code, the intersection between two-deletion balls of $\boldsymbol{x}$ and $\boldsymbol{y}$ is empty, i.e., $\mathcal{B}_{0,2,0}(\boldsymbol{x}) \cap \mathcal{B}_{0,2,0}(\boldsymbol{y})= \emptyset$. Moreover, by Lemma \ref{lem:eq_2D}, the intersection between two-insertion balls of $\boldsymbol{x}$ and $\boldsymbol{y}$ is empty, i.e., $\mathcal{B}_{2,0,0}(\boldsymbol{{x}}) \cap \mathcal{B}_{2,0,0}(\boldsymbol{{y}})= \emptyset$, and the intersection between single-insertion single-deletion balls of $\boldsymbol{x}$ and $\boldsymbol{y}$ is also empty, i.e., $\mathcal{B}_{1,1,0}(\boldsymbol{x}) \cap \mathcal{B}_{1,1,0}(\boldsymbol{y})= \emptyset$.
  \item If $\mathcal{C}$ is a single-deletion single-substitution correcting code, the intersection between single-deletion single-substitution balls of $\boldsymbol{x}$ and $\boldsymbol{y}$ is empty, i.e., $\mathcal{B}_{0,1,1}(\boldsymbol{x}) \cap \mathcal{B}_{0,1,1}(\boldsymbol{y})= \emptyset$. Moreover, by Lemma \ref{lem:eq_DS}, the intersection between single-insertion single-substitution balls of $\boldsymbol{x}$ and $\boldsymbol{y}$ is also empty, i.e.,  $\mathcal{B}_{1,0,1}(\boldsymbol{x}) \cap \mathcal{B}_{1,0,1}(\boldsymbol{y})= \emptyset$.
  \item If $\mathcal{C}$ is a two-substitution correcting code, the intersection between two-substitution balls of $\boldsymbol{x}$ and $\boldsymbol{y}$ is empty, i.e., $\mathcal{B}_{0,0,2}(\boldsymbol{x}) \cap \mathcal{B}_{0,0,2}(\boldsymbol{y})= \emptyset$.
\end{itemize}

Notice that for any sequence $\boldsymbol{x} \in \Sigma^n$, we can always partition its two-edit ball $\mathcal{B}(\boldsymbol{x})$ into the following five disjoint parts: $\mathcal{B}_{2,0,0}(\boldsymbol{{x}})$, $\mathcal{B}_{1,0,1}(\boldsymbol{{x}})$, $\mathcal{B}_{1,1,0}(\boldsymbol{{x}}) \cup \mathcal{B}_{0,0,2}(\boldsymbol{x})$, $\mathcal{B}_{0,1,1}(\boldsymbol{{x}})$, and $\mathcal{B}_{0,2,0}(\boldsymbol{{x}})$.
Therefore, to show that the necessary condition is also sufficient, it remains to show that $\mathcal{B}_{0,0,2}(\boldsymbol{x}) \cap \mathcal{B}_{1,1,0}(\boldsymbol{y})= \emptyset$ and
$\mathcal{B}_{1,1,0}(\boldsymbol{x}) \cap \mathcal{B}_{0,0,2}(\boldsymbol{y})= \emptyset$.

\begin{lemma}\label{lem:eq_ID&2S}
  For two sequences $\boldsymbol{x},\boldsymbol{y} \in \Sigma^n$, if the intersection between two-substitution ball of $\boldsymbol{x}$ and single-insertion single-deletion ball of $\boldsymbol{y}$ is non-empty, i.e., $\mathcal{B}_{0,0,2}(\boldsymbol{x}) \cap \mathcal{B}_{1,1,0}(\boldsymbol{y}) \neq \emptyset$, then the intersection between single-deletion single-substitution balls of $\boldsymbol{x}$ and $\boldsymbol{y}$ is also non-empty, i.e, $\mathcal{B}_{0,1,1}(\boldsymbol{x}) \cap \mathcal{B}_{0,1,1}(\boldsymbol{y}) \neq \emptyset$.
\end{lemma}

\begin{IEEEproof}
     Assume $\boldsymbol{z} \in \mathcal{B}_{0,0,2}(\boldsymbol{x}) \cap \mathcal{B}_{1,1,0}(\boldsymbol{y})$.
     Since $\boldsymbol{z} \in \mathcal{B}_{1,1,0}(\boldsymbol{y})$, due to the properties of insertions and deletions, there exists a sequence $\boldsymbol{z}'$ such that $\boldsymbol{z}' \in \mathcal{B}_{0,1,0}(\boldsymbol{y})$ and $\boldsymbol{z}' \in \mathcal{B}_{0,1,0}(\boldsymbol{z})\subseteq \mathcal{B}_{0,1,2}(\boldsymbol{x})$.
     Moreover, due to the properties of substitutions, there exists a sequence $\boldsymbol{z}''$ such that $\boldsymbol{z}'' \in  \mathcal{B}_{0,1,1}(\boldsymbol{x})$ and $\boldsymbol{z}''\in \mathcal{B}_{0,0,1}(\boldsymbol{z}') \subseteq \mathcal{B}_{0,1,1}(\boldsymbol{y})$.
     Therefore, we have $\mathcal{B}_{0,1,1}(\boldsymbol{x}) \cap \mathcal{B}_{0,1,1}(\boldsymbol{y}) \neq \emptyset$.
\end{IEEEproof}

Since there is no difference between $\{\boldsymbol{x},\boldsymbol{y}\}$ and $\{\boldsymbol{y},\boldsymbol{x}\}$, it follows by Lemma \ref{lem:eq_ID&2S} that $\mathcal{B}_{0,0,2}(\boldsymbol{x}) \cap \mathcal{B}_{1,1,0}(\boldsymbol{y})= \emptyset$ and
$\mathcal{B}_{1,1,0}(\boldsymbol{x}) \cap \mathcal{B}_{0,0,2}(\boldsymbol{y})= \emptyset$ when $\mathcal{B}_{0,1,1}(\boldsymbol{x}) \cap \mathcal{B}_{0,1,1}(\boldsymbol{y})= \emptyset$.
Therefore, the following theorem holds.
\begin{theorem}\label{thm:equiv}
    A code $\mathcal{C}$ can correct two edits if and only if it can correct the following three types of errors separately: two deletions, up to two substitutions, and one deletion and up to one substitution.
\end{theorem}

\begin{remark}
  Let $\mathcal{C} \subseteq \Sigma^n$ be a two-edit correcting code, and assume $\boldsymbol{x}\in \mathcal{C}$ suffers two edits resulting in $\boldsymbol{y}$.
  If $|\boldsymbol{y}| \in \{n-2,n-1,n+1,n+2\}$, we can determine the type of errors and use the corresponding error-correcting code decoder to recover $\boldsymbol{x}$ correctly.
  If $|\boldsymbol{y}|=n$, even though we cannot determine the type of errors, we can still recover $\boldsymbol{x}$ correctly, as shown below.
  Firstly, by Lemma \ref{lem:eq_ID&2S}, $(\mathcal{B}_{1,1,0}(\boldsymbol{x})\cup \mathcal{B}_{0,0,2}(\boldsymbol{x})) \cap (\mathcal{B}_{1,1,0}(\boldsymbol{x}')\cup \mathcal{B}_{0,0,2}(\boldsymbol{x}'))= \emptyset$ for any $\boldsymbol{x}'\neq \boldsymbol{x} \in \mathcal{C}$. 
  Since $|\boldsymbol{y}|=n$, we have $\boldsymbol{y} \in \mathcal{B}_{1,1,0}(\boldsymbol{x})\cup \mathcal{B}_{0,0,2}(\boldsymbol{x})$.
  We claim that $\mathcal{B}_{1,1,0}(\boldsymbol{y}) \cap \mathcal{C} \subseteq \{\boldsymbol{x}\}$ otherwise there exists $\boldsymbol{x}'\neq \boldsymbol{x}$ such that $\boldsymbol{y} \in (\mathcal{B}_{1,1,0}(\boldsymbol{x})\cup \mathcal{B}_{0,0,2}(\boldsymbol{x})) \cap \mathcal{B}_{1,1,0}(\boldsymbol{x}')$, which leads to a contradiction.
  Moreover, we claim that $\mathcal{B}_{0,0,2}(\boldsymbol{y}) \cap \mathcal{C}= \{\boldsymbol{x}\}$ when $\mathcal{B}_{1,1,0}(\boldsymbol{y}) \cap \mathcal{C}= \emptyset$, otherwise there exists $\boldsymbol{x}'\neq \boldsymbol{x}$ such that $\boldsymbol{y} \in (\mathcal{B}_{1,1,0}(\boldsymbol{x})\cup \mathcal{B}_{0,0,2}(\boldsymbol{x})) \cap \mathcal{B}_{0,0,2}(\boldsymbol{x}')$, which leads to a contradiction.
  Therefore, when we try to correct errors, we first use the single-insertion single-deletion correcting code decoder to check if we can obtain a sequence in $\mathcal{C}$.
  If successful, the obtained sequence is $\boldsymbol{x}$, otherwise we proceed to use the two-substitution correcting code decoder to recover $\boldsymbol{x}$ correctly.
\end{remark}

\section{Codes for correcting two deletions}\label{sec:2D}

In this section, we aim to emphasize the significance of using differential sequences and strong-locally-balanced sequences to correct errors related to deletions.
To accomplish this, we consider the construction of two-deletion correcting codes and provide an alternative proof of a slightly weaker result by Guruswami and H{\aa}stad \cite{Guruswami-21-IT-2D}.

\begin{theorem}\label{thm:2D}
    Set $\epsilon= \frac{1}{18}$, $\ell=1296 \log n$, and $P\geq 6(\ell+3)$.
    For fixed integers $a_0 \in [[4]], a_1 \in [[4n+4]], b_0 \in [[2]], b_1 \in [[2n]], b_2 \in [[2n^2]]$, we define the code $\mathcal{C}_{2D} \subseteq \Sigma^n$ in which each sequence $\boldsymbol{x}$ satisfies the following conditions:
    \begin{itemize}
        \item both $\boldsymbol{x}$ and $\psi(\boldsymbol{x})$ are strong-$(\ell,\epsilon)$-locally-balanced sequences;
        \item $\mathrm{VT}^{(k)}(\psi(\boldsymbol{x})) \equiv a_k \pmod{4(n+1)^k+1}$ for $k \in \{0,1\}$;
        \item $\mathrm{VT}^{(k)}(\boldsymbol{x}) \equiv b_k  \pmod{2n^k+1}$ for $k \in \{0,1,2\}$;
        \item $\boldsymbol{x} \in \mathcal{C}_{2D}^P$, where $\mathcal{C}_{2D}^P$ is a $P$-bounded two-deletion correcting code constructed in Lemma \ref{lem:2D_bounded}.
    \end{itemize}
    $\mathcal{C}_{2D}$ is a two-deletion correcting code.
\end{theorem}

Upon comparing the code constructions in Theorem \ref{thm:2D} and Lemma \ref{lem:2D} (a result by Guruswami and H{\aa}stad \cite{Guruswami-21-IT-2D}), it is evident that the primary distinction lies in the first two conditions of each construction.
For the first condition of each construction, we have the following conclusion.

\begin{lemma}
    If both $\boldsymbol{x}$ and $\psi(\boldsymbol{x})$ are strong-$(\ell,\epsilon)$-locally-balanced sequences, then $\boldsymbol{x}$ is a $d$-regular sequence, where $d=\frac{\ell}{\log n}$.
\end{lemma}

\begin{IEEEproof}
    We prove it by contradiction. 
    Assume that $\boldsymbol{x}$ is not a $d$-regular sequence and $\tilde{\boldsymbol{x}}$ is a consecutive sequence of $\boldsymbol{x}$ of length $\ell'\geq \ell$ that does not contain two consecutive $\sigma$'s, where $\sigma \in \Sigma$. Then, there exist integers $s_1,s_{t+1}\geq 0, s_2, \ldots, s_{t} \geq 1$ such that $\tilde{\boldsymbol{x}} = ((1-\sigma)^{s_1}, \sigma, (1-\sigma)^{s_2}, \sigma, \ldots, (1-\sigma)^{s_{t}}, \sigma, (1-\sigma)^{s_{t+1}})$, where $t$ represents the number of occurrences of $\sigma$ in $\tilde{\boldsymbol{x}}$.
    Since $\boldsymbol{x}$ is a strong-$(\ell,\epsilon)$-locally-balanced sequence, we have $t \geq (\frac{1}{2}-\epsilon)\ell'$.
    Then the number of ones in $\psi(\tilde{\boldsymbol{x}})$ is at least $2(t-1) \geq (1-2\epsilon)\ell'-2> (\frac{1}{2}+\epsilon)\ell'+2$, which implies that $\psi(\boldsymbol{x})$ contains a consecutive subsequence of length $\ell'\geq \ell$ with more than $(\frac{1}{2}+\epsilon)\ell'$ ones. This contradicts the fact that $\psi(\boldsymbol{x})$ is a strong-$(\ell,\epsilon)$-locally-balanced sequence.
\end{IEEEproof}

\begin{remark}
    Notice that not all $d$-regular sequences are strong-$(\ell,\epsilon)$-locally-balanced sequences, where $d=\frac{\ell}{\log n}\geq 7$.
    For example, $\boldsymbol{x}= 1^3 0^{\ell-3} 1^3 0^{\ell-3}\cdots$ is a $d$-regular sequence but not a strong-$(\ell,\epsilon)$-locally-balanced sequence for some $0<\epsilon<\frac{1}{2}$, where $\ell=d\log n$.
    Therefore, the first condition in our construction is stronger than that in Lemma \ref{lem:2D}.
\end{remark}

For the second condition of each construction, we have the following two lemmas.

\begin{lemma}\label{lem:con}
  For any $\boldsymbol{x} \in \Sigma^n$, it holds that $\mathrm{VT}^{(0)}(\psi(\boldsymbol{x}))= r^{\boldsymbol{x}}_{n+1}= |\{1 \leq i \leq n+1: x_i\neq x_{i-1} \}|$, where $x_0=x_{n+1}=0$.
\end{lemma}

\begin{IEEEproof}
    The conclusion follows by the definitions of differential sequence and run sequence directly.
\end{IEEEproof}

\begin{lemma}
  For two distinct sequences $\boldsymbol{x}, \boldsymbol{y} \in \Sigma^n$ with $\mathcal{B}_{0,2,0}(\boldsymbol{x}) \cap \mathcal{B}_{0,2,0}(\boldsymbol{y}) \neq \emptyset$ and $\mathrm{VT}^{(0)}(\psi(\boldsymbol{x})) \equiv \mathrm{VT}^{(0)}(\psi(\boldsymbol{y})) \pmod{5}$, it holds that $\mathrm{VT}^{(1)}(\psi(\boldsymbol{x}))- \mathrm{VT}^{(1)}(\psi(\boldsymbol{y}))= \mathrm{VT}^{(0)}(r^{\boldsymbol{y}})- \mathrm{VT}^{(0)}(r^{\boldsymbol{x}})$.
\end{lemma}

\begin{IEEEproof}
    Suppose $\boldsymbol{z}\in \mathcal{B}_{0,2,0}(\boldsymbol{x}) \cap \mathcal{B}_{0,2,0}(\boldsymbol{y}) \neq \emptyset$, by Observation \ref{obs:deletion}, we have $\mathrm{VT}^{(0)}(\psi(\boldsymbol{x}))- \mathrm{VT}^{(0)}(\psi(\boldsymbol{z}))\in \{0,2,4\}$ and $\mathrm{VT}^{(0)}(\psi(\boldsymbol{y}))- \mathrm{VT}^{(0)}(\psi(\boldsymbol{z}))\in \{0,2,4\}$. Then we obtain $m\triangleq \mathrm{VT}^{(0)}(\psi(\boldsymbol{x}))= \mathrm{VT}^{(0)}(\psi(\boldsymbol{y}))$. 
    Let $i_1<i_2<\cdots<i_m$ be the positions that $\psi(x)_{i_j}=1$, for any $j \in [m]$.
    Moreover, set $i_{m+1}= n+2$, we may compute $\mathrm{VT}^{(1)}(\psi(\boldsymbol{x}))= \sum_{j=1}^{m} i_j$ and $\mathrm{VT}^{(0)}(r^{\boldsymbol{x}})= \sum_{j=1}^{m} j(i_{j+1}-i_j)= mi_{m+1}- \sum_{j=1}^m i_j= m(n+2)- \mathrm{VT}^{(1)}(\psi(\boldsymbol{x}))$.
    Therefore, $\mathrm{VT}^{(0)}(r^{\boldsymbol{y}})- \mathrm{VT}^{(0)}(r^{\boldsymbol{x}})= \mathrm{VT}^{(1)}(\psi(\boldsymbol{x}))- \mathrm{VT}^{(1)}(\psi(\boldsymbol{y}))$.
\end{IEEEproof}

Based on the proof of Lemma \ref{lem:2D} and the property of $P$-bounded two-deletion correcting codes, there are no two distinct sequences $\boldsymbol{x},\boldsymbol{y} \in \mathcal{C}_{2D}$ with $\mathcal{B}_{0,2,0}(\boldsymbol{x}) \cap \mathcal{B}_{0,2,0}(\boldsymbol{y})\neq \emptyset$ such that $r^{\boldsymbol{x}}_{n+1}=r^{\boldsymbol{y}}_{n+1}$, $\mathrm{VT}^{(0)}(r^{\boldsymbol{x}})= \mathrm{VT}^{(0)}(r^{\boldsymbol{y}})$, and $\mathrm{VT}^{(k)}(\boldsymbol{x})=\mathrm{VT}^{(k)}(\boldsymbol{y})$ for $k \in \{0,1,2\}$.
Therefore, the above connections will support the correctness of Theorem \ref{thm:2D}.
Below we provide an alternative proof of Theorem \ref{thm:2D} by using the properties of strong-locally-balanced sequences and differential sequences.
 
To prove Theorem \ref{thm:2D}, we assume by contradiction that there are two distinct sequences $\boldsymbol{x}$ and $\boldsymbol{y}$ in $\mathcal{C}_{2D}$ such that the intersection of their two-deletion balls is non-empty. 
In other words, there are indices $g_1<h_1, g_2< h_2$, such that $\boldsymbol{x}_{[n] \setminus \{g_1,h_1\}}= \boldsymbol{y}_{[n] \setminus \{g_2,h_2\}}$. 
Since there is no difference between $\{\boldsymbol{x}, \boldsymbol{y}\}$ and $\{\boldsymbol{y},\boldsymbol{x}\}$, it is reasonable to assume that $g_1 \leq g_2$.
To show that $\mathcal{C}_{2D}$ is a two-deletion correcting code, our strategy is to demonstrate that $\boldsymbol{x}$ and $\boldsymbol{y}$ do not satisfy some of the constraints imposed in Theorem \ref{thm:2D}.

\subsection{Claims}

We start with some simple but useful claims regarding $\boldsymbol{x}$ and $\boldsymbol{y}$.

\subsubsection{The Sequences Themselves}

\begin{claim}\label{cla:2D}
  $\mathrm{VT}^{(k)}(\boldsymbol{x})= \mathrm{VT}^{(k)}(\boldsymbol{y})$ for $k \in \{0,1,2\}$.
\end{claim}

\begin{IEEEproof}
    By the definition of VT syndrome, one deletion in $\boldsymbol{x}$ decreases the value of the $k$-th order VT syndrome of $\boldsymbol{x}$ within the range of $0$ to $n^k$, for $k \in \{0,1,2\}$.
    Let $\boldsymbol{z}\triangleq \boldsymbol{x}_{[n] \setminus \{g_1,h_1\}}= \boldsymbol{y}_{[n] \setminus \{g_2,h_2\}}$, we have $\mathrm{VT}^{(k)}(\boldsymbol{x})- \mathrm{VT}^{(k)}(\boldsymbol{z}) \in [[2n^k]]$ and $\mathrm{VT}^{(k)}(\boldsymbol{y})- \mathrm{VT}^{(k)}(\boldsymbol{z}) \in [[2n^k]]$.
    It follows by $\mathrm{VT}^{(k)}(\boldsymbol{x}) \equiv \mathrm{VT}^{(k)}(\boldsymbol{y}) \pmod{2n^k+1}$ that $\mathrm{VT}^{(k)}(\boldsymbol{x})= \mathrm{VT}^{(k)}(\boldsymbol{y})$.
\end{IEEEproof}

For simplicity, for any $k\in \{1,2\}$, let $\Delta^{(k)} \triangleq \mathrm{VT}^{(k)}(\boldsymbol{x})- \mathrm{VT}^{(k)}(\boldsymbol{y})$.

\begin{claim}[Lemma 5.2 of \cite{Sun-23-IT-DR}]\label{cla:Sima}
    If $g_1 \leq g_2< h_2 \leq h_1$ or $g_1 \leq g_2 < h_1 \leq h_2$, $\Delta^{(k)} \neq 0$ for some $k \in \{1,2\}$.
\end{claim}

\begin{IEEEproof}
    Let $\mathcal{A}_1\triangleq \{\boldsymbol{z}: \boldsymbol{z}= \boldsymbol{x}_{[n]\setminus \{g_1,h_1\}}= \boldsymbol{y}_{[n]\setminus \{g_2,h_2\}}, g_1 \leq g_2< h_2 \leq h_1\}$ and $\mathcal{A}_2\triangleq \{\boldsymbol{z}: \boldsymbol{z}= \boldsymbol{x}_{[n]\setminus \{g_1,h_1\}}= \boldsymbol{y}_{[n]\setminus \{g_2,h_2\}}, g_1 \leq g_2 < h_1 \leq h_2\}$.
    Lemma 5.2 of \cite{Sun-23-IT-DR} showed that $|\mathcal{A}_1|= |\mathcal{A}_2|=0$ when $\mathrm{VT}^{(1)}(\boldsymbol{x})= \mathrm{VT}^{(1)}(\boldsymbol{y})$ and $\mathrm{VT}^{(2)}(\boldsymbol{x})= \mathrm{VT}^{(2)}(\boldsymbol{y})$.
    Then the conclusion follows.
\end{IEEEproof}

\begin{claim}[Lemma 5 of \cite{Guruswami-21-IT-2D}]\label{cla:Guruswami}
    If $x_{g_1}= x_{h_1}= y_{g_2}= y_{h_2}$, $\Delta^{(k)} \neq 0$ for some $k \in \{1,2\}$.
\end{claim}

\begin{IEEEproof}
    Lemma 5 of \cite{Guruswami-21-IT-2D} showed that when $x_{g_1}= x_{h_1}$, we can recover $\boldsymbol{x}$ from $\boldsymbol{x}_{[n]\setminus \{g_1,h_1\}}$ using $\mathrm{VT}^{(1)}(\boldsymbol{x})$ and $\mathrm{VT}^{(2)}(\boldsymbol{x})$.
    Combining this with $\boldsymbol{x}_{[n]\setminus \{g_1,h_1\}}= \boldsymbol{y}_{[n]\setminus \{g_2,h_2\}}$, the conclusion follows.
\end{IEEEproof}

The cases involved in the above two claims are relatively easy to handle, so we will not prove them again in this paper.
Now we proceed to calculate $\Delta^{(k)}$ when $g_1 < h_1 \leq g_2 < h_2$.

\begin{claim}\label{cla:2D_VT}
  If $g_1 < h_1 \leq g_2 < h_2$, for $k \in \{1,2\}$, it holds that
  \begin{equation}\label{eq:delta_k}
  \begin{aligned}
    \Delta^{(k)}
    &= c_{g_1}^{(k)} x_{g_1}+ c_{h_1}^{(k)} x_{h_1}- c_{g_2}^{(k)} y_{g_2}- c_{h_2}^{(k)} y_{h_2} \\
    &~~     + \sum_{i \in [g_1+1,h_1-1] \cup [g_2+2,h_2]} i^{k-1} x_i \\
    &~~+ \sum_{i \in [h_1+1, g_2+1]} ((i-1)^{k-1}+ i^{k-1}) x_i,
  \end{aligned}
  \end{equation}
  where $c_i^{(k)}= \sum_{j=1}^i j^{k-1}$ for $i \in \{g_1,g_2,h_1,h_2\}$.
\end{claim}

\begin{figure*}
\center
\begin{tikzpicture}[line width=1pt]
  \draw (-5,0) node []{$\boldsymbol{y}$}
        (-4,0)   node []{$\cdots$}
        (-2.5,0)   node []{$\cdots$}
        (-0.5,0)    node []{$\cdots$}
        (1,0)    node []{$g_2$}
        (2,0)    node []{$\cdots$}
        (3,0)    node []{$h_2$}
        (4,0)    node []{$\cdots$}
        
        (-5,1) node []{$\boldsymbol{x}$}
        (-4,1)   node []{$\cdots$}
        (-3,1)   node []{$g_1$}
        (-2,1)   node []{$\cdots$}
        (-1,1)   node []{$h_1$}
        (0.5,1)    node []{$\cdots$}
        (2.5,1)    node []{$\cdots$}
        (4,1)    node []{$\cdots$};

  \filldraw       (0,0)  circle [radius=1pt]
                  (0.5,0)  circle [radius=1pt]
                  (-0.5,1)  circle [radius=1pt]
                  (0,1)     circle [radius=1pt]
                  (-1,0)  circle [radius=1pt]
                  (-1.5,0)  circle [radius=1pt]
                  (-3.5,0)  circle [radius=1pt]
                  (-2,0)  circle [radius=1pt]
                  (-3,0)    circle [radius=1pt]
                  (-4.5,0)  circle [radius=1pt]
                  (3,1)   circle [radius=1pt]
                  (3.5,0)     circle [radius=1pt]
                  (4.5,0)     circle [radius=1pt]
                  (1,1)     circle [radius=1pt]
                  (1.5,0)     circle [radius=1pt]
                  (1.5,1)     circle [radius=1pt]
                  (2,1)     circle [radius=1pt]
                  (-1.5,1)  circle [radius=1pt]
                  (-2.5,1)  circle [radius=1pt]
                  (-3.5,1)  circle [radius=1pt]
                  (-4.5,1)  circle [radius=1pt]
                  (2.5,0)   circle [radius=1pt]
                  (3.5,1)   circle [radius=1pt]
                  (4.5,1)   circle [radius=1pt];

  \draw        (-4.5,0.9)--(-4.5,0.1)
               (-4,0.9)--(-4,0.1)
               (-3.5,0.9)--(-3.5,0.1)
               (-2.5,0.9)--(-3,0.1)
               (-2,0.9)--(-2.5,0.1)
               (-1.5,0.9)--(-2,0.1)
               (-0.5,0.9)--(-1.5,0.1)
               (0,0.9)--(-1,0.1)
               (1,0.9)--(0,0.1)
               (0.5,0.9)--(-0.5,0.1)
               (1.5,0.9)--(0.5,0.1)
               (2.5,0.9)--(2,0.1)
               (2,0.9)--(1.5,0.1)
               (3,0.9)--(2.5,0.1)
               (3.5,0.9)--(3.5,0.1)
               (4,0.9)--(4,0.1)
               (4.5,0.9)--(4.5,0.1);
\end{tikzpicture}
\caption{Illustrations of $\boldsymbol{x}$ and $\boldsymbol{y}$ when $\boldsymbol{x}_{[n] \setminus \{g_1,h_1\}}= \boldsymbol{y}_{[n] \setminus \{g_2,h_2\}}$, where $g_1<h_1\leq g_2<h_2$. Each pair of bits connected by a solid line are of equal value.}
\label{fig:2D_1}
\end{figure*}
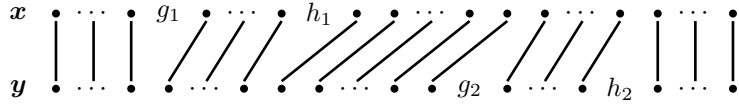

\begin{IEEEproof}
    By Figure \ref{fig:2D_1}, it is routine to check that 
    \begin{equation*}
    x_i=
      \begin{cases}
        y_i, & \mbox{if } i \in [1,g_1-1] \cup [h_2+1,n]; \\
        y_{i-1}, & \mbox{if } i \in [g_1+1,h_1-1] \cup [g_2+2,h_2]; \\
        y_{i-2}, & \mbox{if } i \in [h_1+1, g_2+1].
      \end{cases}
    \end{equation*}
    Then the conclusion follows.
\end{IEEEproof}

\subsubsection{The Differential Sequences}

\begin{claim}\label{cla:2D_psi}
  $\mathrm{VT}^{(k)}(\psi(\boldsymbol{x}))= \mathrm{VT}^{(k)}(\psi(\boldsymbol{y}))$ for $k \in \{0,1\}$.
\end{claim}

\begin{IEEEproof}
    By the definition of VT syndrome, one deletion in $\boldsymbol{x}$ decreases the value of the $k$-th order VT syndrome of $\psi(\boldsymbol{x})$ within the range of $0$ to $2(n+1)^k$, for $k \in \{0,1\}$.
    Let $\boldsymbol{z}\triangleq \boldsymbol{x}_{[n] \setminus \{g_1,h_1\}}= \boldsymbol{y}_{[n] \setminus \{g_2,h_2\}}$, we have $\mathrm{VT}^{(k)}(\psi(\boldsymbol{x}))- \mathrm{VT}^{(k)}(\psi(\boldsymbol{z})) \in [[4(n+1)^k]]$ and $\mathrm{VT}^{(k)}(\psi(\boldsymbol{y}))- \mathrm{VT}^{(k)}(\psi(\boldsymbol{z})) \in [[4(n+1)^k]]$.
    It follows by $\mathrm{VT}^{(k)}(\psi(\boldsymbol{x})) \equiv \mathrm{VT}^{(k)}(\psi(\boldsymbol{y})) \pmod{4(n+1)^k+1}$ that $\mathrm{VT}^{(k)}(\psi(\boldsymbol{x}))= \mathrm{VT}^{(k)}(\psi(\boldsymbol{y}))$.
\end{IEEEproof}

For simplicity, let $\Delta^{(1)}_{\psi} \triangleq \mathrm{VT}^{(1)}(\psi(\boldsymbol{x}))- \mathrm{VT}^{(1)}(\psi(\boldsymbol{y}))$. Now we proceed to calculate $\Delta^{(1)}_{\psi}$.

\begin{claim}\label{cla:2D_VT-differential_sequence}
  Suppose $g_1 < h_1 \leq g_2 < h_2$, it holds that
  \begin{equation}\label{eq:delta_psi_k}
  \begin{aligned}
    \Delta^{(1)}_{\psi}
    &= \sum_{i \in [g_1+2, h_1-1] \cup [g_2+3,h_2]} \psi(x)_i+ 2\sum_{i \in [h_1+2,g_2+1]} \psi(x)_i+ \delta \\
    &= \sum_{i \in [g_1+2, g_2+1] \setminus \{h_1,h_1+1\}} \psi(x)_i+ \sum_{i \in [h_1+2,h_2] \setminus \{g_2+2\}} \psi(x)_i+ \delta,
  \end{aligned}
  \end{equation}
  where $\delta$ can be calculated as follows.
  \begin{itemize}
    \item If $h_1=g_1+1$ and $h_2= g_2+1$, 
        \begin{align*}
          \delta
            &= \sum_{i \in [g_1,g_1+2]} i \psi(x)_i- g_1 (\psi(x)_{g_1} \oplus \psi(x)_{g_1+1} \oplus \psi(x)_{g_1+2})  \\
            &~~+ (g_2+2) (\psi(y)_{g_2} \oplus \psi(y)_{g_2+1} \oplus \psi(y)_{g_2+2})- \sum_{i \in [g_2,g_2+2]} i \psi(y)_i.
        \end{align*}
    \item If $h_1=g_1+1$ and $h_2> g_2+1$, 
        \begin{align*}
          \delta
            &= \sum_{i \in [g_1,g_1+2]} i \psi(x)_i- g_1 (\psi(x)_{g_1} \oplus \psi(x)_{g_1+1} \oplus \psi(x)_{g_1+2})  \\
            &~~+ (g_2+2) (\psi(y)_{g_2} \oplus \psi(y)_{g_2+1})- \sum_{i \in [g_2,g_2+1]} i \psi(y)_i \\
            &~~+ (h_2+1) (\psi(y)_{h_2} \oplus \psi(y)_{h_2+1})- \sum_{i \in [h_2,h_2+1]} i \psi(y)_i.
        \end{align*}
    \item If $h_1>g_1+1$ and $h_2= g_2+1$,
        \begin{align*}
          \delta
            &= \sum_{i \in [g_1,g_1+1]} i \psi(x)_i- g_1 (\psi(x)_{g_1} \oplus \psi(x)_{g_1+1})  \\
            &~~+ \sum_{i \in [h_1,h_1+1]} i \psi(x)_i- (h_1-1) (\psi(x)_{h_1} \oplus \psi(x)_{h_1+1})  \\
            &~~+ (g_2+2) (\psi(y)_{g_2} \oplus \psi(y)_{g_2+1} \oplus \psi(y)_{g_2+2})- \sum_{i \in [g_2,g_2+2]} i \psi(y)_i.
        \end{align*}
    \item If $h_1>g_1+1$ and $h_2> g_2+1$,
        \begin{align*}
          \delta
            &= \sum_{i \in [g_1,g_1+1]} i \psi(x)_i- g_1 (\psi(x)_{g_1} \oplus \psi(x)_{g_1+1})  \\
            &~~+ \sum_{i \in [h_1,h_1+1]} i \psi(x)_i- (h_1-1) (\psi(x)_{h_1} \oplus \psi(x)_{h_1+1})  \\
            &~~+ (g_2+2) (\psi(y)_{g_2} \oplus \psi(y)_{g_2+1})- \sum_{i \in [g_2,g_2+1]} i \psi(y)_i \\
            &~~+ (h_2+1) (\psi(y)_{h_2} \oplus \psi(y)_{h_2+1})- \sum_{i \in [h_2,h_2+1]} i \psi(y)_i.
        \end{align*}
  \end{itemize}
\end{claim}

\begin{figure*}
\center
\begin{tikzpicture}[line width=1pt]   
  \fill[fill=green!20] (-3.8,7.2) -- (-2,7.2) -- (-3.5,5.8) -- (-3.8,5.8);
  
  \fill[fill=green!20] (-0.5,7.2) -- (-0.2,7.2) -- (-0.2,5.8) -- (-2.1,5.8);
  
  \draw (-9,6.5) node []{Case 1. $h_1= g_1+1, h_2= g_2+1$}
        (-5.7,6) node []{$\psi(\boldsymbol{y})$}
        (-4.5,6)   node []{$\cdots$}
        (-2.5,6)    node []{$\cdots$}
        (-1.5,6)    node []{$g_2$}
        (-1,6)    node []{$h_2$}
        (0.5,6)    node []{$\cdots$}
        
        (-5.7,7) node []{$\psi(\boldsymbol{x})$}
        (-4.5,7)   node []{$\cdots$}
        (-3.5,7)   node []{$g_1$}
        (-3,7)   node []{$h_1$}
        (-1.5,7)    node []{$\cdots$}
        (0.5,7)    node []{$\cdots$};

  \filldraw       (-2,6)  circle [radius=1pt]
                  (-2.5,7)  circle [radius=1pt]
                  (-2,7)     circle [radius=1pt]
                  (-3,6)  circle [radius=1pt]
                  (-4,6)  circle [radius=1pt]
                  (-3.5,6)     circle [radius=1pt]
                  (-5,6)  circle [radius=1pt]
                  (-1,7)   circle [radius=1pt]
                  (-0.5,6)     circle [radius=1pt]
                  (0,6)     circle [radius=1pt]
                  (1,6)     circle [radius=1pt]
                  (-2,7)     circle [radius=1pt]
                  (-2,7)     circle [radius=1pt]
                  (-4,7)  circle [radius=1pt]
                  (-5,7)  circle [radius=1pt]
                  (-0.5,7)     circle [radius=1pt]
                  (0,7)   circle [radius=1pt]
                  (1,7)   circle [radius=1pt];

  \draw        (-5,6.9)--(-5,6.1)
               (-4.5,6.9)--(-4.5,6.1)
               (-4,6.9)--(-4,6.1)
               (-2,6.9)--(-3,6.1)
               (-1,6.9)--(-2,6.1)
               (-1.5,6.9)--(-2.5,6.1)
               (1,6.9)--(1,6.1)
               (0,6.9)--(0,6.1)
               (0.5,6.9)--(0.5,6.1);

  \fill[fill=green!20] (-3.8,5.2) -- (-2,5.2) -- (-3.5,3.8) -- (-3.8,3.8);

  \fill[fill=green!20] (-0.5,5.2) -- (0,5.2) -- (-0.9,3.8) -- (-2.1,3.8);

  \fill[fill=green!20] (1.8,3.8) -- (0.6,3.8) -- (1.4,5.2) -- (1.8,5.2);
  
  \draw (-9,4.5) node []{Case 2. $h_1= g_1+1, h_2> g_2+1$}
        (-5.7,4) node []{$\psi(\boldsymbol{y})$}
        (-4.5,4)   node []{$\cdots$}
        (-2.5,4)    node []{$\cdots$}
        (-1.5,4)    node []{$g_2$}
        (0,4)    node []{$\cdots$}
        (1,4)    node []{$h_2$}
        (2.5,4)    node []{$\cdots$}
        
        (-5.7,5) node []{$\psi(\boldsymbol{x})$}
        (-4.5,5)   node []{$\cdots$}
        (-1.5,5)   node []{$\cdots$}
        (-3.5,5)   node []{$g_1$}
        (-3,5)   node []{$h_1$}
        (0.5,5)    node []{$\cdots$}
        (2.5,5)    node []{$\cdots$};

  \filldraw       (-2,4)  circle [radius=1pt]
                  (-1,4)  circle [radius=1pt]
                  (-2.5,5)  circle [radius=1pt]
                  (-2,5)     circle [radius=1pt]
                  (-3,4)  circle [radius=1pt]
                  (-4,4)  circle [radius=1pt]
                  (-3.5,4)     circle [radius=1pt]
                  (-5,4)  circle [radius=1pt]
                  (1,5)   circle [radius=1pt]
                  (1.5,4)     circle [radius=1pt]
                  (2,4)     circle [radius=1pt]
                  (3,4)     circle [radius=1pt]
                  (-1,5)     circle [radius=1pt]
                  (-0.5,4)     circle [radius=1pt]
                  (-0.5,5)     circle [radius=1pt]
                  (0,5)     circle [radius=1pt]
                  (-4,5)  circle [radius=1pt]
                  (-5,5)  circle [radius=1pt]
                  (0.5,4)   circle [radius=1pt]
                  (1.5,5)     circle [radius=1pt]
                  (2,5)   circle [radius=1pt]
                  (3,5)   circle [radius=1pt];

  \draw        (-5,4.9)--(-5,4.1)
               (-4.5,4.9)--(-4.5,4.1)
               (-4,4.9)--(-4,4.1)
               (-2,4.9)--(-3,4.1)
               (-1,4.9)--(-2,4.1)
               (-1.5,4.9)--(-2.5,4.1)
               (0.5,4.9)--(0,4.1)
               (0,4.9)--(-0.5,4.1)
               (1,4.9)--(0.5,4.1)
               (3,4.9)--(3,4.1)
               (2,4.9)--(2,4.1)
               (2.5,4.9)--(2.5,4.1);

  \fill[fill=green!20] (-3.8,3.2) -- (-2.6,3.2) -- (-3.4,1.8) -- (-3.8,1.8);

  \fill[fill=green!20] (-1.1,3.2) -- (0,3.2) -- (-1.5,1.8) -- (-2,1.8);

  \fill[fill=green!20] (1.5,3.2) -- (1.8,3.2) -- (1.8,1.8) -- (-0.1,1.8);
  
  \draw (-9,2.5) node []{Case 3. $h_1> g_1+1, h_2= g_2+1$}
        (-5.7,2) node []{$\psi(\boldsymbol{y})$}
        (-4.5,2)   node []{$\cdots$}
        (-2.5,2)   node []{$\cdots$}
        (-0.5,2)    node []{$\cdots$}
        (0.5,2)    node []{$g_2$}
        (1,2)    node []{$h_2$}
        (2.5,2)    node []{$\cdots$}
        
        (-5.7,3) node []{$\psi(\boldsymbol{x})$}
        (-4.5,3)   node []{$\cdots$}
        (-3.5,3)   node []{$g_1$}
        (-2,3)   node []{$\cdots$}
        (-1,3)   node []{$h_1$}
        (0.5,3)    node []{$\cdots$}
        (2.5,3)    node []{$\cdots$};

  \filldraw       (0,2)  circle [radius=1pt]
                  (-0.5,3)  circle [radius=1pt]
                  (0,3)     circle [radius=1pt]
                  (-1,2)  circle [radius=1pt]
                  (-1.5,2)  circle [radius=1pt]
                  (-4,2)  circle [radius=1pt]
                  (-3.5,2)     circle [radius=1pt]
                  (-2,2)  circle [radius=1pt]
                  (-3,2)    circle [radius=1pt]
                  (-5,2)  circle [radius=1pt]
                  (1,3)   circle [radius=1pt]
                  (1.5,2)     circle [radius=1pt]
                  (2,2)     circle [radius=1pt]
                  (3,2)     circle [radius=1pt]
                  (1,3)     circle [radius=1pt]
                  (1.5,2)     circle [radius=1pt]
                  (1.5,3)     circle [radius=1pt]
                  (-1.5,3)  circle [radius=1pt]
                  (-2.5,3)  circle [radius=1pt]
                  (-3,3)  circle [radius=1pt]
                  (-4,3)  circle [radius=1pt]
                  (-5,3)  circle [radius=1pt]
                  (1.5,3)     circle [radius=1pt]
                  (2,3)   circle [radius=1pt]
                  (3,3)   circle [radius=1pt];

  \draw        (-5,2.9)--(-5,2.1)
               (-4.5,2.9)--(-4.5,2.1)
               (-4,2.9)--(-4,2.1)
               (-2.5,2.9)--(-3,2.1)
               (-2,2.9)--(-2.5,2.1)
               (-1.5,2.9)--(-2,2.1)
               (0,2.9)--(-1,2.1)
               (1,2.9)--(0,2.1)
               (0.5,2.9)--(-0.5,2.1)
               (3,2.9)--(3,2.1)
               (2,2.9)--(2,2.1)
               (2.5,2.9)--(2.5,2.1);

  \fill[fill=green!20] (-3.8,1.2) -- (-2.6,1.2) -- (-3.4,-0.2) -- (-3.8,-0.2);

  \fill[fill=green!20] (-1.1,1.2) -- (0,1.2) -- (-1.5,-0.2) -- (-2,-0.2);

  \fill[fill=green!20] (1.5,1.2) -- (2.0,1.2) -- (1.1,-0.2) -- (-0.1,-0.2);

  \fill[fill=green!20] (3.8,-0.2) -- (2.6,-0.2) -- (3.4,1.2) -- (3.8,1.2);
  
  \draw (-9,0.5) node []{Case 4. $h_1> g_1+1, h_2> g_2+1$}
        (-5.7,0) node []{$\psi(\boldsymbol{y})$}
        (-4.5,0)   node []{$\cdots$}
        (-2.5,0)   node []{$\cdots$}
        (-0.5,0)    node []{$\cdots$}
        (0.5,0)    node []{$g_2$}
        (2,0)    node []{$\cdots$}
        (3,0)    node []{$h_2$}
        (4.5,0)    node []{$\cdots$}
        
        (-5.7,1) node []{$\psi(\boldsymbol{x})$}
        (-4.5,1)   node []{$\cdots$}
        (-3.5,1)   node []{$g_1$}
        (-2,1)   node []{$\cdots$}
        (-1,1)   node []{$h_1$}
        (0.5,1)    node []{$\cdots$}
        (2.5,1)    node []{$\cdots$}
        (4.5,1)    node []{$\cdots$};

  \filldraw       (0,0)  circle [radius=1pt]
                  (1,0)  circle [radius=1pt]
                  (-0.5,1)  circle [radius=1pt]
                  (0,1)     circle [radius=1pt]
                  (-1,0)  circle [radius=1pt]
                  (-1.5,0)  circle [radius=1pt]
                  (-4,0)  circle [radius=1pt]
                  (-3.5,0)     circle [radius=1pt]
                  (-2,0)  circle [radius=1pt]
                  (-3,0)    circle [radius=1pt]
                  (-5,0)  circle [radius=1pt]
                  (3,1)   circle [radius=1pt]
                  (3.5,0)     circle [radius=1pt]
                  (4,0)     circle [radius=1pt]
                  (5,0)     circle [radius=1pt]
                  (1,1)     circle [radius=1pt]
                  (1.5,0)     circle [radius=1pt]
                  (1.5,1)     circle [radius=1pt]
                  (2,1)     circle [radius=1pt]
                  (-1.5,1)  circle [radius=1pt]
                  (-2.5,1)  circle [radius=1pt]
                  (-3,1)  circle [radius=1pt]
                  (-4,1)  circle [radius=1pt]
                  (-5,1)  circle [radius=1pt]
                  (2.5,0)   circle [radius=1pt]
                  (3.5,1)     circle [radius=1pt]
                  (4,1)   circle [radius=1pt]
                  (5,1)   circle [radius=1pt];

  \draw        (-5,0.9)--(-5,0.1)
               (-4.5,0.9)--(-4.5,0.1)
               (-4,0.9)--(-4,0.1)
               (-2.5,0.9)--(-3,0.1)
               (-2,0.9)--(-2.5,0.1)
               (-1.5,0.9)--(-2,0.1)
               (0,0.9)--(-1,0.1)
               (1,0.9)--(0,0.1)
               (0.5,0.9)--(-0.5,0.1)
               (2.5,0.9)--(2,0.1)
               (2,0.9)--(1.5,0.1)
               (3,0.9)--(2.5,0.1)
               (5,0.9)--(5,0.1)
               (4,0.9)--(4,0.1)
               (4.5,0.9)--(4.5,0.1);
\end{tikzpicture}
\caption{Illustrations of $\psi(\boldsymbol{x})$ and $\psi(\boldsymbol{y})$ when $\boldsymbol{x}_{[n] \setminus \{g_1,h_1\}}= \boldsymbol{y}_{[n] \setminus \{g_2,h_2\}}$, where $g_1<h_1\leq g_2<h_2$. Each pair of bits connected by a solid line are of equal value, while those connected by a green undertone signify a connection resulted from deletions.}
\label{fig:2D_2}
\end{figure*}
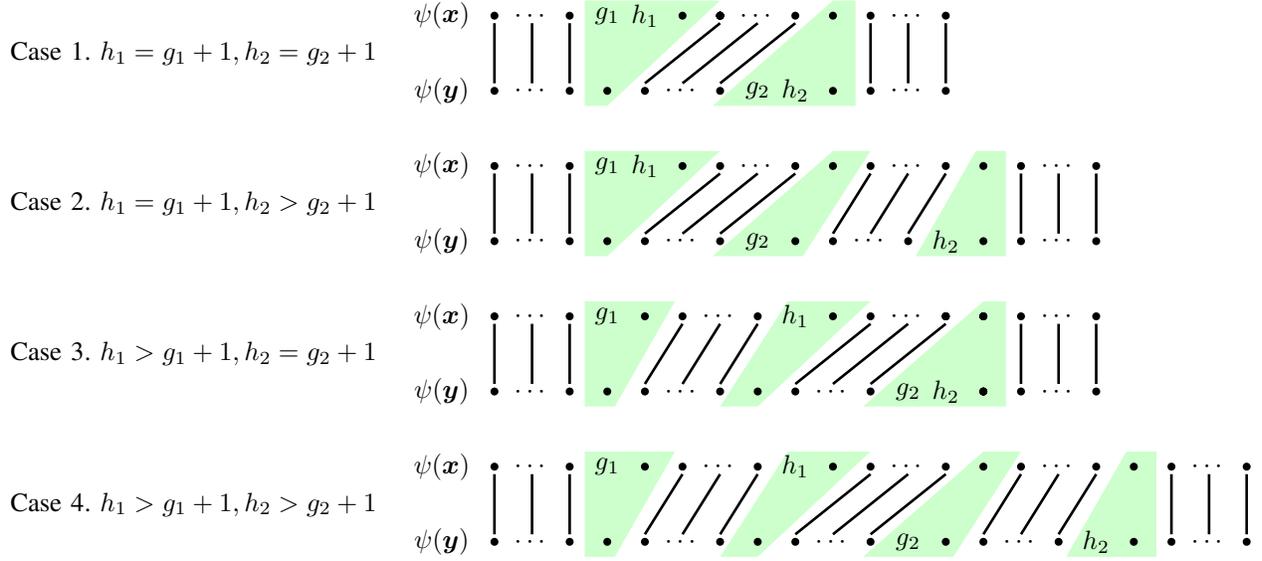

\begin{IEEEproof}
    By Observation \ref{obs:deletion} and Figure \ref{fig:2D_2}, the following holds.
    \begin{itemize}
    \item If $h_1=g_1+1$ and $h_2= g_2+1$, then $\psi(x)_{g_1} \oplus \psi(x)_{g_1+1} \oplus \psi(x)_{g_1+2}= \psi(y)_{g_1}$, $\psi(x)_{g_2+2}= \psi(y)_{g_2} \oplus \psi(y)_{g_2+1} \oplus \psi(y)_{g_2+2}$, and 
        \begin{equation*}
        \psi(x)_i=
          \begin{cases}
            \psi(y)_i, & \mbox{if } i \in [1,g_1-1] \cup [g_2+3,n]; \\
            \psi(y)_{i-2}, & \mbox{if } i \in [h_1+2, g_2+1].
        \end{cases}
        \end{equation*}
    \item If $h_1=g_1+1$ and $h_2> g_2+1$, then $\psi(x)_{g_1} \oplus \psi(x)_{g_1+1} \oplus \psi(x)_{g_1+2}= \psi(y)_{g_1}$, $\psi(x)_{g_2+2}= \psi(y)_{g_2} \oplus \psi(y)_{g_2+1}$, $\psi(x)_{h_2+1}= \psi(y)_{h_2} \oplus \psi(y)_{h_2+1}$, and 
        \begin{equation*}
        \psi(x)_i=
          \begin{cases}
            \psi(y)_i, & \mbox{if } i \in [1,g_1-1] \cup [h_2+2,n]; \\
            \psi(y)_{i-1}, & \mbox{if } i \in [g_2+3, h_2]; \\
            \psi(y)_{i-2}, & \mbox{if } i \in [h_1+2, g_2+1].
        \end{cases}
        \end{equation*}
    \item If $h_1>g_1+1$ and $h_2= g_2+1$, then $\psi(x)_{g_1} \oplus \psi(x)_{g_1+1}= \psi(y)_{g_1}$, $\psi(x)_{h_1} \oplus \psi(x)_{h_1+1}= \psi(y)_{h_1-1}$, $\psi(x)_{g_2+2}= \psi(y)_{g_2} \oplus \psi(y)_{g_2+1} \oplus \psi(y)_{g_2+2}$, and 
        \begin{equation*}
        \psi(x)_i=
          \begin{cases}
            \psi(y)_i, & \mbox{if } i \in [1,g_1-1] \cup [h_2+2,n]; \\
            \psi(y)_{i-1}, & \mbox{if } i \in [g_1+2, h_1-1]; \\
            \psi(y)_{i-2}, & \mbox{if } i \in [h_1+2, g_2+1].
        \end{cases}
        \end{equation*}
    \item If $h_1>g_1+1$ and $h_2> g_2+1$, then $\psi(x)_{g_1} \oplus \psi(x)_{g_1+1}= \psi(y)_{g_1}$, $\psi(x)_{h_1} \oplus \psi(x)_{h_1+1}= \psi(y)_{h_1-1}$, $\psi(x)_{g_2+2}= \psi(y)_{g_2} \oplus \psi(y)_{g_2+1}$, $\psi(x)_{h_2+1}= \psi(y)_{h_2} \oplus \psi(y)_{h_2+1}$, and 
        \begin{equation*}
        \psi(x)_i=
          \begin{cases}
            \psi(y)_i, & \mbox{if } i \in [1,g_1-1] \cup [h_2+2,n]; \\
            \psi(y)_{i-1}, & \mbox{if } i \in [g_1+2, h_1-1] \cup [g_2+3,h_2]; \\
            \psi(y)_{i-2}, & \mbox{if } i \in [h_1+2,g_2+1].
        \end{cases}
        \end{equation*}
    \end{itemize}
    Then the conclusion follows.
\end{IEEEproof}

\subsubsection{The Complement Sequences}

In the proof of Theorem \ref{thm:2D}, sometimes we consider the complement sequences to reduce unnecessary discussions, thus we need to explore some properties regarding $\overline{\boldsymbol{x}}$ and $\overline{\boldsymbol{y}}$.

\begin{claim}\label{cla:complement}
$\mathrm{VT}^{(k)}(\overline{\boldsymbol{x}})= \mathrm{VT}^{(k)}(\overline{\boldsymbol{y}})$ if $\mathrm{VT}^{(k)}(\boldsymbol{x})= \mathrm{VT}^{(k)}(\boldsymbol{y})$ for any non-negative integer $k$.
\end{claim}

\begin{IEEEproof}
    Since $\mathrm{VT}^{(k)}(\overline{\boldsymbol{x}})= \sum_{i=1}^{n} c_i^{(k)} (1-x_i)= \sum_{i=1}^{n} c_i^{(k)}- \sum_{i=1}^{n}  c_i^{(k)} x_i= \sum_{i=1}^{n} c_i^{(k)}- \mathrm{VT}^{(k)}(\boldsymbol{x})$, the claim holds.
\end{IEEEproof}

\begin{claim}
    When a sequence is a strong-$(\ell,\epsilon)$-locally-balanced sequence, its complement is also a strong-$(\ell,\epsilon)$-locally-balanced sequence.
\end{claim}

\begin{IEEEproof}
    Assume $\boldsymbol{x}$ is a strong-$(\ell,\epsilon)$-locally-balanced sequence, then for any $1 \leq i \leq n$ and $\ell' \geq \ell$ with $i+\ell'-1 \leq n$, we have $\mathrm{VT}^{(0)}(\boldsymbol{x}_{[i,i+\ell'-1]}) \in [ (\frac{1}{2}-\epsilon) \ell', (\frac{1}{2}+\epsilon) \ell']$. 
    We may compute $\mathrm{VT}^{(0)}(\overline{\boldsymbol{x}}_{[i,i+\ell'-1]})= \ell'- \mathrm{VT}^{(0)}(\boldsymbol{x}_{[i,i+\ell'-1]}) \in [ (\frac{1}{2}-\epsilon) \ell', (\frac{1}{2}+\epsilon) \ell']$.
    Therefore, $\overline{\boldsymbol{x}}$ is also a strong-$(\ell,\epsilon)$-locally-balanced sequence, which completes the proof.
\end{IEEEproof}

\subsection{Proof of Theorem \ref{thm:2D}}

Assume there exist two distinct sequences $\boldsymbol{x}, \boldsymbol{y} \in \mathcal{C}_{2D}$, such that $\boldsymbol{x}_{[n] \setminus \{g_1,h_1\}}= \boldsymbol{y}_{[n] \setminus \{g_2,h_2\}}$ for some indices $g_1,h_1,g_2,h_2 \in [n]$ with $g_1<h_1$, $g_2< h_2$, and $g_1\leq g_2$. 
We first distinguish between the following three cases based on the size relationship of $g_1,h_1,g_2,h_2$.
\begin{enumerate}
  \item $g_1 \leq g_2 < h_2 \leq h_1$;
  \item $g_1 \leq g_2 < h_1 \leq h_2$;
  \item $g_1 < h_1 \leq g_2 < h_2$.
\end{enumerate}
The first two cases can be easily handled and can be ruled out using Claims \ref{cla:2D} and \ref{cla:Sima}.
Our main focus is on the last case, where we have $g_1 < h_1 \leq g_2 < h_2$. 
This case is the most challenging one compared to the previous two cases.
In this case, it is reasonable to assume that $h_2-g_1 \geq P$ since $\mathcal{C}_{2D}$ is a $P$-bounded two-deletion correcting code.

Firstly, we consider the number of ones that changed in $\psi(\boldsymbol{x})$.
Assume that two deletions in $\boldsymbol{x}$ decrease the number of ones in $\psi(\boldsymbol{x})$ by $N$. By Observation \ref{obs:deletion}, it can be deduced that $N \in \{0, 2, 4\}$.
Then, the value of $N$ can be determined by $\mathrm{VT}^{(0)}(\psi(\boldsymbol{x})) \pmod{5}$ and $\mathrm{VT}^{(0)}(\psi(\boldsymbol{x})_{[n] \setminus \{g_1,h_1\}})$.
Now we distinguish between the following three cases according to the value of $N$.

If $N=0$, by Observation \ref{obs:deletion}, we may compute $\delta \geq 0$, where $\delta$ is defined in Claim \ref{cla:2D_VT-differential_sequence}.
It follows that
\begin{align*}
  \Delta^{(1)}_{\psi}
  & \geq \sum_{i \in [g_1+2, h_1-1] \cup [g_2+3,h_2]} \psi(x)_i+ 2\sum_{i \in [h_1+2,g_2+1]} \psi(x)_i \\
  & \geq \sum_{i \in [g_1+2,h_2] \setminus \{h_1,h_1+1,g_2+2\}} \psi(x)_i \\
  & \overset{(\ast)}{\geq} \left(  \frac{1}{2}-\epsilon \right) (h_2-g_1)- 4> 0,
\end{align*}
where ($\ast$) holds since $\psi(\boldsymbol{x})$ is a strong-$(\ell,\epsilon)$-locally-balanced sequence and $h_2-g_1 \geq P >\ell$.
 
If $N=4$, by Observation \ref{obs:deletion}, we have $h_1>g_1+1$, $h_2>g_2+1$, and $\psi(x)_i=1$ for $i \in \{g_1,g_1+1,h_1,h_1+1,g_2,g_2+1,h_2,h_2+1\}$. Then by Claim \ref{cla:2D_VT-differential_sequence}, we may compute $\delta= -2(h_2-g_1)- 2(g_2-h_1)$.
It follows that
\begin{align*}
  \Delta^{(1)}_{\psi} 
  &= -2(h_2-g_1)- 2(g_2-h_1)+ \sum_{i \in [g_1+2, h_1-1] \cup [g_2+3,h_2]} \psi(x)_i+ 2\sum_{i \in [h_1+2,g_2+1]} \psi(x)_i \\
  &\leq -2(h_2-g_1)- 2(g_2-h_1)+ (h_2-g_1)+ 2(g_2-h_1) \\
  &= -(h_2-g_1)< 0.
\end{align*}

If $N=2$, we further consider the number of ones that changed in $\boldsymbol{x}$.
Assume that two deletions in $\boldsymbol{x}$ decrease the number of ones in $\boldsymbol{x}$ by $N'$, then $N' \in \{0,1,2\}$.
Moreover, the value of $N'$ can be determined by $\mathrm{VT}^{(0)}(\boldsymbol{x}) \pmod{3}$ and $\mathrm{VT}^{(0)}(\boldsymbol{x}_{[n] \setminus \{g_1,h_1\}})$.
The case where $N' \in \{0,2\}$ can be ruled out by Claims \ref{cla:2D} and \ref{cla:Guruswami}.
Therefore, we need to focus on the case where $N'=1$.
In this case, $(x_{g_1}, x_{h_1}),(y_{g_2}, y_{h_2}) \in \{(0,1), (1,0)\}$.
Below we distinguish between the following two cases based on the value of $(x_{g_1}, x_{h_1})$ and $(y_{g_2}, y_{h_2})$
and in both cases we will show that either $\Delta^{(1)}_{\psi} \neq 0$ or $\Delta^{(k)} \neq 0$ for some $k \in \{1,2\}$.
Once this is established, we complete the proof for that $\mathcal{C}_{2D}$ is a two-deletion correcting code.

\begin{itemize}
  \item If $\{(x_{g_1}, x_{h_1}), (y_{g_2}, y_{h_2})\}= \{(0,1), (1,0)\}$, then by Equation (\ref{eq:delta_k}),
      \begin{itemize}
        \item when $(x_{g_1}, x_{h_1})= (1,0)$ and $(y_{g_2}, y_{h_2})= (0,1)$, we have $\Delta^{(1)}= g_1-h_2+ \sum_{i \in [g_1+1, h_2] \setminus \{h_1\}} x_i+ \sum_{i \in [h_1+1,g_2+1]} x_i$. It follows by Claim \ref{cla:2D} that 
            \begin{align}\label{eq:h_2g_1}
              h_2-g_1= \sum_{i \in [g_1+1, h_2] \setminus \{h_1\}} x_i+ \sum_{i \in [h_1+1,g_2+1]} x_i.
            \end{align}
        \item when $(x_{g_1}, x_{h_1})= (0,1)$ and $(y_{g_2}, y_{h_2})= (1,0)$, we consider $\overline{\boldsymbol{x}}$ and $\overline{\boldsymbol{y}}$ and have $\mathrm{VT}^{(1)}(\overline{\boldsymbol{x}})- \mathrm{VT}^{(1)}(\overline{\boldsymbol{y}})= g_1-h_2+ \sum_{i \in [g_1+1, h_2] \setminus \{h_1\}} \overline{x}_i+ \sum_{i \in [h_1+1,g_2+1]} \overline{x}_i$. It follows by Claims \ref{cla:2D} and \ref{cla:complement} that 
            \begin{align}\label{eq:h_2g_1'}
              h_2-g_1= \sum_{i \in [g_1+1, h_2] \setminus \{h_1\}} \overline{x}_i+ \sum_{i \in [h_1+1,g_2+1]} \overline{x}_i.
            \end{align}
      \end{itemize}
    In the following discussion, we only care about whether a sequence is a strong-locally-balanced sequence, regardless of what the sequence actually is. 
    Therefore, Equations (\ref{eq:h_2g_1}) and (\ref{eq:h_2g_1'}) are essentially the same.
    Without loss of generality assume Equation (\ref{eq:h_2g_1}) holds, i.e., $(x_{g_1}, x_{h_1})= (1,0)$ and $(y_{g_2}, y_{h_2})= (0,1)$.
    Since $\boldsymbol{x}$ is a strong-$(\ell,\epsilon)$-locally-balanced sequence and $h_2- g_1 \geq P > \ell$, we have $(\frac{1}{2}-\epsilon)(h_2-g_1)-1 \leq \sum_{i \in [g_1+1, h_2] \setminus \{h_1\}} x_i \leq (\frac{1}{2}+\epsilon)(h_2-g_1)$. 
    It follows by Equation (\ref{eq:h_2g_1}) that
    \begin{align*}
      \sum_{i \in [h_1+1,g_2+1]} x_i
      &= h_2- g_1- \sum_{i \in [g_1+1, h_2] \setminus \{h_1\}} x_i  \\
      &\geq \left(  \frac{1}{2}-\epsilon \right) (h_2-g_1) \geq \left(  \frac{1}{2}-\epsilon \right) P \geq \ell.
    \end{align*}
    Then, $g_2-h_1 \geq \ell$ and $\sum_{i \in [h_1+1,g_2+1]} x_i \leq (\frac{1}{2}+\epsilon)(g_2-h_1+1)$.
    Therefore, $(\frac{1}{2}-\epsilon)(h_2-g_1) \leq (\frac{1}{2}+\epsilon)(g_2-h_1+1)$, i.e., $h_2-g_1 \leq \frac{\frac{1}{2}+\epsilon}{\frac{1}{2}-\epsilon}(g_2-h_1+1)$.
    
    Now we move on to calculate Equation (\ref{eq:delta_psi_k}). By Observation \ref{obs:deletion}, we may compute $\delta \leq -2(g_2-h_1)+2$, where $\delta$ is defined in Claim \ref{cla:2D_VT-differential_sequence}. 
    Then it follows that
    \begin{align*}
      \Delta^{(1)}_{\psi}
      & \leq -2(g_2-h_1)+2 + \sum_{i \in [g_1+2, g_2+1] \setminus \{h_1,h_1+1\}} \psi(x)_i+ \sum_{i \in [h_1+2,h_2] \setminus \{g_2+2\}} \psi(x)_i \\
      & \leq -2(g_2-h_1) + 2\sum_{i \in [g_1+1, h_2]} \psi(x)_i \\
      & \overset{(\ast)}{\leq} -2(g_2-h_1)+ 2 \left(  \frac{1}{2}+\epsilon \right)(h_2-g_1) \\
      & \leq -2(g_2-h_1)+ 2 \frac{(\frac{1}{2}+\epsilon)^2}{\frac{1}{2}-\epsilon}(g_2-h_1+1) \\
      &= -\frac{11}{18}(g_2-h_1)+ \frac{25}{18}<0,
    \end{align*}
    where $(\ast)$ holds since $\psi(\boldsymbol{x})$ is a strong-$(\ell,\epsilon)$-locally-balanced sequence and $h_2-g_1 \geq P> \ell$.
  \item If $(x_{g_1}, x_{h_1})=(y_{g_2}, y_{h_2}) \in \{(1,0), (0,1)\}$, similar to the previous case, we can assume $(x_{g_1}, x_{h_1})= (y_{g_2}, y_{h_2})= (1,0)$.
      We first calculate Equation (\ref{eq:delta_k}) for $k=1$ and have the following two expressions:
  \begin{equation}\label{eq:Delta_1}
    \begin{aligned}
      \Delta^{(1)}
      &= 
        g_1-g_2+ \sum_{i \in [g_1+1, h_2] \setminus \{h_1\}} x_i+ \sum_{i \in [h_1+1,g_2+1]} x_i\\
      &=
        -\sum_{i \in [g_1+1, g_2+1] \setminus \{h_1\}} (1-x_i)+ \sum_{i \in [h_1+1,h_2]} x_i.
    \end{aligned}
    \end{equation}
    Recall that $\boldsymbol{x}$ is a strong-$(\ell,\epsilon)$-locally-balanced sequence and $h_2- g_1 \geq P > \ell$, we have $(\frac{1}{2}-\epsilon)(h_2-g_1)-1 \leq \sum_{i \in [g_1+1, h_2] \setminus \{h_1\}} x_i \leq (\frac{1}{2}+\epsilon)(h_2-g_1)$. 
    Since $\Delta^{(1)}=0$ (by Claim \ref{cla:2D}), it follows by the first expression of Equation (\ref{eq:Delta_1}) that
    \begin{equation}\label{eq:g_1g_2}
    \begin{aligned}
      g_2-g_1 \geq \sum_{i \in [g_1+1, h_2] \setminus \{h_1\}} x_i \geq \left(  \frac{1}{2}-\epsilon \right)(h_2-g_1)-1 \geq \left(  \frac{1}{2}-\epsilon \right) P-1 > \ell.
    \end{aligned}
    \end{equation}
    Notice that $\sum_{i \in [g_1+1, g_2+1] \setminus \{h_1\}} (1-x_i)$ equals the number of zeros in $\boldsymbol{x}_{[g_1+1,g_2+1] \setminus \{h_1\}}$, we have $\ell< (\frac{1}{2}-\epsilon)(g_2-g_1+1)-1 \leq \sum_{i \in [g_1+1, g_2+1] \setminus \{h_1\}} (1-x_i) \leq (\frac{1}{2}+\epsilon)(g_2-g_1+1)$. It follows by the second expression of Equation (\ref{eq:Delta_1}) that $\sum_{i \in [h_1+1,h_2]} x_i= \sum_{i \in [g_1+1, g_2+1] \setminus \{h_1\}} (1-x_i) \geq \ell$. Then $h_2-h_1 > \ell$ and $(\frac{1}{2}-\epsilon)(h_2-h_1) \leq \sum_{i \in [h_1+1,h_2]} x_i \leq (\frac{1}{2}+\epsilon)(h_2-h_1)$.
    We may obtain the inequalities
    \begin{equation}\label{eq:gh}
      \frac{(\frac{1}{2}-\epsilon)(g_2-g_1+1)-1}{\frac{1}{2}+\epsilon} \leq h_2-h_1 \leq \frac{\frac{1}{2}+\epsilon}{\frac{1}{2}-\epsilon}(g_2-g_1+1),
    \end{equation}
    and
    \begin{equation}\label{eq:gh'}
      \frac{\frac{1}{2}-\epsilon}{\frac{1}{2}+\epsilon}(h_2-h_1) \leq g_2-g_1+1 \leq \frac{(\frac{1}{2}+\epsilon)(h_2-h_1)+1}{\frac{1}{2}-\epsilon}.
    \end{equation}
    Now we move on to calculate Equation (\ref{eq:delta_psi_k}) and further distinguish between the following four cases based on which deletion decreases the number of ones in $\psi(\boldsymbol{x})$ and $\psi(\boldsymbol{y})$, respectively. 
    \begin{itemize}
      \item Deleting $x_{g_1}$ and $y_{g_2}$ decreases the number of ones by two in $\psi(\boldsymbol{x})$ and $\psi(\boldsymbol{y})$, respectively. 
          Recall that $h_2-h_1 > \ell$ and $g_2-g_1 > \ell$ (from above arguments).
          Now by Observation \ref{obs:deletion}, we may compute $\delta \leq -2(g_2-g_1)+2$, where $\delta$ is defined in Claim \ref{cla:2D_VT-differential_sequence}.
          Then it follows that
          \begin{equation}\label{eq:Delta_psi}
          \begin{aligned}
              \Delta^{(1)}_{\psi}
              &\leq -2(g_2-g_1)+2 + \sum_{i \in [g_1+2, g_2+1] \setminus \{h_1,h_1+1\}} \psi(x)_i+ \sum_{i \in [h_1+2,h_2] \setminus \{g_2+2\}} \psi(x)_i \\
              &\overset{(\ast)}{\leq} -2(g_2-g_1)+2 + \left(  \frac{1}{2}+\epsilon \right) (g_2-g_1)+ \left(  \frac{1}{2}+\epsilon \right)(h_2-h_1) \\
              &\overset{(\star)}{\leq} -2(g_2-g_1)+2 + \left(  \frac{1}{2}+\epsilon \right)(g_2-g_1)+ \frac{(\frac{1}{2}+\epsilon)^2}{\frac{1}{2}-\epsilon}(g_2-g_1+1) \\
              &= -\frac{3}{4}(g_2-g_1)+\frac{97}{36}< 0,
          \end{aligned}
          \end{equation}
          where $(\ast)$ holds since $\psi(\boldsymbol{x})$ is a strong-$(\ell,\epsilon)$-locally-balanced sequence and $(\star)$ holds by Equation (\ref{eq:gh}).
      
      \item Deleting $x_{h_1}$ and $y_{h_2}$ decreases the number of ones by two in $\psi(\boldsymbol{x})$ and $\psi(\boldsymbol{y})$, respectively. 
          This case is similar to the previous one.
          We present it for completeness.
          Recall that $h_2-h_1 > \ell$ and $g_2-g_1 > \ell$, by Observation \ref{obs:deletion}, we may compute $\delta \leq  -2(h_2-h_1)+2$, where $\delta$ is defined in Claim \ref{cla:2D_VT-differential_sequence}.
      Then it follows that
      \begin{align*}
          \Delta^{(1)}_{\psi}
          &\leq -2(h_2-h_1)+2 + \sum_{i \in [g_1+2, g_2+1] \setminus \{h_1,h_1+1\}} \psi(x)_i+ \sum_{i \in [h_1+2,h_2] \setminus \{g_2+2\}} \psi(x)_i \\
          &\overset{(\ast)}{\leq} -2(h_2-h_1)+2 + \left(  \frac{1}{2}+\epsilon \right)(g_2-g_1)+ \left(  \frac{1}{2}+\epsilon \right)(h_2-h_1) \\
          &\overset{(\star)}{\leq} -2(h_2-h_1)+2+ \frac{\frac{1}{2}+\epsilon}{\frac{1}{2}-\epsilon}\left(\left( \frac{1}{2}+\epsilon \right)(h_2-h_1)+1 \right) + \left(  \frac{1}{2}+\epsilon \right)(h_2-h_1) \\
          &= -\frac{3}{4}(h_2-h_1)+\frac{13}{4}< 0,
      \end{align*}
      where $(\ast)$ holds since $\psi(\boldsymbol{x})$ is a strong-$(\ell,\epsilon)$-locally-balanced sequence and $(\star)$ holds by Equation (\ref{eq:gh'}).
      
      \item Deleting $x_{g_1}$ and $y_{h_2}$ decreases the number of ones by two in $\psi(\boldsymbol{x})$ and $\psi(\boldsymbol{y})$, respectively. In this case, by Observation \ref{obs:deletion}, we may compute $\delta \leq -2(h_2-g_1)+2 \leq -2(g_2-g_1)+2$, where $\delta$ is defined in Claim \ref{cla:2D_VT-differential_sequence}. It follows by Equation (\ref{eq:Delta_psi}) that $\Delta^{(1)}_{\psi}<0$.
          
      \item Deleting $x_{h_1}$ and $y_{g_2}$ decreases the number of ones by two in $\psi(\boldsymbol{x})$ and $\psi(\boldsymbol{y})$, respectively (this case is significantly different from the previous three cases). 
          Notice that by the second expression of Equation (\ref{eq:Delta_1}), $m \triangleq |\{i \in [g_1+1,g_2+1] \setminus \{h_1\}: x_i= 0\}|= |\{i \in [h_1+1,h_2]: x_i= 1\}|$.
            Set $m_1 \triangleq |\{i \in [g_1+1,h_1-1]: x_i=0\}|$, it follows that $|\{i \in [h_1+1,g_2+1]: x_i=0\}|= m-m_1$.
            Moreover, set $m_2 \triangleq |\{i \in [h_1+1,g_2+1]: x_i=1\}|$, it follows that $|\{i \in [g_2+2,h_2]: x_i=1\}|= m-m_2$.
            Now, we calculate Equation (\ref{eq:delta_k}) for $k=2$ and obtain
          \begin{equation}\label{eq:m_2m_1}
          \begin{aligned}
            \Delta^{(2)}
            &= -\sum_{i \in [g_1+1,g_2]} i + \sum_{i \in [g_1+1,h_1-1] \cup [g_2+2,h_2]} i x_i+ \sum_{i \in [h_1+1, g_2+1]} (i-1+i) x_i \\
            &= -\sum_{i \in [g_1+1, h_1-1]} i(1-x_i)- \sum_{i \in [h_1+1, g_2+1]} (i-1)(1-x_{i})+ \sum_{i \in [g_2+2,h_2]} ix_i+ \sum_{i \in [h_1+1,g_2+1]} ix_i \\
            &\geq
            \begin{cases}
              m_2^2+ (m-m_2)^2, & \mbox{if } m_1 \geq m_2 \\
              m_1^2- (m_2-m_1)^2+ (m-m_2)^2, & \mbox{if } m_1 < m_2
            \end{cases}.
          \end{aligned}
          \end{equation}
          The last inequality can be justified by considering the extreme case where $\boldsymbol{x}_{[g_1+1,h_1-1]}= 1^{h_1-g_1-1-m_1} 0^{m_1}$, $\boldsymbol{x}_{[h_1+1,g_2+1]}= 1^{m_2} 0^{m-m_1}$, and $\boldsymbol{x}_{[g_2+2,h_2]}= 1^{m-m_2} 0^{h_2-g_2-1-(m-m_2)}$.
          Obviously, $\Delta^{(2)}> 0$ when $m_1 \geq m_2$.
          When $m_1< m_2$, we proceed to bound the values of $m_1$ and $m_2$.
          
          Firstly, by Observation \ref{obs:deletion}, we may compute $-2(g_2-h_1) \leq \delta \leq -2(g_2-h_1)+2$, where $\delta$ is defined in Claim \ref{cla:2D_VT-differential_sequence}. 
          Since $\Delta^{(1)}_{\psi}=0$ (by Claim \ref{cla:2D_psi}), it follows that $-\delta= \sum_{i \in [g_1+2, g_2+1] \setminus \{h_1,h_1+1\}} \psi(x)_i+ \sum_{i \in [h_1+2,h_2] \setminus \{g_2+2\}} \psi(x)_i$. Recall that $h_2-h_1 > \ell$ and $g_2-g_1> \ell$, it follows that
          \begin{equation*}
              \left(  \frac{1}{2}-\epsilon \right)(g_2-g_1)+ \left(  \frac{1}{2}-\epsilon \right) (h_2-h_1)-4 \leq -\delta \leq \left(  \frac{1}{2}+\epsilon \right)(g_2-g_1)+ \left(  \frac{1}{2}+\epsilon \right)(h_2-h_1).
          \end{equation*}
          Combining this with Equation (\ref{eq:gh}), we may compute
          \begin{equation*}
          \begin{aligned}
            \frac{\frac{1}{2}-\epsilon}{\frac{1}{2}+\epsilon}(g_2-g_1)-5 \leq 2(g_2-h_1) \leq \frac{\frac{1}{2}+\epsilon}{\frac{1}{2}-\epsilon}(g_2-g_1)+3.
          \end{aligned}
          \end{equation*}
          Now by Equation (\ref{eq:g_1g_2}), we may compute 
          \begin{align*}
            g_2-h_1 
            &\geq \frac{1}{2}\frac{\frac{1}{2}-\epsilon}{\frac{1}{2}+\epsilon}(g_2-g_1)-3 \\
            &\geq \frac{1}{2}\frac{\frac{1}{2}-\epsilon}{\frac{1}{2}+\epsilon}\left( \left(  \frac{1}{2}-\epsilon \right) P-1 \right)-3 \\
            &\geq \frac{8}{45}P- 4 \geq \ell
          \end{align*}
          and 
            \begin{align*}
                h_1-g_1
                &= (g_2-g_1)- (g_2-h_1) \\
                & \geq (g_2-g_1)- \frac{1}{2}\frac{\frac{1}{2}+\epsilon}{\frac{1}{2}-\epsilon}(g_2-g_1)-2 \\
                &= \left(1- \frac{1}{2}\frac{\frac{1}{2}+\epsilon}{\frac{1}{2}-\epsilon}\right) (g_2-g_1) -2 \\
                &\geq \left(1- \frac{1}{2}\frac{\frac{1}{2}+\epsilon}{\frac{1}{2}-\epsilon}\right) \left( \left(  \frac{1}{2}-\epsilon \right) P-1 \right) - 2 \\
                &\geq \frac{1}{6}P-3 \geq \ell.
            \end{align*}
            Then it follows that
            \begin{align*}
              m_2= \sum_{i \in [h_1+1,g_2+1]} x_i
              &\leq \left(  \frac{1}{2}+ \epsilon \right)(g_2-h_1+1) \\
              &\leq \frac{1}{2}\frac{(\frac{1}{2}+\epsilon)^2}{\frac{1}{2}-\epsilon}(g_2-g_1)+3 \\
              &= \frac{25}{72}(g_2-g_1)+O(1)
            \end{align*}
            and
            \begin{align*}
              m_1= \sum_{i \in [g_1+1,h_1-1]} (1-x_i)
              & \geq \left(  \frac{1}{2}-\epsilon \right)(h_1-g_1)-1 \\
              &\geq \left(  \frac{1}{2}-\epsilon \right) \left( \left( 1-\frac{1}{2}\frac{\frac{1}{2}+\epsilon}{\frac{1}{2}-\epsilon} \right) (g_2-g_1)-2 \right)- 1\\
              &\geq \left( \frac{1}{4}-\frac{3}{2}\epsilon \right) (g_2-g_1)-3 \\
              &=\frac{1}{6}(g_2-g_1)+O(1).
            \end{align*}
            Notice that $m \geq (\frac{1}{2}-\epsilon)(g_2-g_1+1)-1= \frac{4}{9}(g_2-g_1)+O(1)$, we may compute
              \begin{align*}
              \Delta^{(2)} 
                &\geq m_1^2- (m_2-m_1)^2+ (m-m_2)^2 \\
                &\geq \left( \frac{1}{6}(g_2-g_1)+O(1) \right)^2- \left( \left( \frac{25}{72}-\frac{1}{6} \right)(g_2-g_1)+O(1) \right)^2+ \left( \left( \frac{4}{9}-\frac{25}{72}\right)(g_2-g_1)+O(1) \right)^2 \\
                &= \left(\frac{1}{36}- \left( \frac{25}{72}-\frac{1}{6} \right)^2+ \left( \frac{4}{9}-\frac{25}{72}\right)^2 \right)(g_2-g_1)^2+O(g_2-g_1)\\
                &= \frac{1}{216}(g_2-g_1)^2+O(g_2-g_1)> 0.
              \end{align*}
    \end{itemize}
\end{itemize}

In all cases, we have shown that either $\Delta^{(1)}_{\psi} \neq 0$ or $\Delta^{(k)} \neq 0$ for some $k \in \{1,2\}$, which contradicts the conclusions of Claims \ref{cla:2D} and \ref{cla:2D_psi}.
Therefore, $\mathcal{C}_{2D}$ is a two-deletion correcting code.

\begin{remark}\label{rmk:2D}
    Notice that in the second and third conditions outlined in Theorem \ref{thm:2D}, when the VT syndrome is used modulo a larger number, Claims \ref{cla:2D} and \ref{cla:2D_psi} still hold.
    Consequently, one can check that Theorem \ref{thm:2D} still holds when the VT syndrome is used modulo a larger number.
\end{remark}

\section{Codes for correcting up to two substitutions}\label{sec:2S}

In this section, we consider the construction of two-substitution correcting codes.
Remarkably, instead of constructing two-substitution correcting codes with the smallest possible redundancy, our objective is to construct codes with the smallest possible redundancy that can correct not only two deletions but also up to two substitutions.
Therefore, the best choice would be to add minimal redundancy to our existing two-deletion correcting codes, enabling them to correct up to two substitutions as well.
Below we will demonstrate that our two-deletion correcting codes, constructed in Theorem \ref{thm:2D}, can correct up to two substitutions as well after making a slight modification.

\begin{theorem}\label{thm:2S}
      For fixed integers $b_0 \in [[4]], b_1 \in [[4n]], b_2 \in [[4n^2]]$, we define the code $\mathcal{C}_{2S} \subseteq \Sigma^n$ in which each sequence $\boldsymbol{x}$ satisfies $\mathrm{VT}^{(k)}(\boldsymbol{x}) \equiv b_k \pmod{4n^k+1}$ for $k \in \{0,1,2\}$.
  $\mathcal{C}_{2S}$ is a two-substitution correcting code.
\end{theorem}

\begin{IEEEproof}
Assume $\boldsymbol{x} \in \mathcal{C}_{2S}$ suffers two substitutions at positions $i$ and $j$, where $i\leq j$, resulting in the sequence $\boldsymbol{z}$. 
In the case where $i=0$ or $j=0$, it indicates that $\boldsymbol{x}$ has suffered fewer than two substitutions.
Notice that to recover $\boldsymbol{x}$ from $\boldsymbol{z}$, it is sufficient to find $i$ and $j$.
    For $k \in \{0,1,2\}$, let $\tilde{\Delta}^{(k)} \triangleq b_k- \mathrm{VT}^{(k)}(\boldsymbol{z}) \pmod{4n^k+1} \in [-2n^k,2n^k]$. Since one substitution in $\boldsymbol{x}$ decreases the value of the $k$-th order VT syndrome of $\boldsymbol{x}$ within the range of $-n^{k}$ to $n^{k}$, we may compute
\begin{equation}\label{eq:1}
   \begin{aligned}
     &\Delta^{(0)}\triangleq \tilde{\Delta}^{(0)}= (2x_i-1)+(2x_j-1), \\
     &\Delta^{(1)}\triangleq \tilde{\Delta}^{(1)} = i(2x_i-1)+j(2x_j-1), \\
     &\Delta^{(2)}\triangleq 2\tilde{\Delta}^{(2)}-\tilde{\Delta}^{(1)} = i^2(2x_i-1)+j^2(2x_j-1).
   \end{aligned}
\end{equation}
Now we distinguish between the following five cases according to the value of $\Delta^{(0)}$.
\begin{itemize}
  \item If $\Delta^{(0)}= 2$, it indicates that $x_i= x_j =1$. Then by Equation (\ref{eq:1}), we may compute $i= \frac{1}{2}(\Delta^{(1)}- \sqrt{2\Delta^{(2)}- (\Delta^{(1)})^2})$ and $j= \frac{1}{2}(\Delta^{(1)}+ \sqrt{2\Delta^{(2)}- (\Delta^{(1)})^2})$.
  \item If $\Delta^{(0)}= 1$, it indicates that $i=0$ and $x_j= 1$. Then by Equation (\ref{eq:1}), we may compute $j= \Delta^{(1)}$.
  \item If $\Delta^{(0)}= 0$, it indicates that $i=j=0$ or $(x_i, x_j) \in \{(0,1), (1,0)\}$. Now we consider $\Delta^{(1)}$.
      \begin{itemize}
        \item If $\Delta^{(1)}> 0$, it indicates that $(x_i, x_j)= (0,1)$. Then by Equation (\ref{eq:1}), we may compute $i= \frac{1}{2}(\frac{\Delta^{(2)}}{\Delta^{(1)}}- \Delta^{(1)})$ and  $j= \frac{1}{2}(\frac{\Delta^{(2)}}{\Delta^{(1)}}+ \Delta^{(1)})$.
        \item If $\Delta^{(1)}= 0$, it indicates that $i=j=0$.
        \item If $\Delta^{(1)}< 0$, it indicates that $(x_i, x_j)= (1,0)$. Then by Equation (\ref{eq:1}), we may compute $i= \frac{1}{2}(\frac{\Delta^{(2)}}{\Delta^{(1)}}+ \Delta^{(1)})$ and  $j= \frac{1}{2}(\frac{\Delta^{(2)}}{\Delta^{(1)}}- \Delta^{(1)})$.
      \end{itemize}
  \item If $\Delta^{(0)}= -1$, it indicates that $i=0$ and $x_j= 0$. Then by Equation (\ref{eq:1}), we may compute $j= -\Delta^{(1)}$.
  \item If $\Delta^{(0)}= -2$, it indicates that $x_i= x_j =0$. Then by Equation (\ref{eq:1}), we may compute $i= \frac{1}{2}(-\Delta^{(1)}- \sqrt{-2\Delta^{(2)}- (\Delta^{(1)})^2})$ and $j= \frac{1}{2}(-\Delta^{(1)}+ \sqrt{-2\Delta^{(2)}- (\Delta^{(1)})^2})$.
\end{itemize}
In all cases, we can determine the values of $i$ and $j$, which implies that $\boldsymbol{x}$ can be recovered.
Therefore, $\mathcal{C}_{2S}$ is a two-substitution correcting code.
\end{IEEEproof}

\section{codes for correcting one deletion and one substitution}\label{sec:1D1S}

In this section, we consider the construction of single-deletion single-substitution correcting codes.
Notice that the best previously known result requires a redundancy of $6 \log n+O(1)$. Our main contribution in this section is to improve the redundancy to $4 \log n+O(\log\log n)$ by using the ideas developed in Section \ref{sec:2D}.

\begin{theorem}\label{thm:DS}
    Set $\epsilon= \frac{1}{18}$, $\ell= 1296 \log n$, and $P= 6(\ell+3)$.
    For fixed sequences $\boldsymbol{a}= a_0a_1, \boldsymbol{b}= b_0b_1b_2, \boldsymbol{c}= c_1c_2c_3, \boldsymbol{c}'= c_1'c_2'c_3'$, where $a_0 \in [[3]], a_1 \in [[3n]], b_0 \in [[6]], b_1 \in [[6n+6]], b_2 \in [[6(n+1)^2]], c_1,c_1' \in [[6P-1]], c_2,c_2' \in [[12P^2-1]], c_3,c_3' \in [[24P^3-1]]$, we define the code $\mathcal{C}_{DS} \triangleq \mathcal{C}_{DS}(\boldsymbol{a},\boldsymbol{b},\boldsymbol{c},\boldsymbol{c}') \subseteq \Sigma^n$ in which each sequence $\boldsymbol{x}$ satisfies the following conditions:
    \begin{itemize}
        \item both $\boldsymbol{x}$ and $\psi(\boldsymbol{x})$ are strong-$(\ell,\epsilon)$-locally-balanced sequences;
        \item $\mathrm{VT}^{(k)}(\boldsymbol{x}) \equiv a_k \pmod{3n^{k}+1}$ for $k \in \{0,1\}$;
        \item $\mathrm{VT}^{(k)}(\psi(\boldsymbol{x})) \equiv b_k \pmod{6(n+1)^{k}+1}$ for $k \in \{0,1,2\}$;
        \item $\boldsymbol{x} \in \mathcal{C}_{DS}^P$, where $\mathcal{C}_{DS}^P$ is a $P$-bounded single-deletion single-substitution correcting code constructed in Lemma \ref{lem:bounded-DS}.
    \end{itemize}
    $\mathcal{C}_{DS}$ is a single-deletion single-substitution correcting code.
\end{theorem}

\begin{remark}\label{rmk:edit}
  Since $\mathrm{VT}^{(1)}(\boldsymbol{x}) \equiv \mathrm{VT}^{(1)}(\boldsymbol{y}) \pmod{3n+1}$ for any $\boldsymbol{x}, \boldsymbol{y} \in \mathcal{C}_{DS}$, it follows by Lemma \ref{lem:levenstein} that $\mathcal{C}_{DS}$ is a single-edit correcting code.
\end{remark}

\begin{corollary}\label{cor:size}
     By choosing appropriate sequences $\boldsymbol{a},\boldsymbol{b}, \boldsymbol{c}, \boldsymbol{c}'$, the redundancy of the code constructed in Theorem \ref{thm:DS} is at most $4 \log n+ 12\log\log n+O(1)$.
\end{corollary}

\begin{IEEEproof}
    Let $\mathcal{S} \triangleq \{\boldsymbol{x} \in \Sigma^n:~\text{both $\boldsymbol{x}$ and $\psi(\boldsymbol{x})$ are strong-$(\ell,\epsilon)$-locally-balanced sequences}\}$, it follows by Corollary \ref{cor:balance} that $|\mathcal{S}| \geq 2^{n-1}$. Since $\mathcal{S}= \sqcup_{\boldsymbol{a},\boldsymbol{b},\boldsymbol{c},\boldsymbol{c}'} \mathcal{C}_{DS}(\boldsymbol{a},\boldsymbol{b},\boldsymbol{c},\boldsymbol{c}')$, it follows by the pigeonhole principle that there exists a choice of $\boldsymbol{a},\boldsymbol{b},\boldsymbol{c},\boldsymbol{c}'$ such that 
    \begin{equation*}
      |\mathcal{C}_{DS}(\boldsymbol{a},\boldsymbol{b},\boldsymbol{c},\boldsymbol{c}')| 
      \geq \frac{|\mathcal{S}|}{ \prod_{k=0}^1 (3n^{k}+1) \prod_{k=0}^2 (6(n+1)
      ^{k}+1) \prod_{k=1}^3 (3(2P)^{k})^2}= O\left(\frac{2^n}{n^4(\log n)^{12}}\right).
    \end{equation*}
    In other words, $r(\mathcal{C}_{DS}(\boldsymbol{a},\boldsymbol{b},\boldsymbol{c},\boldsymbol{c}')) \leq 4 \log n+ 12\log\log n+O(1)$, which completes the proof.
\end{IEEEproof}

For simplicity, for any sequence $\boldsymbol{x}$ and two indices $d,e$, let $\boldsymbol{x}(d,e)$ be the error-sequence received from $\boldsymbol{x}$ by deleting $x_d$ and substituting $x_e$, and let $\boldsymbol{x}(e)$ be the error-sequence received from $\boldsymbol{x}$ by substituting $x_e$.

To prove Theorem \ref{thm:DS}, we assume by contradiction that there are two distinct sequences $\boldsymbol{x}$ and $\boldsymbol{y}$ in $\mathcal{C}_{DS}$ such that the intersection of their single-deletion single-substitution balls is non-empty. 
In other words, there are indices $d_x, e_x, d_y, e_y$ such that $\boldsymbol{x}(d_x,e_x)= \boldsymbol{y}(d_y,e_y)$.
To show that $\mathcal{C}_{DS}$ is a single-deletion single-substitution correcting code, our strategy is to demonstrate that $\boldsymbol{x}$ and $\boldsymbol{y}$ do not satisfy some of the constraints imposed in Theorem \ref{thm:DS}.

\subsection{Claims}

We start with some simple but useful claims regarding $\boldsymbol{x}$ and $\boldsymbol{y}$.
\subsubsection{The Sequences Themselves}

Since $\boldsymbol{x}$ and $\boldsymbol{y}$ are strong-locally-balanced sequences, it holds that $\boldsymbol{x}, \boldsymbol{y} \notin \{0^n, 1^n\}$. 
Then by Lemma \ref{cla:1}, it is reasonable to assume that $1\leq d_x\neq e_x, d_y\neq e_y\leq n$ and the deletion always occurs after the substitution.

\begin{lemma}[Remark 1, \cite{Song-22-arXiv-DS}]\label{lem:error-symbol}
    Suppose $\boldsymbol{x} \in \Sigma^n \setminus \{0^n, 1^n\}$ suffers one deletion and one substitution and the resultant sequence is $\boldsymbol{z}$, then the values of the deleted and substituted symbols can be determined by $\mathrm{VT}^{(0)}(\boldsymbol{x}) \pmod{4}$ and $\mathrm{VT}^{(0)}(\boldsymbol{z})$.
\end{lemma}

\begin{claim}\label{cla:2}
  It is reasonable to assume that $x_{d_x}= y_{d_y}$ and $x_{e_x}= y_{e_y}$.
\end{claim}

\begin{IEEEproof}
    Combining Lemma \ref{lem:error-symbol} with $\mathrm{VT}^{(0)}(\boldsymbol{x}) \equiv \mathrm{VT}^{(0)}(\boldsymbol{y}) \pmod{4}$ and $\boldsymbol{x}(d_x,e_x)= \boldsymbol{y}(d_y,e_y)$, the claim holds.
\end{IEEEproof}
For simplicity, let $\delta_x= 1$ if $e_x \in [d_x+1,d_y]$ and $\delta_x= 0$ otherwise, and let $\delta_y= 1$ if $e_y \in [d_x,d_y-1]$ and $\delta_y= 0$ otherwise. 
When $d_x\leq d_y$, due to the properties of substitutions, it is routine to check that $\boldsymbol{x}(d_x,e_y+\delta_y)= \boldsymbol{y}(d_y,e_x-\delta_x)$ if $\boldsymbol{x}(d_x,e_x)= \boldsymbol{y}(d_y,e_y)$.

\begin{claim}\label{cla:3}
  It is reasonable to assume that $d_x \leq d_y$ and $e_x \leq e_y$.
\end{claim}

\begin{IEEEproof}
  Since there is no difference between $\{\boldsymbol{x},\boldsymbol{y}\}$ and $\{\boldsymbol{y},\boldsymbol{x}\}$, we can always assume $d_x \leq d_y$, then $\boldsymbol{x}(d_x,e_y+\delta_y)= \boldsymbol{y}(d_y,e_x-\delta_x)$ if $\boldsymbol{x}(d_x,e_x)= \boldsymbol{y}(d_y,e_y)$.
  When $e_x> e_y$, one of the following holds.
  \begin{itemize}
    \item $e_y+\delta_y \leq e_x- \delta_x$. In this case, $\boldsymbol{x}(d_x,e_y+\delta_y)= \boldsymbol{y}(d_y,e_x-\delta_x)$ with $d_x \leq d_y$ and $e_y+\delta_y \leq e_x- \delta_x$, which satisfies our assumptions.
    \item $e_y+\delta_y > e_x- \delta_x$. In this case, $\delta_x= \delta_y= 1$ and $e_x= e_y+1$. It is routine to check that $\boldsymbol{x}_{[n] \setminus \{d_x\}}= \boldsymbol{y}_{[n]\setminus \{d_y\}}$, which contradicts the fact that $\mathcal{C}_{DS}$ is a single-edit correcting code.
  \end{itemize}
   Therefore, it is reasonable to assume $e_x \leq e_y$.
\end{IEEEproof}

\begin{claim}
    It is reasonable to assume $e_x \neq e_y+\delta_y$ and $e_x-\delta_x \neq e_y$.
\end{claim}

\begin{IEEEproof}
    By the definitions of $\delta_x$ and $\delta_y$, it is routine to check that $e_x= e_y+\delta_y$ if and only if $e_x-\delta_x= e_y$. 
    When $e_x= e_y+\delta_y$, it follows by $\boldsymbol{x}(d_x,e_x)= \boldsymbol{y}(d_y,e_y)$ that $\boldsymbol{x}_{[n] \setminus \{d_x\}}= \boldsymbol{y}_{[n] \setminus \{d_y\}}$, which contradicts the fact that $\mathcal{C}_{DS}$ is a single-edit correcting code, then the claim holds.
\end{IEEEproof}

\begin{claim}\label{cla:DSVT1}
  $\mathrm{VT}^{(k)}(\boldsymbol{x})= \mathrm{VT}^{(k)}(\boldsymbol{y})$ for $k \in \{0,1\}$.
\end{claim}

\begin{IEEEproof}
    By the definition of VT syndrome, for $k \in \{0,1\}$, one deletion in $\boldsymbol{x}$ decreases the value of the $k$-th order VT syndrome of $\boldsymbol{x}$ within the range of $0$ to $n^k$, while one substitution in $\boldsymbol{x}$ decreases the value of the $k$-th order VT syndrome of $\boldsymbol{x}$ within the range of $-n^k$ to $n^k$.
    Let $\boldsymbol{z}\triangleq \boldsymbol{x}(d_x,e_x)= \boldsymbol{y}(d_y,e_y)$, we have $\mathrm{VT}^{(k)}(\boldsymbol{x})- \mathrm{VT}^{(k)}(\boldsymbol{z}) \in [-n^k,2n^k]$ and $\mathrm{VT}^{(k)}(\boldsymbol{y})- \mathrm{VT}^{(k)}(\boldsymbol{z}) \in [-n^k,2n^k]$.
    It follows by $\mathrm{VT}^{(k)}(\boldsymbol{x}) \equiv \mathrm{VT}^{(k)}(\boldsymbol{y}) \pmod{3n^k+1}$ that $\mathrm{VT}^{(k)}(\boldsymbol{x})= \mathrm{VT}^{(k)}(\boldsymbol{y})$.
\end{IEEEproof}

For simplicity, let $\Delta^{(1)} \triangleq \mathrm{VT}^{(1)}(\boldsymbol{x})- \mathrm{VT}^{(1)}(\boldsymbol{y})$.
Now we proceed to calculate $\Delta^{(1)}$.

\begin{claim}\label{cla:DS_VT}
    $\Delta^{(1)}= \sum_{i \in [d_x+1,d_y] \setminus \{e_x, e_y+\delta_y\}} x_i
    + \sum_{i \in \{e_x, e_y+\delta_y \}} ix_i- \sum_{i \in \{e_x-\delta_x, e_y \}} iy_i
    + d_x x_{d_x}- d_y y_{d_y}$, where $y_{e_x-\delta_x}= 1-x_{e_x}$ and $x_{e_y+\delta_y}= 1- y_{e_y}$.
\end{claim}

\begin{IEEEproof}
    Since $\boldsymbol{x}(d_x,e_x)= \boldsymbol{y}(d_y,e_y)$ with $d_x \leq d_y$ and $e_x\neq e_y+\delta_y$, it is routine to check that $y_{e_x-\delta_x}= 1-x_{e_x}$, $x_{e_y+\delta_y}= 1- y_{e_y}$, and 
    \begin{equation*}
        x_i=
      \begin{cases}
        y_i, & \mbox{if } i \in [1,d_x-1] \cup [d_y+1,n] \setminus \{e_x, e_y+\delta_y\}; \\
        y_{i-1}, & \mbox{if } i \in [d_x+1,d_y] \setminus \{e_x, e_y+\delta_y\}.
      \end{cases}
    \end{equation*}
    If $e_y \notin [d_x,d_y-1]$, then $\delta_y=0$ and $e_y+\delta_y \neq d_x$.
    If $e_y \in [d_x,d_y-1]$, then $\delta_y=1$ and $e_y+\delta_y \neq d_x$.
    Combining this with $e_x \neq d_x$, we have $d_x \notin \{e_x,e_y+\delta_y\}$.
    Similarly, $d_y \notin \{e_y, e_x-\delta_x\}$.
    Then the conclusion follows.
\end{IEEEproof}

\subsubsection{The Differential Sequences}

\begin{claim}\label{cla:DSVT2}
  $\mathrm{VT}^{(k)}(\psi(\boldsymbol{x}))= \mathrm{VT}^{(k)}(\psi(\boldsymbol{y}))$ for $k \in \{0,1,2\}$.
\end{claim}

\begin{IEEEproof}
    By Observations \ref{obs:deletion} and \ref{obs:substitution}, for $k \in \{0,1,2\}$, one deletion in $\boldsymbol{x}$ decreases the value of the $k$-th order VT syndrome of $\psi(\boldsymbol{x})$ within the range of $0$ to $2(n+1)^k$, while one substitution in $\boldsymbol{x}$ decreases the value of the $k$-th order VT syndrome of $\psi(\boldsymbol{x})$ within the range of $-2(n+1)^k$ to $2(n+1)^k$.
    Let $\boldsymbol{z}\triangleq \boldsymbol{x}(d_x,e_x)= \boldsymbol{y}(d_y,e_y)$, we have $\mathrm{VT}^{(k)}(\psi(\boldsymbol{x}))- \mathrm{VT}^{(k)}(\psi(\boldsymbol{z})) \in [-2(n+1)^k,4(n+1)^k]$ and $\mathrm{VT}^{(k)}(\psi(\boldsymbol{y}))- \mathrm{VT}^{(k)}(\psi(\boldsymbol{z})) \in [-2(n+1)^k,4(n+1)^k]$.
    It follows by $\mathrm{VT}^{(k)}(\psi(\boldsymbol{x})) \equiv \mathrm{VT}^{(k)}(\psi(\boldsymbol{y})) \pmod{6(n+1)^k+1}$ that $\mathrm{VT}^{(k)}(\psi(\boldsymbol{x}))= \mathrm{VT}^{(k)}(\psi(\boldsymbol{y}))$.
\end{IEEEproof}

For simplicity, for $k \in \{1,2\}$, let $\Delta^{(k)}_{\psi} \triangleq \mathrm{VT}^{(k)}(\psi(\boldsymbol{x}))- \mathrm{VT}^{(k)}(\psi(\boldsymbol{y}))$.
Now we proceed to calculate $\Delta^{(k)}_{\psi}$.

\begin{claim}\label{cla:Delta_psi}
    For $k \in \{1,2\}$, let $\zeta_s^{(k)} \triangleq \sum_{i \in [d_x+2,d_y]} i^{k-1} \psi(x(e_x))_i$, $\zeta_e^{(k)} \triangleq \eta(x)_e^{(k)}- \eta(y)_e^{(k)}$ where $\eta(x)_e^{(k)}= c_{e_x}^{(k)} (2\psi(x)_{e_x}-1)+ c_{e_x+1}^{(k)} (2\psi(x)_{e_x+1}-1)$ and $\eta(y)_e^{(k)}= c_{e_y}^{(k)} (2\psi(y)_{e_y}-1)+ c_{e_y+1}^{(k)} (2\psi(y)_{e_y+1}-1)$, and $\zeta_d^{(k)} \triangleq \eta(x)_d^{(k)}- \eta(y)_d^{(k)}$, where $\eta(x)_d^{(k)} \triangleq c_{d_x}^{(k)} \psi(x(e_x))_{d_x}+ c_{d_x+1}^{(k)} \psi(x(e_x))_{d_x+1}- c_{d_x}^{(k)} \psi(y(e_y))_{d_x}$ and $\eta(y)_d^{(k)} \triangleq c_{d_y}^{(k)} \psi(y(e_y))_{d_y}+ c_{d_y+1}^{(k)} \psi(y(e_y))_{d_y+1}- c_{d_y+1}^{(k)} \psi(x(e_x))_{d_y+1}$, then $\Delta^{(k)}_{\psi}= \zeta_s^{(k)}+ \zeta_d^{(k)}+ \zeta_e^{(k)}$.
\end{claim}

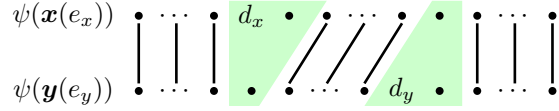
\begin{figure*}
\center
\begin{tikzpicture}[line width=1pt]   
  \fill[fill=green!20] (-1.3,1.2) -- (0,1.2) -- (-0.9,-0.2) -- (-1.3,-0.2);
  
  \fill[fill=green!20] (1.4,1.2) -- (1.8,1.2) -- (1.8,-0.2) -- (0.5,-0.2);
  
  \draw (-3.5,1) node []{$\psi(\boldsymbol{x}(e_x))$}
        (-2,1)   node []{$\cdots$}
        (-1,1)    node []{$d_x$}
        (0.5,1)    node []{$\cdots$}
        (2.5,1)    node []{$\cdots$}
        
        (-3.5,0) node []{$\psi(\boldsymbol{y}(e_y))$}
        (-2,0)   node []{$\cdots$}
        (0,0)    node []{$\cdots$}
        (1,0)    node []{$d_y$}
        (2.5,0)    node []{$\cdots$};

  \filldraw       (-2.5,1)  circle [radius=1pt]
                  (-1.5,1)  circle [radius=1pt]
                  (0,1)     circle [radius=1pt]
                  (-0.5,1)  circle [radius=1pt]
                  (1,1)  circle [radius=1pt]
                  (1.5,1)  circle [radius=1pt]
                  (2,1)  circle [radius=1pt]
                  (3,1)     circle [radius=1pt]
                  (-2.5,0)  circle [radius=1pt]
                  (-1.5,0)  circle [radius=1pt]
                  (-1,0)  circle [radius=1pt]
                  (-0.5,0)     circle [radius=1pt]
                  (0.5,0)  circle [radius=1pt]
                  (1.5,0)  circle [radius=1pt]
                  (2,0)  circle [radius=1pt]
                  (3,0)     circle [radius=1pt];

  \draw        (-2.5,0.9)--(-2.5,0.1)
               (-2,0.9)--(-2,0.1)
               (-1.5,0.9)--(-1.5,0.1)
               (0,0.9)--(-0.5,0.1)
               (0.5,0.9)--(0,0.1)
               (1,0.9)--(0.5,0.1)
               (2,0.9)--(2,0.1)
               (2.5,0.9)--(2.5,0.1)
               (3,0.9)--(3,0.1);
\end{tikzpicture}
\caption{Illustrations of $\psi(\boldsymbol{x}(e_x))$ and $\psi(\boldsymbol{y}(e_y))$ when $\boldsymbol{x}(e_x)_{[n] \setminus \{d_x\}}= \boldsymbol{y}(e_y)_{[n] \setminus \{d_y\}}$, where $d_x<d_y$. Each pair of bits connected by a solid line are of equal value, while those connected by a green undertone signify a connection resulted from deletions.}
\label{fig:DS}
\end{figure*}

\begin{IEEEproof}
    Recall that $\boldsymbol{x}(e_x)$ represents the sequence obtained from $\boldsymbol{x}$ by substituting $x_{e_x}$. It follows by Observation \ref{obs:substitution} that 
    \begin{equation*}
        \psi(x(e_x))_i= 
        \begin{cases}
            \psi(x)_i, & \mbox{if } i \notin \{e_x,e_x+1\} \\
            1-\psi(x)_i, & \mbox{if } i \in \{e_x,e_x+1\}
        \end{cases}.
    \end{equation*}
    We may compute
    \begin{align*}
        \mathrm{VT}^{(k)}(\psi(\boldsymbol{x}))- \mathrm{VT}^{(k)}(\psi(\boldsymbol{x}(e_x)))
        &= c_{e_x}^{(k)} (2\psi(x)_{e_x}-1)+ c_{e_x+1}^{(k)} (2\psi(x)_{e_x+1}-1)= \eta(x)_e^{(k)}.
    \end{align*}
    Similarly, we may compute $\mathrm{VT}^{(k)}(\psi(\boldsymbol{y}))- \mathrm{VT}^{(k)}(\psi(\boldsymbol{y}(e_y)))= \eta(y)_e^{(k)}$.

    Recall that $\boldsymbol{x}(d_x,e_x)$ is obtained from $\boldsymbol{x}(e_x)$ by deleting its $d_x$-th bit and $\boldsymbol{y}(d_y,e_y)$ is obtained from $\boldsymbol{y}(e_y)$ by deleting its $d_y$-th bit. Since $\boldsymbol{x}(d_x,e_x)= \boldsymbol{y}(d_y,e_y)$, it follows by Observation \ref{obs:deletion} and Figure \ref{fig:DS} that $\psi(x(e_x))_{d_x} \oplus \psi(x(e_x))_{d_x+1}= \psi(y(e_y))_{d_x}$, $\psi(y(e_y))_{d_y} \oplus \psi((e_y))_{d_y+1}= \psi(x(e_x))_{d_y+1}$, and 
    \begin{equation*}
        \psi(x(e_x))_i=
      \begin{cases}
        \psi(y(e_x))_i, & \mbox{if } i \in [1,d_x-1] \cup [d_y+2,n+1]; \\
        \psi(y(e_x))_{i-1}, & \mbox{if } i \in [d_x+2,d_y].
      \end{cases}
    \end{equation*}
    Now we may compute
    \begin{align*}
        \mathrm{VT}^{(k)}(\psi(\boldsymbol{x}(e_x)))- \mathrm{VT}^{(k)}(\psi(\boldsymbol{y}(e_y)))
        &= c_{d_x}^{(k)} \psi(x(e_x))_{d_x}+ c_{d_x+1}^{(k)} \psi(x(e_x))_{d_x+1}- c_{d_x}^{(k)} \psi(y(e_y))_{d_x} \\
        &~~+ c_{d_y+1}^{(k)} \psi(x(e_x))_{d_y+1} -c_{d_y}^{(k)} \psi(y(e_y))_{d_y}- c_{d_y+1}^{(k)} \psi(y(e_y))_{d_y+1}\\ 
        &~~~~+ \sum_{i \in [d_x+2,d_y]} i^{k-1} \psi(x(e_x))_i \\
        &= \eta(x)_d^{(k)}- \eta(y)_d^{(k)}+ \zeta_s^{(k)}= \zeta_d^{(k)}+ \zeta_s^{(k)}.
    \end{align*}
    
    Therefore, we obtain
    \begin{equation}\label{eq:VT}
    \begin{aligned}
      \Delta^{(k)}_{\psi}
      &\triangleq \mathrm{VT}^{(k)}(\psi(\boldsymbol{x}))- \mathrm{VT}^{(k)}(\psi(\boldsymbol{y})) \\
      &= \mathrm{VT}^{(k)}(\psi(\boldsymbol{x}))- \mathrm{VT}^{(k)}(\psi(\boldsymbol{x}(e_x))) \\
      &~~~~+ \mathrm{VT}^{(k)}(\psi(\boldsymbol{y}(e_y)))- \mathrm{VT}^{(k)}(\psi(\boldsymbol{y})) \\
      &~~~~+ \mathrm{VT}^{(k)}(\psi(\boldsymbol{x}(e_x)))- \mathrm{VT}^{(k)}(\psi(\boldsymbol{y}(e_y)))\\
      &= \eta(x)_e^{(k)}- \eta(y)_e^{(k)}+ \zeta_d^{(k)}+ \zeta_s^{(k)} \\
      &= \zeta_e^{(k)}+ \zeta_d^{(k)}+ \zeta_s^{(k)}.
    \end{aligned}
    \end{equation}
\end{IEEEproof}

\subsubsection{The Reverse Sequences}

In the proof of Theorem \ref{thm:DS}, sometimes we consider the reverse sequences to reduce unnecessary discussions, thus we need to show that $R(\boldsymbol{x})$ and $R(\boldsymbol{y})$ satisfy similar constraints as $\boldsymbol{x}$ and $\boldsymbol{y}$.

\begin{claim}
    $\mathrm{VT}^{(k)}(R(\boldsymbol{x}))= \mathrm{VT}^{(k)}(R(\boldsymbol{y}))$ if $\mathrm{VT}^{(k)}(\boldsymbol{x})= \mathrm{VT}^{(k)}(\boldsymbol{y})$ for $k \in \{0,1,2\}$.
\end{claim}

\begin{IEEEproof}
    Since for any sequence $\boldsymbol{x} \in \Sigma^n$, we have $\mathrm{VT}^{(0)}(R(\boldsymbol{x}))= \sum_{i=1}^{n} x_{n-i}= \sum_{i=1}^{n} x_i= \mathrm{VT}^{(0)}(\boldsymbol{x})$, $\mathrm{VT}^{(1)}(R(\boldsymbol{x}))= \sum_{i=1}^{n} i x_{n-i}= \sum_{i=1}^{n} (n-i) x_i= n \mathrm{VT}^{(0)}(\boldsymbol{x})- \mathrm{VT}^{(1)}(\boldsymbol{x})$, and $\mathrm{VT}^{(2)}(R(\boldsymbol{x}))
    = \sum_{i=1}^{n} \sum_{j=1}^i j x_{n-i}= \sum_{i=1}^{n} \sum_{j=1}^{n-i} j x_{i}= \sum_{i=1}^{n} \left(\binom{n+1}{2}-i (n+1) + \binom{i+1}{2}\right) x_i= \binom{n+1}{2}\mathrm{VT}^{(0)}(\boldsymbol{x})- (n+1)\mathrm{VT}^{(1)}(\boldsymbol{x})+ \mathrm{VT}^{(2)}(\boldsymbol{x})$, then the conclusion follows.
\end{IEEEproof}

\begin{claim}
    $\mathrm{VT}^{(k)}(\psi(R(\boldsymbol{x})))= \mathrm{VT}^{(k)}(\psi(R(\boldsymbol{y})))$ if $\mathrm{VT}^{(k)}(\psi(\boldsymbol{x}))= \mathrm{VT}^{(k)}(\psi(\boldsymbol{y}))$ for $k \in \{0,1,2\}$.
\end{claim}

\begin{IEEEproof}
    Notice that $\psi(\boldsymbol{x})= (x_1,x_1 \oplus x_2, \ldots, x_{n-1} \oplus x_n, x_n)$, it is routine to check that $\psi(R(\boldsymbol{x}))=R(\psi(\boldsymbol{x}))= (x_n,x_n \oplus x_{n-1}, \ldots, x_2 \oplus x_1, x_1)$, then the claim holds.
\end{IEEEproof}

\begin{claim}\label{cla:reverse}
  $R(\boldsymbol{x})$ and $R(\boldsymbol{y})$ satisfy similar constraints as $\boldsymbol{x}$ and $\boldsymbol{y}$.
\end{claim}

\begin{IEEEproof}
    Since for any strong-$(\ell,\epsilon)$-locally-balanced sequence, its reverse is also a strong-$(\ell,\epsilon)$-locally-balanced sequence. 
    Notice that $R(\boldsymbol{x})(n+1-d_x,n+1-e_x)= R(\boldsymbol{y})(n+1-d_y,n+1-e_y)$ when $\boldsymbol{x}(d_x,e_x)= \boldsymbol{y}(d_y,e_y)$, it follows that $R(\mathcal{C}) \triangleq \{R(\boldsymbol{x}): \boldsymbol{x} \in \mathcal{C}\}$ is a $P$-bounded single-deletion single-substitution correcting code when $\mathcal{C}$ is a $P$-bounded single-deletion single-substitution correcting code.
    Combining this with the above two claims, the conclusion follows.
\end{IEEEproof}

\subsection{Proof of Theorem \ref{thm:DS}}

Assume there are two distinct sequences $\boldsymbol{x}$ and $\boldsymbol{y}$ in $\mathcal{C}_{DS}$ such that the intersection of their single-deletion single-substitution balls is non-empty. 
By Lemma \ref{cla:1} and Claim \ref{cla:3}, it holds that there are indices $d_x \neq e_x, d_y \neq e_y \in [n]$ with $d_x \leq d_y$ and $e_x\leq e_y$ such that $\boldsymbol{x}(d_x,e_x)= \boldsymbol{y}(d_y,e_y)$.
Then by Claim \ref{cla:2}, the deleted and substituted symbols can be determined.
Therefore, we have $x_{e_x}= y_{e_y}$ and $x_{d_x}= y_{d_y}$.

\begin{lemma}\label{lem:1}
  If $x_{d_x}=x_{e_x}$, then $\boldsymbol{x}= \boldsymbol{y}$.
\end{lemma}

\begin{IEEEproof}
    When $x_{d_x}= x_{e_x}= 1$, it follows by Claim \ref{cla:DS_VT} that $\mathrm{VT}^{(1)}(\boldsymbol{x})- \mathrm{VT}^{(1)}(\boldsymbol{y})= \sum_{i \in [d_x+1,d_y] \setminus \{e_x, e_y+\delta_y\}} x_i+ e_x- e_y+ d_x-d_y \leq  e_x- e_y \leq 0$. Since by Claim \ref{cla:DSVT1} $\mathrm{VT}^{(1)}(\boldsymbol{x})- \mathrm{VT}^{(1)}(\boldsymbol{y})= 0$, it follows that $e_x= e_y \notin [d_x,d_y]$ and $\boldsymbol{x}_{[d_x,d_y]}= \boldsymbol{y}_{[d_x,d_y]}= 1^{d_y-d_x+1}$, i.e., $\boldsymbol{x}= \boldsymbol{y}$.
    
    Similarly, when $x_{d_x}= x_{e_x}= 0$, it follows by Claim \ref{cla:DS_VT} that $\mathrm{VT}^{(1)}(\overline{\boldsymbol{x}})- \mathrm{VT}^{(1)}(\overline{\boldsymbol{y}})= \sum_{i \in [d_x+1,d_y] \setminus \{e_x, e_y+\delta_y\}} \overline{x}_i+ e_x- e_y+ d_x-d_y \leq  e_x- e_y \leq 0$. Since by Claims \ref{cla:complement} and \ref{cla:DSVT1} $\mathrm{VT}^{(1)}(\overline{\boldsymbol{x}})- \mathrm{VT}^{(1)}(\overline{\boldsymbol{y}})= 0$, it follows that $e_x= e_y \notin [d_x,d_y]$ and $\boldsymbol{x}_{[d_x,d_y]}= \boldsymbol{y}_{[d_x,d_y]}= 0^{d_y-d_x+1}$, i.e., $\boldsymbol{x}= \boldsymbol{y}$.
\end{IEEEproof}

\begin{lemma}\label{lem:0}
If $x_{d_x} \neq x_{e_x}$, then $e_y- e_x \in \{\mathrm{VT}^{(0)}(\boldsymbol{x}_{[d_x+1, d_y] \setminus \{e_x, e_y+\delta_y\}}), \mathrm{VT}^{(0)}(\overline{\boldsymbol{x}}_{[d_x+1, d_y] \setminus \{e_x, e_y+\delta_y\}})\}$.
\end{lemma}

\begin{IEEEproof}
    When $x_{d_x}= 0, x_{e_x}= 1$, it follows by Claim \ref{cla:DS_VT} that $\mathrm{VT}^{(1)}(\boldsymbol{x})- \mathrm{VT}^{(1)}(\boldsymbol{y})= e_x- e_y+ \sum_{i \in [d_x+1,d_y] \setminus \{e_x, e_y+\delta_y\}} x_i$. Since by Claim \ref{cla:DSVT1} $\mathrm{VT}^{(1)}(\boldsymbol{x})- \mathrm{VT}^{(1)}(\boldsymbol{y})= 0$, it follows that $e_y- e_x= \mathrm{VT}^{(0)}(\boldsymbol{x}_{[d_x+1, d_y] \setminus \{e_x, e_y+\delta_y\}})$.
    
    Similarly, when $x_{d_x}= 1, x_{e_x}= 0$, it follows by Claim \ref{cla:DS_VT} that $\mathrm{VT}^{(1)}(\overline{\boldsymbol{x}})- \mathrm{VT}^{(1)}(\overline{\boldsymbol{y}})= e_x- e_y+ \sum_{i \in [d_x+1,d_y] \setminus \{e_x, e_y+\delta_y\}} \overline{x}_i$. Since by Claims \ref{cla:complement} and \ref{cla:DSVT1} $\mathrm{VT}^{(1)}(\overline{\boldsymbol{x}})- \mathrm{VT}^{(1)}(\overline{\boldsymbol{y}})= 0$, it follows that $e_y- e_x= \mathrm{VT}^{(0)}(\overline{\boldsymbol{x}}_{[d_x+1, d_y] \setminus \{e_x, e_y+\delta_y\}})$.
\end{IEEEproof}

\begin{lemma}\label{lem:size}
Following Lemma \ref{lem:0}, when we consider the size relationship between $d_x, d_y, e_x, e_y$, it is sufficient to consider three cases: namely $e_x \leq e_y < d_x \leq d_y$, $e_x < d_x \leq e_y < d_y$, and $d_x < e_x \leq e_y < d_y$.
Moreover,
\begin{itemize}
  \item when $e_x \leq e_y < d_x \leq d_y$, it is reasonable to assume that $d_y-e_x \geq P$;
  \item when $e_x < d_x \leq e_y < d_y$, it is reasonable to assume that $d_y-d_x \geq \frac{1}{2}P$ and $(\frac{1}{2}-\epsilon)(d_y-d_x)-2 \leq e_y-e_x \leq (\frac{1}{2}+\epsilon)(d_y-d_x)$;
  \item when $d_x < e_x \leq e_y < d_y$, it is reasonable to assume that $d_y-d_x \geq P$ and $(\frac{1}{2}-\epsilon)(d_y-d_x)-2 \leq e_y-e_x \leq (\frac{1}{2}+\epsilon)(d_y-d_x)$.
\end{itemize}
\end{lemma}

\begin{IEEEproof}
Recall that $e_x \neq d_x$, $d_y \neq e_y$, $e_x \leq e_y$, and $d_x \leq d_y$, we can distinguish between the following six cases according to the size relationship between $d_x, d_y, e_x, e_y$:
    \begin{enumerate}
        \item $e_x \leq e_y < d_x \leq d_y$;
        \item $e_x < d_x \leq e_y < d_y$;
        \item $e_x < d_x \leq d_y < e_y$;
        \item $d_x < e_x \leq e_y < d_y$;
        \item $d_x < e_x \leq d_y < e_y$;
        \item $d_x \leq d_y < e_x \leq e_y$.
    \end{enumerate}
    Since $R(\boldsymbol{y})$ and $R(\boldsymbol{x})$ satisfy similar constraints as $\boldsymbol{x}$ and $\boldsymbol{y}$ (by Claim \ref{cla:reverse}), it follows that Cases 6 and 5 are equivalent to Cases 1 and 2, respectively.
    Moreover, Case 3 can be ruled out by Lemma \ref{lem:0} since $e_y-e_x > d_y-d_x$.
    Therefore, it is sufficient to consider Cases 1, 2, and 4. 
    Recall that $\mathcal{C}_{DS}$ is a $P$-bounded single-deletion single-substitution correcting code, the following holds.
    \begin{itemize}
      \item When $e_x \leq e_y < d_x \leq d_y$, it is reasonable to assume that $d_y-e_x \geq P$.
      \item When $e_x < d_x \leq e_y < d_y$, it is reasonable to assume $d_y- e_x \geq P$.
            Since by Lemma \ref{lem:0} $e_y- e_x \leq d_y- d_x$, it follows that $d_y-d_x \geq \frac{1}{2}(d_y-d_x+ e_y-e_x) \geq \frac{1}{2}(d_y-e_x) \geq \frac{1}{2}P$.
            Recall that both $\boldsymbol{x}$ and $\overline{\boldsymbol{x}}$ are strong-$(\ell,\epsilon)$-locally-balanced sequences, it follows by Lemma \ref{lem:0} that $(\frac{1}{2}-\epsilon)(d_y-d_x)-2 \leq e_y- e_x \leq (\frac{1}{2}+\epsilon)(d_y-d_x)$.
      \item When $d_x < e_x \leq e_y < d_y$, it is reasonable to assume $d_y- d_x \geq P$.
            Recall that both $\boldsymbol{x}$ and $\overline{\boldsymbol{x}}$ are strong-$(\ell,\epsilon)$-locally-balanced sequences, it follows by Lemma \ref{lem:0} that $(\frac{1}{2}-\epsilon)(d_y-d_x)-2 \leq e_y-e_x \leq (\frac{1}{2}+\epsilon)(d_y-d_x)$.
    \end{itemize}
\end{IEEEproof}

Now, to complete the proof of Theorem \ref{thm:DS}, it is sufficient to show that $\Delta^{(k)}_{\psi} \neq 0$ (defined in Equation (\ref{eq:VT})), for some $k \in \{1,2\}$, under the conclusions of Lemmas \ref{lem:0} and \ref{lem:size}.
To do this, we first analyze $\psi(\boldsymbol{x})$ and $\psi(\boldsymbol{y})$. 
For simplicity, let $N(\boldsymbol{x}) \triangleq (N(x)_d,N(x)_e)$ denote that one substitution in $\boldsymbol{x}$ decreases the number of ones in $\psi(\boldsymbol{x})$ by $N(x)_e$, while one deletion in $\boldsymbol{x}(e_x)$ decreases the number of ones in $\psi(\boldsymbol{x}(e_x))$ by $N(x)_d$.
By Observations \ref{obs:deletion} and \ref{obs:substitution}, $N(x)_d \in \{0,2\}$ and $N(x)_e \in \{-2,0,2\}$. 
Then the value of $N \triangleq N(x)_d+N(x)_e= N(y)_d+N(y)_e \in \{-4,-2,0,2\}$ can be determined by $\mathrm{VT}^{(0)}(\psi(\boldsymbol{x})) \pmod{7}$ and $\mathrm{VT}^{(0)}(\psi(\boldsymbol{x}(d_x,e_x)))$.
Now we distinguish between the following four cases based on the value of $N$.
Before we do it, we calculate $\eta(x)_d^{(k)}$ and $\eta(y)_d^{(k)}$ for all possible values of $N(x)_d$ and $N(y)_d$ (summarized in Table \ref{tab:deletion}), and calculate $\eta(x)_e^{(k)}$ and $\eta(y)_e^{(k)}$ for all possible values of $N(x)_e$ and $N(y)_e$ (summarized in Table \ref{tab:substitution}).
Moreover, we present the following lemma to be used later.
\begin{lemma}\label{lem:set}
    For two distinct multi-sets $A, B$ with $a \leq b$ for any $a \in A, b\in B$, if $\sum_{i \in A} 1- \sum_{i \in B} 1= 0$, i.e.,$|A|=|B|$, then $\sum_{i \in A} i- \sum_{i \in B} i< 0$.
\end{lemma}

\begin{IEEEproof}
    Since $|A|=|B|$ and $a \leq b$ for any $a \in A, b\in B$, it holds that $\sum_{i \in A} i- \sum_{i \in B} i \leq 0$, where the inequality holds for equality if and only if $A= B= \{a,a,\ldots,a\}$ for some $a$. 
    Combining this with $A \neq B$, the lemma holds. 
\end{IEEEproof}

Recall that in Claim \ref{cla:Delta_psi}, for $k\in \{1,2\}$, we have $\Delta^{(k)}_{\psi}= \zeta_s^{(k)}+ \zeta_d^{(k)}+ \zeta_e^{(k)}$ where $\zeta_s^{(k)}= \sum_{i \in [d_x+2,d_y]} i^{k-1} \psi(x(e_x))_i$, $\zeta_e^{(k)}= \eta(x)_e^{(k)}- \eta(y)_e^{(k)}$, and $\zeta_d^{(k)} \triangleq \eta(x)_d^{(k)}- \eta(y)_d^{(k)}$.

\subsubsection{\texorpdfstring{$N=2$}{}}
In this case, $N(\boldsymbol{x})= N(\boldsymbol{y})= (0,2)$. Then by Tables \ref{tab:deletion} and \ref{tab:substitution}, it is routine to check that $\zeta_d^{(1)} \in \{0,1,2\}$ and $\zeta_e^{(1)}= 2(e_y-e_x) \geq 0$. 
Notice that $\zeta_s^{(1)} \geq 0$, then
\begin{align*}
    \Delta^{(1)}_{\psi}
    &= \zeta_s^{(1)}+ \zeta_d^{(1)}+ \zeta_e^{(1)}
    \geq 0+0+0= 0.
\end{align*}
Moreover, when $\Delta^{(1)}_{\psi}= 0$, the following holds:
\begin{itemize}
  \item $\zeta_s^{(1)}= 0$, then $\psi(\boldsymbol{x}(e_x))_{[d_x+2,d_y]}= \psi(\boldsymbol{y}(e_y))_{[d_x+1,d_y-1]}= 0^{d_y-d_x-1}$;
  \item $\zeta_d^{(1)}= 0$, then $\psi(x(e_x))_{d_x}= \psi(y(e_y))_{d_x}$, $\psi(x(e_x))_{d_x+1}= 0$, $\psi(y(e_y))_{d_y}= 0$, and $\psi(x(e_x))_{d_y+1}= \psi(y(e_y))_{d_y+1}$ (by Table \ref{tab:deletion}).
\end{itemize}
In other words, when $\Delta^{(1)}_{\psi}= 0$, it follows by Figure \ref{fig:DS} that $\psi(\boldsymbol{x}(e_x))= \psi(\boldsymbol{y}(e_y))$, i.e., $\boldsymbol{x}(e_x)= \boldsymbol{y}(e_y)$. This is impossible since $\mathcal{C}_{DS}$ is a single-edit correcting code (by Remark \ref{rmk:edit}).
Therefore, $\Delta^{(1)}_{\psi} \neq 0$.

\subsubsection{\texorpdfstring{$N=-4$}{}}

In this case $N(\boldsymbol{x})= N(\boldsymbol{y})= (-2,-2)$. Then by Tables \ref{tab:deletion} and \ref{tab:substitution}, it is routine to check that $\zeta_d^{(1)}= -2(d_y-d_x)$ and $\zeta_e^{(1)}= -2(e_y-e_x)$. 
Notice that $\zeta_s^{(1)}< d_y-d_x$, then
\begin{align*}
    \Delta^{(1)}_{\psi}
    &= \zeta_s^{(1)}+ \zeta_d^{(1)}+ \zeta_e^{(1)} \\
    &< (d_y-d_x)- 2(d_y-d_x)- 2(e_y-e_x)< 0.
\end{align*}

\subsubsection{\texorpdfstring{$N=0$}{}}

In this case $N(\boldsymbol{x}), N(\boldsymbol{y}) \in \{(0,0),(-2,2)\}$. 
Next, we further distinguish between four cases according to the values of $N(\boldsymbol{x})$ and $N(\boldsymbol{y})$.

\paragraph{$N(\boldsymbol{x})= (0,0)$ and $N(\boldsymbol{y})= (0,0)$}
    By Tables \ref{tab:deletion} and \ref{tab:substitution}, it is routine to check that $\zeta_d^{(1)} \in \{0,1,2\}$ and $\zeta_e^{(1)} \in \{-2,0,2\}$. 
    Moreover, when $\Delta^{(1)}_{\psi}= 0$, it holds that $\zeta_e^{(1)} \in \{-2,0\}$ and $\zeta_s^{(1)} \leq 2$. 
    Recall that $\psi(\boldsymbol{x})$ is a strong-locally-balanced sequence, it follows by $\zeta_s^{(1)} \leq 2$ that $d_y-d_x < \ell$.
    Combining this with $e_y-e_x \leq d_y-d_x$ (by Lemma \ref{lem:0}), we obtain $e_y-e_x < \ell$.
    If $e_y \geq d_x$, we may compute $\max\{d_y-e_x,d_y-d_x\} \leq d_y-d_x+ e_y-e_x < 2 \ell< P$, this case can be forbidden since $\mathcal{C}_{DS}$ is a $P$-bounded single-deletion single-substitution correcting code.
    Therefore, it is reasonable to assume that  $e_y< d_x$.
    Now, we consider the value of $\zeta_e^{(1)}$.
    \begin{itemize}
        \item If $\zeta_e^{(1)}= 0$, then $\zeta_s^{(1)}= 0$ and $\zeta_d^{(1)}= 0$, which is impossible (the reason is the same as the case of $N(\boldsymbol{x})=N(\boldsymbol{y})= (0,2)$). 
        \item If $\zeta_e^{(1)}= -2$, we obtain $\zeta_e^{(2)}= -(e_x+1)- (e_y+1)$ by Table \ref{tab:substitution}. Below we consider the value of $\zeta_d^{(1)}$.
        \begin{itemize}
            \item If $\zeta_d^{(1)}= 0$, we obtain $\zeta_d^{(2)}= 0$ (by Table \ref{tab:deletion}) and $\zeta_s^{(2)}> 2(d_x+2)$ (by the fact that $\zeta_s^{(1)}= - \zeta_d^{(1)}- \zeta_e^{(1)}= 2$ and the definition of $\zeta_s^{(2)}$).
            \item If $\zeta_d^{(1)}= 1$, we obtain $\zeta_d^{(2)}\in \{d_x+1,d_y+1\}$ (by Table \ref{tab:deletion}) and $\zeta_s^{(2)} \geq d_x+2$ (by the fact that $\zeta_s^{(1)}= - \zeta_d^{(1)}- \zeta_e^{(1)}= 1$ and the definition of $\zeta_s^{(2)}$).
            \item If $\zeta_d^{(1)}= 2$, we obtain $\zeta_d^{(2)}= (d_x+1)+(d_y+1)$ (by Table \ref{tab:deletion}) and $\zeta_s^{(2)}= 0$ (by the fact that $\zeta_s^{(1)}= - \zeta_d^{(1)}- \zeta_e^{(1)}= 0$ and the definition of $\zeta_s^{(2)}$).
        \end{itemize}
        In all cases, we may compute $\Delta^{(2)}_{\psi}= \zeta_s^{(2)}+\zeta_d^{(2)}+\zeta_e^{(2)}> 0$.
    \end{itemize}
        
\paragraph{$N(\boldsymbol{x})= (-2,2)$ and $N(\boldsymbol{y})= (-2,2)$}

By Tables \ref{tab:deletion} and \ref{tab:substitution}, for $k\in \{1,2\}$, it is routine to check that $\zeta_d^{(k)}= -\sum_{i \in [d_x+1,d_y]} i^{k-1}- \sum_{i \in [d_x+2,d_y+1]} i^{k-1}$ and $\zeta_e^{(k)}= \sum_{i \in [e_x+1,e_y]} i^{k-1}+ \sum_{i \in [e_x+2,e_y+1]} i^{k-1}$. 
Below we consider the size relationship between $e_x,d_x,e_y,d_y$.
\begin{itemize}
    \item If $e_x \leq e_y < d_x \leq d_y$, we define the multi-sets $A \triangleq [e_x+1,e_y] \cup [e_x+2,e_y+1]$ and $B \triangleq [d_x+1,d_y+1] \cup \{i \in [d_x+2,d_y]: \psi(x(e_x))_i= 0 \}$. For $k \in \{1,2\}$, it follows that $\zeta_e^{(k)}= \sum_{i \in A} i^{k-1}$ and $\zeta_s^{(k)}+ \zeta_d^{(k)}= - \sum_{i \in B} i^{k-1}$. Therefore, 
        \begin{align*}
            \Delta^{(k)}_{\psi}
            &= \zeta_s^{(k)} + \zeta_d^{(k)}+ \zeta_e^{(k)} 
            = \sum_{i \in A} i^{k-1}- \sum_{i \in B} i^{k-1}.
        \end{align*}
        Notice that $a < b$ for any $a \in A$ and $b \in B$, it follows by Lemma \ref{lem:set} that $\Delta^{\psi}_2 \neq 0$ when $\Delta^{\psi}_1= 0$.

    \item If $e_x< d_x \leq e_y< d_y$ or $d_x< e_x \leq e_y< d_y$, it follows by Lemma \ref{lem:size} that $e_y-e_x \leq (\frac{1}{2}+\epsilon)(d_y-d_x)$ and $d_y-d_x > \ell$. 
        Since $\psi(\boldsymbol{x})$ is a strong-$(\ell,\epsilon)$-locally-balanced sequence, then  $\zeta_s^{(1)} \leq (\frac{1}{2}+\epsilon)(d_y-d_x)+2$.
        We may compute
        \begin{align*}
            \Delta^{(1)}_{\psi}
            &= \zeta_s^{(1)} + \zeta_d^{(1)}+ \zeta_e^{(1)}
            \leq \left(  \frac{1}{2}+\epsilon \right)(d_y-d_x)+2-2(d_y-d_x)+ 2\left(  \frac{1}{2}+\epsilon \right)(d_y-d_x)
            = \left( 3\epsilon- \frac{1}{2}\right) (d_y-d_x)+2< 0.
        \end{align*}
\end{itemize}

\begin{table*}
\renewcommand\arraystretch{1.5}
    \caption{The values of $\eta(x)_d^{(k)}$ and $\eta(y)_d^{(k)}$ are calculated for $k \in \{1,2\}$. In the calculation of $\eta(x)_d^{(k)}$, $N_d$ represents the possible value of $N(x)_d$, and the error represents the transformation $\psi(x(e_x))_{d_x}\psi(x(e_x))_{d_x+1} \rightarrow \psi(y(e_y))_{d_x}$. On the other hand, when calculating $\eta(y)_d^{(k)}$, $N_d$ represents the possible value of $N(y)_d$, and the error represents the transformation $\psi(y(e_y))_{d_y}\psi(y(e_y))_{d_y+1} \rightarrow \psi(x(e_x))_{d_y+1}$.}
    \centering
    \begin{tabular}{c|c|c|c}
        \hline
        $N_d$ & Error & $\eta(x)_d^{(k)}$ & $\eta(y)_d^{(k)}$ \\
        \hline

        \multirow{3}*{$0$} & $00 \rightarrow 0$ & $0$ & 0 \\
        \cline{2-4}
        
        & $10 \rightarrow 1$ & $0$ & $-(d_y+1)^{k-1}$ \\
        \cline{2-4}

        & $01 \rightarrow 1$ & $(d_x+1)^{k-1}$ & 0 \\
        \hline

        $-2$ & $11 \rightarrow 0$ & $\sum_{i \in [d_x]} i^{k-1}+ \sum_{i \in [d_x+1]} i^{k-1}$ & $\sum_{i \in [d_y]} i^{k-1}+ \sum_{i \in [d_y+1]} i^{k-1}$\\
        \hline
    \end{tabular}
    \label{tab:deletion}
\end{table*}

\begin{table*}
\renewcommand\arraystretch{1.5}
    \caption{The values of $\eta(x)_e^{(k)}$ and $\eta(y)_e^{(k)}$ are calculated for $k \in \{1,2\}$. In the calculation of $\eta(x)_e^{(k)}$, $N_e$ represents the possible value of $N(x)_e$, and the error represents the transformation $\psi(x)_{e_x}\psi(x)_{e_x+1} \rightarrow \overline{\psi(x)_{e_x}\psi(x)_{e_x+1}}$. On the other hand, when calculating $\eta(y)_e^{(k)}$, $N_e$ represents the possible value of $N(y)_e$, and the error represents the transformation $\psi(y)_{e_y}\psi(y)_{e_y+1} \rightarrow \overline{\psi(y)_{e_y}\psi(y)_{e_y+1}}$.}
    \centering
    \begin{tabular}{c|c|c|c}
        \hline
        $N_e$ & Error & $\eta(x)_e^{(k)}$ & $\eta(y)_e^{(k)}$ \\
        \hline

        $2$ & $00 \rightarrow 11$ & $-\sum_{i \in [e_x]} i^{k-1}- \sum_{i \in [e_x+1]} i^{k-1}$ & $-\sum_{i \in [e_y]} i^{k-1}- \sum_{i \in [e_y+1]} i^{k-1}$ \\
        \hline
        
        \multirow{2}*{0} & $01 \rightarrow 10$ & $(e_x+1)^{k-1}$ & $(e_y+1)^{k-1}$ \\
        \cline{2-4}

        & $10 \rightarrow 01$ & $-(e_x+1)^{k-1}$ & $-(e_y+1)^{k-1}$\\
        \hline

        -2 & $11 \rightarrow 00$ & $\sum_{i \in [e_x]} i^{k-1}+ \sum_{i \in [e_x+1]} i^{k-1}$ & $\sum_{i \in [e_y]} i^{k-1}+ \sum_{i \in [e_y+1]} i^{k-1}$ \\
        \hline
    \end{tabular}
    \label{tab:substitution}
\end{table*}

\paragraph{$N(\boldsymbol{x})= (-2,2)$ and $N(\boldsymbol{y})= (0,0)$}
By Tables \ref{tab:deletion} and \ref{tab:substitution}, for $k \in \{1,2\}$, it is routine to check that $\eta(x)_d^{(k)}= \sum_{i \in [d_x]} i^{k-1}+ \sum_{i \in [d_x+1]} i^{k-1}$, $\eta(y)_d^{(k)} \in \{0, -(d_y+1)^{k-1} \}$, $\eta(x)_e^{(k)}= -\sum_{i \in [e_x]} i^{k-1}- \sum_{i \in [e_x+1]} i^{k-1}$, and $\eta(y)_e^{(k)} \in \{-(e_y+1)^{k-1}, (e_y+1)^{k-1}\}$. 
Now we consider the size relationship between $e_x,d_x,e_y,d_y$.
\begin{itemize}
  \item If $e_x \leq e_y < d_x \leq d_y$ or $e_x< d_x \leq e_y< d_y$, by the definitions of $\zeta_s^{(1)}, \zeta_d^{(1)}, \zeta_e^{(1)}$, we may compute 
      \begin{align*}
        \Delta^{(1)}_{\psi}= \zeta_s^{(1)}+ \zeta_d^{(1)}+ \zeta_e^{(1)} \geq 0+ (2d_x+1)- (2e_x+2)> 0.
      \end{align*}

  \item If $d_x< e_x \leq e_y< d_y$, we define the set $A \triangleq \{i \in [e_x+2,d_y]: \psi(x(e_x))_i= 1 \}$ and the multi-set $B \triangleq [d_x+1,e_x] \cup \{i \in [d_x+2,e_x+1] : \psi(x(e_x))_i= 0 \}$. For $k \in \{1,2\}$, by the definitions of $\zeta_s^{(k)}, \zeta_d^{(k)}, \zeta_e^{(k)}$, we compute $\Delta_{\psi}^{(k)}$ according to the values of $\eta(y)_d^{(k)}$ and $\eta(y)_e^{(k)}$.
      \begin{itemize}
        \item If $\eta(y)_d^{(k)}= 0$ and $\eta(y)_e^{(k)}= -(e_y+1)^{k-1}$, we may compute
        \begin{align*}
            \Delta_{\psi}^{(k)}= \zeta_s^{(k)} + \zeta_d^{(k)}+ \zeta_e^{(k)}= \sum_{i \in A \cup \{e_y+1\}} i^{k-1}- \sum_{i \in B} i^{k-1}.
        \end{align*}
        Notice that $A \cup \{e_y+1\} \neq B$ and $a \geq b$ for any $a \in A \cup \{e_y+1\}$ and $b \in B$, it follows by Lemma \ref{lem:set} that $\Delta^{\psi}_2 \neq 0$ when $\Delta^{\psi}_1= 0$. 
        \item If $\eta(y)_d^{(k)}= -(d_y+1)^{k-1}$ and $\eta(y)_e^{(k)}= -(e_y+1)^{k-1}$, we may compute
        \begin{align*}
            \Delta_{\psi}^{(k)}= \zeta_s^{(k)} + \zeta_d^{(k)}+ \zeta_e^{(k)}= \sum_{i \in A \cup \{e_y+1,d_y+1\}} i^{k-1}- \sum_{i \in B} i^{k-1}.
        \end{align*}
        Notice that $A \cup \{e_y+1,d_y+1\} \neq B$ and $a \geq b$ for any $a \in A \cup \{e_y+1,d_y+1\}$ and $b \in B$, it follows by Lemma \ref{lem:set} that $\Delta^{\psi}_2 \neq 0$ when $\Delta^{\psi}_1= 0$. 
        
        \item If $\eta(y)_d^{(k)}= -(d_y+1)^{k-1}$ and $\eta(y)_e^{(k)}= (e_y+1)^{k-1}$, we may compute
        \begin{align*}
            \Delta_{\psi}^{(k)}= \zeta_s^{(k)} + \zeta_d^{(k)}+ \zeta_e^{(k)}= \sum_{i \in A} i^{k-1}- \sum_{i \in B} i^{k-1}+ (d_y+1)^{k-1}- (e_y+1)^{k-1}.
        \end{align*}
        Notice that $a > b$ for any $a \in A$ and $b \in B$, it follows by Lemma \ref{lem:set} that $\sum_{i \in A} i- \sum_{i \in B} i> 0$ when $|A|= |B|$. Therefore, we may compute $\Delta^{\psi}_2> 0$ when $\Delta^{\psi}_1= 0$.
        \item If $\eta(y)_d^{(k)}= 0$ and $\eta(y)_e^{(k)}= (e_y+1)^{k-1}$ (this case is different from the previous three cases), we may compute
        \begin{align*}
            \Delta_{\psi}^{(k)}= \zeta_s^{(k)} + \zeta_d^{(k)}+ \zeta_e^{(k)}= \sum_{i \in A} i^{k-1}- \sum_{i \in B \cup \{e_y+1\}} i^{k-1}.
        \end{align*}
        It follows by $\Delta_{\psi}^{(1)}= 0$ that $|A|= |B|+1$. 
        Since $d_y-e_x> e_y- e_x > \ell$ (by Lemma \ref{lem:size}) and $\psi(\boldsymbol{x})$ is a strong-$(\ell,\epsilon)$-locally-balanced sequence, we may compute $|A| \geq (\frac{1}{2}-\epsilon)(e_y-e_x)$.
        Let $j_1$ and $j_2$ be the smallest and second smallest elements in $A$, respectively, then $j_1 \geq e_x+2$, $j_2 \geq e_x+3 \geq d_x+1$, and $a \geq e_x+4$ for any $a \in A \setminus \{j_1,j_2\}$. 
        Notice that $a \geq b+3$ for any $a \in A \setminus \{j_1,j_2\}$ and $b \in B \setminus \{d_x+1\}$, then we may compute 
        \begin{align*}
          \Delta_{\psi}^{(2)}
          &= \sum_{i \in A} i- \sum_{i \in B \cup \{e_y+1\}} i \\
          &\geq (e_x+2)+(d_x+1)+ \sum_{i \in A \setminus \{j_1,j_2\}} i- (e_y+1)- (d_x+1)- \sum_{i \in B \setminus \{d_x+1\}} i\\
          &\geq -(e_y-e_x)+ 3(|A|-2)  \\
          &\geq -(e_y-e_x)+ 3\left( \frac{1}{2}-\epsilon \right) (e_y-e_x)-6 \\
          &= \left(  \frac{1}{2}-3\epsilon \right)(e_y-e_x)-6> 0.
        \end{align*}
      \end{itemize}

\end{itemize}

\paragraph{$N(\boldsymbol{x})= (0,0)$ and $N(\boldsymbol{y})= (-2,2)$}
By Tables \ref{tab:deletion} and \ref{tab:substitution}, for $k \in \{1,2\}$, it is routine to check that $\eta(x)_d^{(k)}=0$ if $\psi(x(e_x))_{d_x+1}= 0$ and $\eta(x)_d^{(k)}= (d_x+1)^{k-1}$ if $\psi(x(e_x))_{d_x+1}= 1$, $\eta(y)_d^{(k)}= \sum_{i \in [d_y]} i^{k-1}+ \sum_{i \in [d_y+1]} i^{k-1}$, $\eta(x)_e^{(k)} \in \{-(e_x+1)^{k-1},(e_x+1)^{k-1}\}$, and $\eta(y)_e^{(k)}= -\sum_{i \in [e_y]} i^{k-1}- \sum_{i \in [e_y+1]} i^{k-1}$.
Now we consider the size relationship between $e_x,d_x,e_y,d_y$.
\begin{itemize}
  \item If $e_x \leq e_y < d_x \leq d_y$, we may compute 
  \begin{align*}
    \Delta^{(1)}_{\psi}= \zeta_s^{(1)}+ \zeta_d^{(1)}+ \zeta_e^{(1)}\leq (d_y-d_x-1)- 2d_y+ (2e_y+2)= -d_x-d_y-1+ 2(e_y+1)< 0.
    \end{align*}
  \item If $e_x< d_x \leq e_y< d_y$, since $d_y-d_x > \ell$ (by Lemma \ref{lem:size}) and $\psi(\boldsymbol{x})$ is a strong-$(\ell,\epsilon)$-locally-balanced sequence, we may compute $\zeta_s^{(1)} \leq (\frac{1}{2}+\epsilon)(d_y-d_x)+2$. Again by Lemma \ref{lem:size} $e_y-e_x \leq (\frac{1}{2}+\epsilon)(d_y-d_x)$, we obtain
    \begin{align*}
        \Delta^{(1)}_{\psi}
        &= \zeta_s^{(1)}+ \zeta_d^{(1)}+ \zeta_e^{(1)} \\
        &\leq \left(  \frac{1}{2}+\epsilon \right)(d_y-d_x)+2- 2d_y+ (2e_y+2) \\
        &\leq \left(  \frac{1}{2}+\epsilon \right)(d_y-d_x)- 2(d_y-d_x)+ 2(e_y-e_x)+ 4 \\
        &\leq \left( \left(  \frac{1}{2}+\epsilon \right) - 2+ 2\left( \frac{1}{2}+\epsilon \right) \right)(d_y-d_x)+ 4 \\
        &= \left( 3\epsilon- \frac{1}{2} \right)(d_y-d_x) +4< 0.
    \end{align*}

  \item If $d_x< e_x \leq e_y< d_y$, we define the set $A \triangleq \{i \in [d_x+1,e_y]: \psi(x(e_x))_i= 1\}$ and the multi-set $B \triangleq [e_y+2,d_y+1] \cup \{i \in [e_y+1,d_y]: \psi(x(e_x))_i= 0\}$. Now we consider the value of $\eta(x)_e^{(k)}$.
     \begin{itemize}
       \item If $\eta(x)_e^{(k)}= (e_x+1)^{k-1}$, for $k \in \{1,2\}$, we may compute 
        \begin{align*}
            \Delta_{\psi}^{(k)}= \zeta_s^{(k)} + \zeta_d^{(k)}+ \zeta_e^{(k)}= \sum_{i \in A \cup \{e_x+1\}} i^{k-1}- \sum_{i \in B} i^{k-1}.
        \end{align*}
        Notice that $A \cup \{e_x+1\} \neq B$ and $a\leq b$ for any $a \in A \cup \{e_x+1\}$ and $b \in B$, it follows by Lemma \ref{lem:set} that $\Delta^{\psi}_2 \neq 0$ when $\Delta^{\psi}_1= 0$.
       \item If $\eta(x)_e^{(k)}= -(e_x+1)^{k-1}$, for $k \in \{1,2\}$, we may compute 
        \begin{align*}
            \Delta_{\psi}^{(k)}= \zeta_s^{(k)} + \zeta_d^{(k)}+ \zeta_e^{(k)}= \sum_{i \in A} i^{k-1}- \sum_{i \in B \cup \{e_x+1\}} i^{k-1}.
        \end{align*}
        It follows by $\Delta_{\psi}^{(1)}= 0$ that $|A|= |B|+1$. 
        Since $e_y-d_x> e_y- e_x> \ell$ (by Lemma \ref{lem:size}) and $\psi(\boldsymbol{x})$ is a strong-$(\ell,\epsilon)$-locally-balanced sequence, we may compute $|A| \geq (\frac{1}{2}-\epsilon)(e_y-d_x)-2 \geq (\frac{1}{2}-\epsilon)(e_y-e_x)-2$.
        Let $j_1$ and $j_2$ be the largest and second largest elements in $A$, respectively, then $j_1 \leq e_y$, $j_2 \leq e_y-1 \leq d_y+1$, and $a \leq e_y-2$ for any $a \in A \setminus \{j_1,j_2\}$. 
        Notice that $a \leq b-3$ for any $a \in A \setminus \{j_1,j_2\}$ and $b \in B \setminus \{d_y+1\}$, then we may compute 
        \begin{align*}
          \Delta_{\psi}^{(2)}
          &= \sum_{i \in A} i- \sum_{i \in B \cup \{e_x+1\}} i \\
          &\leq e_y+(d_y+1)+ \sum_{i \in A \setminus \{j_1,j_2\}} i- (e_x+1)- (d_y+1)- \sum_{i \in B \setminus \{d_y+1\}} i\\
          &\leq e_y-e_x- 3(|A|-2)  \\
          &\leq e_y-e_x- 3\left( \left(  \frac{1}{2}-\epsilon \right) (e_y-e_x)-4\right) \\
          &= \left(3\epsilon-\frac{1}{2}\right)(e_y-e_x)+12< 0.
        \end{align*}
     \end{itemize}

\end{itemize}

\subsubsection{\texorpdfstring{$N=-2$}{}}

In this case $N(\boldsymbol{x}), N(\boldsymbol{y}) \in \{(0,-2),(-2,0)\}$. 
Next, we further distinguish between four cases according to the values of $N(\boldsymbol{x})$ and $N(\boldsymbol{y})$.

\paragraph{$N(\boldsymbol{x})= (0,-2)$ and $N(\boldsymbol{y})= (0,-2)$}
By Tables \ref{tab:deletion} and \ref{tab:substitution}, for $k \in \{1,2\}$, it is routine to check that $\eta(x)_d^{(k)}=0$ if $\psi(x(e_x))_{d_x+1}= 0$ and $\eta(x)_d^{(k)}= (d_x+1)^{k-1}$ if $\psi(x(e_x))_{d_x+1}= 1$, $\eta(y)_d^{(k)} \in \{0, -(d_y+1)^{k-1}\}$, and $\zeta_e^{(k)}= -\sum_{i \in [e_x+1,e_y]} i^{k-1}- \sum_{i \in [e_x+2,e_y+1]} i^{k-1}$. 
Now we consider the size relationship between $e_x,d_x,e_y,d_y$.
\begin{itemize}
  \item If $e_x \leq e_y < d_x \leq d_y$, we define the multi-set $B \triangleq [e_x+1,e_y] \cup [e_x+2,e_y+1]$. Moreover, we define the set $A$ based on the value of $\eta(y)_d^{(k)}$. 
      \begin{itemize}
        \item If $\eta(y)_d^{(k)}= 0$, let $A \triangleq \{i \in [d_x+1,d_y]: \psi(x(e_x))_i= 1\}$;
        \item If $\eta(y)_d^{(k)}= -(d_y+1)^{k-1}$, let $A \triangleq \{d_y+1\} \cup \{i \in [d_x+1,d_y]: \psi(x(e_x))_i= 1\}$.
      \end{itemize}
      Then for $k \in \{1,2\}$, it is routine to check that
    \begin{align*}
      \Delta^{(k)}_{\psi}
      &= \zeta_s^{(k)} + \zeta_d^{(k)}+ \zeta_e^{(k)}
      = \sum_{i \in A} i^{k-1}- \sum_{i \in B} i^{k-1}.
    \end{align*}
    Notice that $a> b$ for any $a \in A$ and $b \in B$, it follows by Lemma \ref{lem:set} that $\Delta^{\psi}_2 \neq 0$ when $\Delta^{\psi}_1= 0$.

  \item If $e_x < d_x \leq e_y \leq d_y$ or $d_x< e_x \leq e_y \leq d_y$, it follows by Lemma \ref{lem:size} that $e_y-e_x \geq (\frac{1}{2}-\epsilon)(d_y-d_x)-2$ and $d_y-d_x >\ell$.
      Since $\psi(\boldsymbol{x})$ is a strong-$(\ell,\epsilon)$-locally-balanced sequence, then $\zeta_s^{(1)} \leq (\frac{1}{2}+\epsilon)(d_y-d_x)+2$.
      We may compute
        \begin{align*}
            \Delta^{(1)}_{\psi}
            &= \zeta_s^{(1)} + \zeta_d^{(1)}+ \zeta_e^{(1)} \\
            &\leq \left(  \frac{1}{2}+\epsilon \right)(d_y-d_x)+2+2- 2\left(  \frac{1}{2}-\epsilon \right)(d_y-d_x)+4 \\
            &= \left( 3\epsilon- \frac{1}{2}\right)(d_y-d_x)+8< 0.
        \end{align*}
\end{itemize}

\paragraph{$N(\boldsymbol{x})= (-2,0)$ and $N(\boldsymbol{y})= (-2,0)$}
By Tables \ref{tab:deletion} and \ref{tab:substitution}, it is routine to check that $\zeta_d^{(1)}= -2(d_y-d_x)$ and $\zeta_e^{(1)} \in \{-2,0,2\}$, we may compute 
\begin{align*}
    \Delta^{(1)}_{\psi}
    &= \zeta_s^{(1)}+ \zeta_d^{(1)}+ \zeta_e^{(1)}
    \leq (d_y-d_x-1)- 2(d_y-d_x)+ 2= -(d_y-d_x)+1 \leq 0.
\end{align*}
When $\Delta^{(1)}_{\psi}=0$, $d_y=d_x+1$ and $\zeta_e^{(1)}=2$.
In this case, we have $\zeta_s^{(2)}=0$, $\zeta_d^{(2)}=-2d_y-1$, and $\zeta_e^{(2)}=e_x+e_y+2$.
Since by Lemma \ref{lem:size} $e_x,e_y< d_y$, we may compute
\begin{align*}
    \Delta^{(2)}_{\psi}
    &= \zeta_s^{(2)}+ \zeta_d^{(2)}+ \zeta_e^{(2)}
    =-2d_y-1+ e_x+e_y+2<0.
\end{align*}

\paragraph{$N(\boldsymbol{x})= (0,-2)$ and $N(\boldsymbol{y})= (-2,0)$}

By Tables \ref{tab:deletion} and \ref{tab:substitution}, it is routine to check that $\zeta_d^{(1)} \in \{-2d_y-1,-2d_y\}$ and $\zeta_e^{(1)} \in \{ 2e_x, 2e_x+2\}$. 
Now we consider the size relationship between $e_x,d_x,e_y,d_y$.
\begin{itemize}
  \item If $e_x \leq e_y < d_x \leq d_y$ or $e_x< d_x \leq e_y< d_y$, notice that $\zeta_s^{(1)} \leq d_y-d_x-1$, we may compute 
      \begin{align*}
        \Delta^{(1)}_{\psi}= \zeta_s^{(1)}+ \zeta_d^{(1)}+ \zeta_e^{(1)}\leq (d_y-d_x-1)- 2d_y+ (2e_x+2)= -d_x-d_y-1+ 2(e_x+1)< 0.
      \end{align*}
  \item If $d_x< e_x \leq e_y< d_y$, by Lemma \ref{lem:size}, $d_y-d_x >\ell$ and $e_y-e_x \geq (\frac{1}{2}-\epsilon)(d_y-d_x)-2$.
      Since $\psi(\boldsymbol{x})$ is a strong-$(\ell,\epsilon)$-locally-balanced sequence, it follows that $\zeta_s^{(1)} \leq (\frac{1}{2}+\epsilon)(d_y-d_x)+2$.
      We may compute
        \begin{align*}
          \Delta_{\psi}^{(1)}
          &= \zeta_s^{(1)}+ \zeta_d^{(1)}+ \zeta_e^{(1)} \\
          &\leq \left(  \frac{1}{2}+\epsilon \right)(d_y-d_x)+2-2d_y+ 2e_x+2 \\
          &\leq \left(  \frac{1}{2}+\epsilon \right)(d_y-d_x)- 2(e_y-e_x)+2 \\
          &\leq \left(  \frac{1}{2}+\epsilon \right)(d_y-d_x)- 2\left(  \frac{1}{2}-\epsilon \right)(d_y-d_x)+6 \\
          &= \left( 3\epsilon- \frac{1}{2}\right) (d_y-d_x)+6< 0.
        \end{align*}
\end{itemize}

\paragraph{$N(\boldsymbol{x})= (-2,0)$ and $N(\boldsymbol{y})= (0,-2)$}

By Tables \ref{tab:deletion} and \ref{tab:substitution}, for $k \in \{1,2\}$, it is routine to check that $\eta(x)_d^{(k)}= \sum_{i\in [d_x]}i^{k-1}+ \sum_{i \in [d_x+1]}i^{k-1}$, $\eta(y)_d^{(k)} \in \{0,-(d_y+1)^{k-1}\}$, $\eta(x)_e^{(k)} \in \{(e_x+1)^{k-1}, -(e_x+1)^{k-1}\}$, and $\eta(y)_e^{(k)}= \sum_{i\in [e_y]}i^{k-1}+ \sum_{i \in [e_y+1]}i^{k-1}$. 
Now we consider the size relationship between $e_x,d_x,e_y,d_y$.
\begin{itemize}
  \item If $e_x \leq e_y < d_x \leq d_y$, since $\zeta_s^{(1)} \geq 0$, we may compute 
  \begin{align*}
  \Delta_{\psi}^{(1)}
  &= \zeta_s^{(1)} + \zeta_d^{(1)}+ \zeta_e^{(1)}
  \geq 0 + (2d_x+1)-2e_y-2> 0.
  \end{align*}
  \item If $e_x < d_x \leq e_y< d_y$, we define the multi-set $A \triangleq [d_x+1,e_y] \cup \{i \in [d_x+2,e_y+1]: \psi(x(e_x))_i= 0 \}$.
      Moreover, we define the multi-set $B$ according to the value of $\eta(y)_d^{(k)}$.
      \begin{itemize}
        \item If $\eta(y)_d^{(k)}= 0$, let $B \triangleq \{i \in [e_y+2,d_y] : \psi(x(e_x))_i= 1 \}$.
        \item If $\eta(y)_d^{(k)}= -(d_y+1)^{k-1}$, let $B \triangleq \{d_y+1\} \cup \{i \in [e_y+2,d_y] : \psi(x(e_x))_i= 1 \}$.
      \end{itemize}
      Now we consider the value of $\eta(x)_e^{(k)}$.
      \begin{itemize}
        \item If $\eta(x)_e^{(k)}= -(e_x+1)^{k-1}$, it is routine to check that 
            \begin{align*}
                \Delta_{\psi}^{(k)}
                &= \zeta_s^{(k)} + \zeta_d^{(k)}+ \zeta_e^{(k)}
                = -\sum_{i \in A \cup \{e_x+1\}} i^{k-1}+ \sum_{i \in B} i^{k-1}.
            \end{align*}
            Notice that $a< b$ for any $a \in A \cup \{e_x+1\}$ and $b \in B$, it follows by Lemma \ref{lem:set} that $\Delta_{\psi}^{(2)} \neq 0$ when $\Delta_{\psi}^{(1)}= 0$.
        \item If $\eta(x)_e^{(k)}= (e_x+1)^{k-1}$, it is routine to check that 
            \begin{align*}
                \Delta_{\psi}^{(k)}
                &= \zeta_s^{(k)} + \zeta_d^{(k)}+ \zeta_e^{(k)}
                = -\sum_{i \in A } i^{k-1}+ \sum_{i \in B} i^{k-1}+ (e_x+1)^{k-1}.
            \end{align*}
            It follows by $\Delta_{\psi}^{(1)}= 0$ that $|A|= |B|+1$.
            Now, we bound the value of $|B|$.
            Since by Lemma \ref{lem:size} $e_y- e_x \leq (\frac{1}{2}+ \epsilon)(d_y-d_x)$, it follows that $d_y- e_y \geq (d_y-d_x)- (e_y-e_x) \geq (\frac{1}{2}- \epsilon)(d_y-d_x) > \ell$. We may calculate $|B| \geq (\frac{1}{2}-\epsilon)(d_y-e_y)-3 \geq (\frac{1}{2}-\epsilon)^2(d_y-d_x)-3$.
            Let $j_1$ and $j_2$ be the largest and second largest elements in $A$, respectively, then $j_1 \leq e_y+1$, $j_2 \leq e_y$, and $a \leq e_y$ for any $a \in A \setminus \{j_1,j_2\}$. 
            Moreover, let $j_3$ be the smallest element in $B$, then $j_3 \geq e_y+2$ and $b \geq e_y+3$ for any $b \in B \setminus \{j_3\}$. 
            Notice that $e_y-e_x \leq (\frac{1}{2}+\epsilon)(d_y-d_x)$ and $b-a \geq 3$ for any $a \in A$ and $b\in B$, we may compute
            \begin{align*}
                \Delta_{\psi}^{(2)}
                &= -\sum_{i \in A } i+ \sum_{i \in B} i+ (e_x+1) \\
                &\geq -e_y-(e_y+1)- \sum_{i \in A \setminus \{j_1,j_2\}} i+ (e_y+2)+ \sum_{i \in B \setminus \{j_3\}} i+  (e_x+1) \\
                &\geq -(e_y-e_x)+ 3(|B|-1)+2 \\
                &\geq -\left(  \frac{1}{2}+\epsilon \right)(d_y-d_x)+ 3\left(  \frac{1}{2}-\epsilon \right)^2(d_y-d_x)-10 \\
                &= \left( \frac{1}{4}- 4\epsilon+3\epsilon^2 \right) (d_y-d_x)-10> 0.
            \end{align*}
        \end{itemize}
  \item If $d_x< e_x \leq e_y \leq d_y$, by Lemma \ref{lem:size}, $d_y-d_x > \ell$ and $e_y-e_x \geq (\frac{1}{2}-\epsilon)(d_y-d_x)-2$. Since $\psi(\boldsymbol{x})$ is a strong-$(\ell,\epsilon)$-locally-balanced sequence, it follows that $\zeta_s^{(1)} \leq (\frac{1}{2}+\epsilon)(d_y-d_x)+2$. We may compute 
      \begin{align*}
          \Delta_{\psi}^{(1)}
          &= \zeta_s^{(1)} + \zeta_d^{(1)}+ \zeta_e^{(1)} \\
          &\leq \left(  \frac{1}{2}+ \epsilon \right)(d_y-d_x)+2+ 2d_x+2-2e_y \\
          &\leq \left(  \frac{1}{2}+ \epsilon \right)(d_y-d_x) -2(e_y-e_x)+2 \\
          &\leq \left(  \frac{1}{2}+\epsilon \right)(d_y-d_x) -2\left( \frac{1}{2}-\epsilon \right) (d_y-d_x)+6 \\
          &= \left( 3\epsilon- \frac{1}{2}\right)(d_y-d_x)+6< 0.
        \end{align*}
\end{itemize}

In all cases, we have shown that $\Delta^{(k)}_{\psi} \neq 0$, for some $k \in \{1,2\}$, under the conclusions of Lemmas \ref{lem:0} and \ref{lem:size}.
Therefore, $\mathcal{C}_{DS}$ is indeed a single-deletion single-substitution correcting code.

\begin{remark}\label{rmk:DS}
    Notice that in the second and third conditions outlined in Theorem \ref{thm:DS}, when the VT syndrome is used modulo a larger number, Claims \ref{cla:DSVT1} and \ref{cla:DSVT2} still hold.
    Consequently, one can check that Theorem \ref{thm:DS} still holds when the VT syndrome is used modulo a larger number.
\end{remark}

\begin{remark}
    In Section \ref{sec:2D}, we have discussed the connection between our two-deletion correcting codes and the ones presented by Guruswami and H{\aa}stad \cite{Guruswami-21-IT-2D}. 
    However, our approach uses different techniques, which may have unique potential in constructing other types of error-correcting codes compared with the approach presented in \cite{Guruswami-21-IT-2D}. 
    One notable example is the work of Gabrys et al. \cite{Gabrys-23-IT-DS}, who used the approach developed by Guruswami and H{\aa}stad \cite{Guruswami-21-IT-2D} to construct a single-deletion single-substitution list-decodable code with a list size of two employing $(4+o(1)) \log n$ redundant bits. 
    However in this section, we use our approach to construct a single-deletion single-substitution correcting code that also employs $(4+o(1)) \log n$ redundant bits.
    This result is better than the one achieved in \cite{Gabrys-23-IT-DS}.
\end{remark}

\section{Codes for correcting two edits}\label{sec:2E}

In this section, we consider the construction of two-edit correcting codes and two-edit list-decodable codes in Subsections \ref{subsec:correcting} and \ref{subsec:list-decoding}, respectively.

\subsection{Two-Edit Correcting Codes}\label{subsec:correcting}

Recall that in Theorem \ref{thm:equiv}, we have shown that a code can correct two edits if and only if it can correct two deletions, up to two substitutions, and one deletion and up to one substitution, separately. Combining this with Theorems \ref{thm:2D}, \ref{thm:2S}, and \ref{thm:DS}, we obtain the following construction of two-edit correcting codes.

\begin{theorem}\label{thm:2E}
    Set $\epsilon= \frac{1}{18}$, $\ell= 1296 \log n$, and $P= 6(\ell+3)$.
    For fixed sequences $\boldsymbol{a}= a_0a_1a_2, \boldsymbol{b}= b_0b_1b_2$, where $a_k \in [[4n^{k}]], b_k \in [[6(n+1)^{k}]]$ for $k \in \{0,1,2\}$, we define the code $\mathcal{C}_{2E} \subseteq \Sigma^n$ in which each sequence $\boldsymbol{x}$ satisfies the following conditions:
    \begin{itemize}
        \item both $\boldsymbol{x}$ and $\psi(\boldsymbol{x})$ are strong-$(\ell,\epsilon)$-locally-balanced sequences;
        \item $\mathrm{VT}^{(k)}(\boldsymbol{x}) \equiv a_k \pmod{4n^{k}+1}$ for $k \in \{0,1,2\}$;
        \item $\mathrm{VT}^{(k)}(\psi(\boldsymbol{x})) \equiv b_k \pmod{6(n+1)^{k}+1}$ for $k \in \{0,1,2\}$;
        \item $\boldsymbol{x} \in \mathcal{C}_{2D}^P$, where $\mathcal{C}_{2D}^P$ is a $P$-bounded two-deletion correcting code constructed in Lemma \ref{lem:2D_bounded};
        \item $\boldsymbol{x} \in \mathcal{C}_{DS}^P$, where $\mathcal{C}_{DS}^P$ is a $P$-bounded single-deletion single-substitution correcting code constructed in Lemma \ref{lem:bounded-DS}.
    \end{itemize}
    $\mathcal{C}$ is a two-edit correcting code.
     Moreover, by choosing appropriate parameters, the redundancy of the code is at most $6 \log n+O(\log\log n)$.
\end{theorem}

\begin{IEEEproof}
    When the VT syndrome is used modulo a larger number, Theorems \ref{thm:2D} and \ref{thm:DS} still hold, as indicated in Remarks \ref{rmk:2D} and \ref{rmk:DS}. It follows by Theorems \ref{thm:2D}, \ref{thm:2S}, and \ref{thm:DS} that $\mathcal{C}_{2E}$ can correct two deletions, up to two substitutions, and one deletion and up to one substitution, separately.
    Then by Theorem \ref{thm:equiv}, $\mathcal{C}_{2E}$ is a two-edit correcting code.
    Finally, similar to the proof of Corollary \ref{cor:size}, there exists a choice of parameters such that $r(\mathcal{C}_{2E}) \leq 6\log n+O(\log\log n)$.
\end{IEEEproof}

\subsection{Two-Edit List-Decodable Codes}\label{subsec:list-decoding}

We now move on to the construction of two-edit list-decodable codes.
Our construction makes use of the construction of two-deletion correcting codes presented in Theorem \ref{thm:2D}, the construction of two-substitution correcting codes presented in Theorem \ref{thm:2S}, and the construction of single-deletion single-substitution list-decodable codes presented in Lemma \ref{lem:DS_L}.

\begin{theorem}\label{thm:2E_L}
    Set $\epsilon= \frac{1}{18}$, $\ell= 1296 \log n$, and $P= 6(\ell+3)$.
    For fixed sequences $\boldsymbol{a}= a_0a_1a_2, \boldsymbol{b}= b_0b_1$, where $a_0 \in [[4]], a_1 \in [[4n]], a_2 \in [[4n^{2}]], b_0 \in [[4]], b_1 \in [[4n+4]]$, we define the code $\mathcal{C}_{2E}^{L} \subseteq \Sigma^n$ in which each sequence $\boldsymbol{x}$ satisfies the following conditions:
    \begin{itemize}
        \item both $\boldsymbol{x}$ and $\psi(\boldsymbol{x})$ are strong-$(\ell,\epsilon)$-locally-balanced sequences;
        \item $\mathrm{VT}^{(k)}(\boldsymbol{x}) \equiv a_k \pmod{4n^{k}+1}$ for $k \in \{0,1,2\}$;
        \item $\mathrm{VT}^{(k)}(\psi(\boldsymbol{x})) \equiv b_k \pmod{4(n+1)^{k}+1}$ for $k \in \{0,1\}$;
        \item $\boldsymbol{x} \in \mathcal{C}_{2D}^P$, where $\mathcal{C}_{2D}^P$ is a $P$-bounded two-deletion correcting code constructed in Lemma \ref{lem:2D_bounded}.
    \end{itemize}
    $\mathcal{C}_{2E}^{L}$ is a two-edit list-decodable code with a list size of two.
    Moreover, by choosing appropriate parameters, the redundancy of the code is at most $4 \log n+O(\log\log n)$.
\end{theorem}

\begin{IEEEproof}
When the VT syndrome is used modulo a larger number, Theorem \ref{thm:2D} still holds, as indicated in Remark \ref{rmk:2D}. Additionally, a similar reasoning supports the correctness of Lemma \ref{lem:DS_L}.
Then by Theorems \ref{thm:2D} and \ref{thm:2S} and Lemmas \ref{lem:DS_L} and \ref{lem:IS_L}, $\mathcal{C}_{2E}^{L}$ is a two-deletion correcting code, a two-substitution correcting code, a single-deletion single-substitution list-decodable code with a list size of two, and also a single-insertion single-substitution list-decodable code with a list size of two.
Suppose a sequence from the code is transmitted through a two-edit channel and we receive $\boldsymbol{y}$.
\begin{itemize}
  \item If the length of $\boldsymbol{y}$ is $n+2$, we know that the original sequence suffers two insertions and we can uniquely determine the original sequence;
  \item If the length of $\boldsymbol{y}$ is $n+1$, we know that the original sequence suffers one insertion and up to one substitution and we can determine a list of size two containing the original sequence;
  \item If the length of $\boldsymbol{y}$ is $n$, it becomes indeterminate whether the type of error is one insertion and one deletion or up to two substitutions.
  \begin{itemize}
    \item Assume the type of error is one insertion and one deletion, we can decode at most one sequence belonging to $\mathcal{C}_{2E}^{L}$;
    \item Assume the type of error is up to two substitutions, we can decode at most one sequence belonging to $\mathcal{C}_{2E}^{L}$.
  \end{itemize}
  Therefore, in worst case, we obtain a list of size two containing the original sequence.
  \item If the length of $\boldsymbol{y}$ is $n-1$, we know that the original sequence suffers one deletion and up to one substitution, then we can determine a list of size two containing the original sequence;
  \item If the length of $\boldsymbol{y}$ is $n-2$, we know that the original sequence suffers two deletions, then we can uniquely determine the original sequence.
\end{itemize}
In all cases, we obtain a list of size at most two containing the original sequence.
Therefore, $\mathcal{C}_{2E}^{L}$ is a two-edit list-decodable code with a list size of two.
Finally, similar to the proof of Corollary \ref{cor:size}, there exists a choice of parameters such that $r(\mathcal{C}_{2E}^L) \leq 4 \log n+O(\log\log n)$.
\end{IEEEproof}

\section{Conclusion}\label{sec:conclusion}

In this paper, we focus on the construction of two-edit correcting codes. Firstly, we present a necessary and sufficient condition for a code to be two-edit correctable. 
Then, we use four useful tools, namely high order VT syndromes, differential sequences, strong-locally-balanced sequences, and $P$-bounded error-correcting codes, to construct two-deletion correcting codes, single-deletion single-substitution correcting codes, and two-edit correcting codes. 
To the best of our knowledge, these three types of error-correcting codes we constructed are currently the best result in terms of the principal term coefficients of redundancy.
The tools used in this paper may have significant potential in addressing the problem of constructing codes that can correct errors related to deletions. 
Further exploration of the applications of these tools would be an intriguing direction for future research.
Another interesting problem is to encode strong-$(\ell,\epsilon)$-locally-balanced codes with one bit of redundancy and smaller $\ell$ when fixing $0<\epsilon<\frac{1}{2}$.


\begin{thebibliography}{}
\bibliographystyle{IEEEtran}
\bibitem{Brakensiek-18-IT-kD}
J.~Brakensiek, V.~Guruswami, and S.~Zbarsky, ``Efficient low-redundancy codes for correcting multiple deletions,'' \emph{IEEE Trans. Inf. Theory}, vol. 64 no. 5, pp. 3403-3410, May 2018.


\bibitem{Lev-23-arXiv-CC}
D.~Bar-Lev, A.~Kobovich, O.~Leitersdorf, and E.~Yaakobi, ``Universal framework for parametric constrained coding,'' \emph{arXiv: 2304.01317}, 2023.


\bibitem{Cheng-18-FOCS-kE}
K.~Cheng, Z.~Jin, X.~Li, and K.~Wu, ``Deterministic document exchange protocols, and almost optimal binary codes for edit errors,'' in \emph{Proc. Annu. Symp. Found. Comput. Sci. (FOCS)}, Paris, France, 2018, pp. 200-211.




\bibitem{Gabrys-18-IT-D_T}
R.~Gabrys, E.~Yaakobi, and O.~Milenkovic, ``Codes in the Damerau distance for deletion and adjacent Transposition Correction,'' \emph{IEEE Trans. Inf. Theory}, vol. 64, no. 4, pp. 2550-2570, Apr. 2018.

\bibitem{Gabrys-19-IT-2D}
R.~Gabrys and F.~Sala, ``Codes correcting two deletions,'' \emph{IEEE Trans. Inf. Theory}, vol. 65, no. 2, pp. 965-974, Feb. 2019.

\bibitem{Gabrys-20-ISIT-LBCC}
R.~Gabrys, H.~M.~Kiah, A.~Vardy, E.~Yaakobi, and Y.~Zhang, ``Locally balanced constraints,'' in {\em Proc. IEEE Int. Symp. Inf. Theory (ISIT)}, Los Angeles, CA, USA, Jun. 2020, pp. 664-669.


\bibitem{Gabrys-23-IT-DS}
R.~Gabrys, V.~Guruswami, J.~Ribeiro, and K.~Wu, ``Beyond single-deletion correcting codes: substitutions and transpositions,'' \emph{IEEE Trans. Inf. Theory}, vol. 69, no. 1, pp. 169-186, Jan. 2023.

\bibitem{Guruswami-21-IT-2D}
V.~Guruswami and J.~H{\aa}stad, ``Explicit two-deletion codes with redundancy matching the existential bound,'' \emph{IEEE Trans. Inf. Theory}, vol. 67, no. 10, pp. 6384-6394, Oct. 2021.


\bibitem{Haeupler-19-FOCS-kE}
B.~Haeupler, ``Optimal document exchange and new codes for insertions and deletions,'' in \emph{Proc. Annu. Symp. Found. Comput. Sci. (FOCS)}, Baltimore, MD, USA, 2019, pp. 334-347.

\bibitem{Heckel-19-background}
R.~Heckel, G.~Mikutis, and R.~N.~Grass, ``A characterization of the DNA data storage channel,''  \emph{Scientific Reports}, vol. 9, no. 1, pp. 9663, 2019.

\bibitem{Khuat-23-2D}
T.~-H.~Khuat, H.~Park, and S.~Kim, ``New binary code design to correct one deletion and one insertion error,'' \emph{IEEE Trans. Comm.}, vol. 71, no. 7, pp. 3807-3820, Jul. 2023.

\bibitem{Levenshtein-66-SPD-1D}
V.~I.~Levenshtein, ``Binary codes capable of correcting deletions, insertions, and reversals,'' \emph{Soviet Physics Doklady}, vol. 10, no. 8, pp. 707-710, Aug. 1966.


\bibitem{Li-23-DS}
Y,~Li and F.~Farnoud, ``Linial's algorithm and systematic deletion-correcting codes,'' in {\em Proc. IEEE Int. Symp. Inf. Theory (ISIT)}, Taipei, Taiwan, Jun. 2023, pp. 2703-2707.

\bibitem{Nguyen-21-IT-LBCC}
T.~T.~Nguyen, K.~Cai,, and K.~A.~S.~Immink, ``Efficient design of subblock energy-constrained codes and sliding window-constrained codes,'' \emph{IEEE Trans. Inf. Theory}, vol. 67, no. 12, pp. 7914-7924, Dec. 2021.

\bibitem{Organick-18-background}
L.~Organick et al., ``Random access in large-scale DNA data storage,'' \emph{Nature Biotechnol.}, vol. 36, no. 3, pp. 242-248, 2018.

\bibitem{Schoeny-17-IT-BD}
C.~Schoeny, A.~Wachter-Zeh, R.~Gabrys, and E.~Yaakobi, ``Codes correcting a burst of deletions or insertions,'' \emph{IEEE Trans. Inf. Theory}, vol. 63, no. 4, pp. 1971-1985, Apr. 2017.

\bibitem{Smagloy-20-ISIT-DS}
I.~Smagloy, L.~Welter, A.~Wachter-Zeh, and E.~Yaakobi, ``Single-deletion single-substitution correcting codes,'' in {\em Proc. IEEE Int. Symp. Inf. Theory (ISIT)}, Los Angeles, CA, USA, Jun. 2020, pp. 775-780.

\bibitem{Smagloy-23-IT-DS}
I.~Smagloy, L.~Welter, A.~Wachter-Zeh, and E.~Yaakobi, ``Single-deletion single-substitution correcting codes,'' \emph{IEEE Trans. Inf. Theory}, vol. 69, no. 12, pp. 7659-7671, Dec. 2023.

\bibitem{Sima-20-IT-2D}
J.~Sima, N.~Raviv, and J.~Bruck, ``Two deletion correcting codes from indicator vectors,'' \emph{IEEE Trans. Inf. Theory}, vol. 66, no. 4, pp. 2375-2391, Apr. 2020.
 

\bibitem{Sima-20-ISIT-tD}
J.~Sima, R.~Gabrys, and J.~Bruck, ``Optimal systematic $t$-deletion correcting codes,'' in {\em Proc. IEEE Int. Symp. Inf. Theory (ISIT)}, Los Angeles, CA, USA, Jun. 2020, pp. 769-774.

\bibitem{Sima-21-IT-kD}
J.~Sima and J.~Bruck, ``On optimal $k$-deletion correcting codes,'' \emph{IEEE Trans. Inf. Theory}, vol. 67, no. 6, pp. 3360-3375, Jun. 2021.

\bibitem{Song-22-IT-DS}
W.~Song, N.~Polyanskii, K.~Cai, and X.~He, ``Systematic codes correcting multiple-deletion and multiple-substitution errors,'' \emph{IEEE Trans. Inf. Theory}, vol. 68, no. 10, pp. 6402-6416, Oct. 2022.

\bibitem{Song-22-arXiv-DS}
W.~Song, K.~Cai, and T.~T.~Nguyen, ``List-decodable codes for single-deletion single-substitution with list-size two,'' in {\em Proc. IEEE Int. Symp. Inf. Theory (ISIT)}, Espoo, Finland, Jun. 2022, pp. 1004-1009.

\bibitem{Sun-23-IT-DR}
Y.~Sun and G.~Ge, ``Correcting two-deletion with a constant number of reads,'' \emph{IEEE Trans. Inf. Theory}, vol. 69, no. 5, pp. 2969-2982, May 2023.

\bibitem{Sun-23-IT-BD}
Y.~Sun, Y.~Zhang, and G.~Ge, ``Improved constructions of permutation and multi-permutation codes correcting a burst of stable deletions'', \emph{IEEE Trans. Inf. Theory}, vol. 69, no. 7, pp. 4429-4441, Jul. 2023.

\bibitem{Sun-23-IT-BDR}
Y.~Sun, Y.~Xi, and G.~Ge, ``Sequence reconstruction under single-burst-insertion/deletion/edit
channel,'' \emph{IEEE Trans. Inf. Theory}, vol. 69, no. 7, pp. 4466-4483, Jul. 2023.

\bibitem{Tandon-16-IT-SCC}
A.~Tandon, M.~Motani, and L.~R.~Varshney, ``Subblock-constrained codes for real-time simultaneous energy and information transfer,'' \emph{IEEE Trans. Inf. Theory}, vol. 62, no. 7, pp. 4212-4227, Jul. 2016.

\bibitem{Wang-22-ISIT-LBCC}
C.~Wang, Z.~Lu, Z.~Lan, G.~Ge, and Y.~Zhang, ``Coding schemes for locally balanced constraints,'' in {\em Proc. IEEE Int. Symp. Inf. Theory (ISIT)}, Espoo, Finland, Jun. 2022, pp. 1342-1347.

\end{thebibliography}
\end{document}